\documentclass[5p,authoryear]{elsarticle}                                             
\usepackage{natbib}
\usepackage{epsfig} 
\usepackage{amsmath,amssymb,amsfonts} 
\usepackage[caption=false]{subfig}
\usepackage{float, color,csquotes}
\usepackage[dvipsnames]{xcolor}
\def\url#1{\expandafter\string\csname #1\endcsname}

\usepackage{subfloat}
\usepackage{overpic}
\usepackage{tikz}
\usetikzlibrary{shadows,fadings}

\usepackage{epstopdf}

\usepackage{enumitem}
\newcommand{\seb}[1]{%
{\leavevmode\color{black}#1}%
}

\newcommand{\tamer}[1]{%
{\leavevmode\color{blue}#1}%
}

\usepackage{xspace}
\newcommand{\MATLAB}{\textsc{Matlab}\xspace}

\usepackage{theorem}

\newcommand{\R}{{\rm I\!R}}

\def\rep#1{(\ref{#1})}
\def\scr#1{{\cal #1}}
\newtheorem{theorem}{Theorem}
\newtheorem{lemma}{Lemma}
\newtheorem{proposition}{Proposition}
\newtheorem{corollary}{Corollary}

\newtheorem{assumption}{Assumption}
\newtheorem{remark}{Remark}

\def\qed{ \rule{.08in}{.08in}}

\DeclareMathOperator*{\E}{E}

\def\qed{\hfill $\Box$}
\DeclareMathOperator{\diag}{diag}

\def\qed{ \rule{.08in}{.08in}}

\newcommand{\1}{\mathbf{1}}
\newcommand{\0}{\mathbf{0}}
\newcommand{\x}{\mathbf{x}}
\newcommand{\y}{\mathbf{y}}

\allowdisplaybreaks[1]

\begin{document}

\thispagestyle{empty}
\pagestyle{plain}

\begin{frontmatter}
\title{Modeling and Analysis of a Coupled SIS Bi-Virus Model 
\tnoteref{t2}}

\tnotetext[t2]{The work of SG and HS was supported in part by the Knut and Alice Wallenberg Foundation, Swedish Research Council under Grant 2016-00861; and of KHJ by a  Distinguished Professor Grant from the Swedish Research Council (Org: JRL, project no:  3058). The work of PEP was supported by the National Science Foundation, grant
NSF-ECCS $2032258$. Joint research of CLB and TB was supported by the National Science Foundation Grant NSF-ECCS 2032321}

\author[First]{Sebin Gracy}
\author[Second]{Philip E. Par\'e}
\author[Third]{Ji Liu}
\author[Fourth]{Henrik Sandberg}
\author[Fifth]{Carolyn L. Beck}
\author[Fourth]{Karl Henrik Johansson}
\author[Fifth]{Tamer Ba\c sar}

\address[First]{Department of Electrical Engineering and Computer Science, South Dakota School of Mines and Technology, SD, USA, sebin.gracy@sdsmt.edu} 
 
\address[Second]{School of Electrical and Computer Engineering, Purdue University, IN, USA, philpare@purdue.edu}
\address[Third]{Department of Electrical and Computer Engineering, Stony Brook University. ji.liu@stonybrook.edu}
\address[Fourth]{
 Division of Decision and Control Systems, School of Electrical Engineering and Computer Science, KTH Royal Institute of Technology, and Digital Futures, Stockholm, Sweden.   hsan@kth.se,
 kallej@kth.se}
\address[Fifth]{Coordinated Science Laboratory, University of Illinois at Urbana-Champaign.beck3@illinois.edu, basar1@illinois.edu}


\begin{abstract} 
The paper deals with the setting 
where two viruses (say  virus~1 and virus~2) 
coexist in a population, and they are not necessarily mutually exclusive, in the sense that infection due to one virus does not preclude the possibility of simultaneous infection due to the other. We develop a coupled bi-virus susceptible-infected-susceptible (SIS) model from a $4^n$-state Markov process,
where $n$ is the number of agents (i.e., individuals or subpopulation) in the population. 
We identify a sufficient condition for both 
viruses to eventually die out, and a sufficient condition for the existence, uniqueness and asymptotic stability of the endemic equilibrium of each virus. We establish a sufficient condition and multiple necessary conditions for local exponential convergence to the boundary equilibrium (i.e., one virus persists, the other one dies out) of each virus. Under mild assumptions on the healing rate, we show that there cannot exist a coexisting equilibrium  where for each node there is a 
nonzero fraction infected only by virus~1; a nonzero fraction infected only by virus~2;  but no fraction that is infected by both viruses~1 and~2. Likewise, assuming that healing rates are strictly positive, a coexisting equilibrium where for each node there is a 
nonzero fraction infected by both viruses~1 and~2, but no fraction is infected only by virus~1 (resp. virus~2) does not exist. Further, we provide a necessary condition for the existence of certain other kinds of coexisting equilibria. We show that, unlike the competitive bivirus model, the coupled bivirus model is not monotone.  Finally, we illustrate our theoretical findings using an extensive set of 
simulations.

\end{abstract}
\end{frontmatter}

\section{Introduction}
The phenomenon of spreading processes has been a key facet of human civilization. Several manifestations of this phenomenon are witnessed in the present day too,  
including the spread of opinions in social networks, 
diseases in contact networks, 
viruses in computer networks, 
products in markets, etc. Given the various ramifications of such processes, researchers across diverse disciplines such as physics \citep{newman}, ecology \citep{munster2009avian}, 
epidemiology \citep{bailey1975mathematical}, computer science \citep{wang2003epidemic}, and economics \citep{bloom2018epidemics} have devoted significant attention to the same. 
\par This paper deals with 
the spread of viruses in human contact networks. The first model to capture the spread of a virus was proposed by Daniel Bernoulli in the $18^{\text{th}}$ century to calculate the gain in life expectancy at birth if smallpox
were to be eliminated as a cause of death 
\citep{bernoulli1760essai}. As a discipline in its own right, mathematical epidemiology witnessed enormous growth in the $20^{\text{th}}$ century, with \citep{bailey1975mathematical,hethcote2000mathematics} being some of the key works.  One of the fundamental research objectives in mathematical epidemiology revolves around 
analyzing the system equilibria and determining the convergence behavior of epidemic processes in the vicinity of isolated equilibria.
Leveraging such analysis enables the design of mitigation (or eradication) strategies. 
 To this end, various models have been studied in the literature: susceptible-infected-recovered (SIR) \citep{mei2017dynamics}; susceptible-exposed-infected-recovered (SEIR) \break \citep{arcede2020accounting};  susceptible–asymptomatic–infected \break –recovered susceptible (SAIRS) \citep{rothe2020transmission,pare2020modeling};
 susceptible-infected (SI) \citep{matouk2020complex}; and 
susceptible-infected-susceptible (SIS) 
\citep{van2009virus,khanafer2016stability} being some of the notable ones. 
%

The focus of this paper is on the susceptible-infected-susceptible (SIS) model. In particular, we are interested in \emph{networked SIS models}.
 Very briefly, in a networked SIS model, a population of individuals is 
partitioned into subpopulations, called  \emph{agents} (or nodes)\footnote{Throughout this paper, the terms agents and nodes are used interchangeably.}. The interconnection between various agents can be represented by a (possibly) directed graph.
\seb{Supposing that there is a virus prevalent in the population, it can spread both
between individuals in a subpopulation and also between individuals belonging to different subpopulations. 
If none of the individuals in a subpopulation are infected, then this subpopulation is said to be in the susceptible state; otherwise, it is said to be in the infected state. Note that when a subhpopulation is in the infected state it does not mean that \emph{every} individual in this subpopulation is infected.
An individual
is either in the susceptible state or in the infected state. 
An otherwise susceptible individual getting infected is conditional on this individual coming into contact with infected neighbors and its own infection rate. An infected individual, depending on its healing rate, recovers from the infection, and becomes susceptible to the virus again.
A susceptible individual, as a consequence of coming in contact with infected neighbors and its own infection rate, gets infected with the virus.} 
 A key feature of the (networked) SIS model is that recovering from the virus does \emph{not necessarily} confer permanent immunity. 
Networked SIS models have been extensively studied in the literature; see, for instance, \citep{FallMMNP07,khanafer2016stability,pare2020modeling}.

\par 
Note that none of the aforementioned papers account for settings where multiple strains of a virus could be \emph{simultaneously} active within a population.
The dynamics in the multi-virus setting are far  richer than 
those in the single-virus setting.
 More specifically, suppose that there are two viruses, say virus~1 and virus~2, prevalent; then these viruses could be either a) competitive, e.g., leprosy and tuberculosis  \citep{sahneh2014competitive, zhang2022networked,castillo1989epidemiological}; or b) co-operative, e.g., human immunodeficiency virus 
 (HIV) and syphilis (resp. herpes simplex virus type 2 (HSV-2)) \citep{beutel2012interacting,xu2012multi,zhao2020dynamics}.
In the competitive regime, an agent can be infected either with virus~1 or with virus~2 or neither, whereas in the co-operative regime (also referred to as co-infection) an agent can be \emph{simultaneously} infected with both virus~1 and virus~2.

\par Bi-virus models that 
capture the possibility of an agent being infected with more than one virus at the same time are broadly referred to as \emph{coupled bi-virus models}. Such scenarios are extremely common during pandemics. In fact, during the current Covid-19 pandemic, there have been reports of coinfections with SARS-Cov-2 and  influenza A virus \citep{wang2020clinical}, while a $90$-year old woman in Belgium was simultaneously infected with both the alpha and beta variants of SARS-Cov-2 \citep{bbccovid}. In a similar vein, co-infections with Zika and Dengue viruses have also been reported in the past \citep{dupont2015co}. Other examples include individuals  being simultaneously infected with both tuberculosis and human immunodeficiency virus (HIV), and such coinfections pose particular challenges both from  therapeutical and diagnostic perspectives \citep{pawlowski2012tuberculosis}; with HIV and malaria \citep{alemu2013effect}; with Hepatitis B and C \citep{chu2008hepatitis}; and with chlamydia and gonorrhea \citep{creighton2003co}. The phenomenon of coinfection is also observed in animals; some examples include coinfection with different strains of foot-and-mouth disease virus in livestock \citep{arzt2021simultaneous}; with H9N2 and H7N9 Avian Influenza Viruses in poultry \citep{bhat2022coinfection}; and with different subtypes of the Hepatitis E virus in swine \citep{de2012hev}. 

 In this paper, our focus is on the development and analysis of a coupled networked SIS bi-virus model. 
The authors in \citep{beutel2012interacting} proposed such a model, but it accounted only for undirected graphs, and each of the nodes are restricted to have the same healing and infection rates with respect to each of the viruses. A probabilistic coupled bi-virus SIS model was proposed and studied in \citep{xu2012multi}, but  the authors 
treated infection and recovery from each virus as probabilities, which implies that those are constrained to stay between zero and one. Differently from \citep{xu2012multi}, the model we propose 
admits nonnegative infection rates and positive healing rates larger than one for each virus. Moreover, we provide a richer analysis of the various equilibria of the co-operative viruses regime. Recently, a coupled bi-virus SIS model that accounts for only 
a \emph{single} population node
has been proposed in \citep{zhao2020dynamics}. Overcoming this limitation, the coupled bi-virus model that we propose admits an arbitrary but finite number of nodes. The two viruses may spread through possibly different directed contact graphs. An agent could be either infected by virus~1 or by virus~2 or, at the same time, by both (i.e., viruses~1 and~2) or by neither. To better capture the possibilities with respect to simultaneous infection by both viruses~$1$ and~$2$, we introduce a coupling parameter $\epsilon^{(m)}$ ($\geq 0$), where $m=1,2$. Specifically, if $\epsilon^{(m)}>1$ for $m=1,2$ then infection with virus~1 (resp. virus~2) increases the possibility of infection with virus~2 (resp. virus~1). Such a scenario is observed with respect to infection with human immunodeficiency virus (HIV) and syphilis (resp. herpes simplex virus type 2 (HSV-2)) \citep{newman2013interacting,chen2013outbreaks}. Similarly, if $\epsilon^{(1)}>1$ and $\epsilon^{(2)}<1$ then infection with virus~2 increases the possibility of simultaneous infection with virus~1, but infection with virus~1 decreases the possibility of simultaneous infection with virus~2,  e.g., the spread of malicious pathogens  and growth of immune cells in living organisms \citep{ahn2006epidemic}. Likewise, if $\epsilon^{(m)}<1$ for $m=1,2$, then infection with virus~1 (resp. virus~2) decreases the possibility of infection with virus~2 (resp. virus~1). Such a  scenario corresponds to the simultaneous circulation of multiple strains of influenza viruses in a community, where the presence of the strain from a previous year(s) (resp. the current year) decreases the possibility of also being simultaneously infected with the strain from the current year (resp.~previous years) \citep{krauland2022impact}.
We classify the 
equilibria into the following classes: a) the healthy state (both 
viruses are eradicated),
b) single-virus endemic equilibrium (the endemic equilibrium corresponding to that of virus~1 (resp. virus~2) if only virus~1 (resp. virus~2) were prevalent in the population), 
c) boundary equilibria 
(one virus is eradicated, the other one is persistent, fraction of population infected by both is zero), and 
d) coexisting equilibria 
(both viruses simultaneously infect possibly same fractions of the population). It turns out that our model generalizes the competitive networked bi-virus model; see Remark~\ref{rem:subsume}.

A particular class of nonlinear systems is \emph{monotone dynamical systems}. Very briefly, 
a nonlinear system $\dot{x}=f(x)$ is monotone, if, for two initial states $x_0$ and $y_0$, $x_0\leq y_0$ implies $x(t) \leq y(t)$ for all $t \in \mathbb{R}_{+}$. It is well known that monotone systems, assuming they generically  have a finite number of equilibria, converge to a stable equilibrium point (assuming one exists) for almost all choices of system parameters, and that any limit cycle (if it exists) is non-attractive; see \citep{smith1988systems,hirsch1988stability}\footnote{The term \enquote{generically} is to be understood as follows: the choices of system parameters for which convergence to a stable equilibrium does not happen lies on a set of measure zero.}. In the context of epidemiology, the notion of monotone dynamical systems plays a key role in the following sense: supposing that the model governing the spread of a disease is monotone and that it has a finite number of equilibria, then the typical behavior that a policymaker will have to contend with is that of convergence to some equilibrium point (disease-free, endemic, coexistence, etc.). More pertinently, it 
says that  existence of limit cycles (i.e., the occurrence of waves of epidemic) is less likely. 
Furthermore, even if a limit cycle were to exist, it would be non-attractive \citep{ye2021convergence}.  More complicated behavior such as chaos could be definitively ruled out \citep{sontag2007monotone}. On the contrary, if the system is not monotone, then  no dynamical behavior, including chaos, can be definitively ruled out without additional analysis \citep{sontag2007monotone}.
 Note that the competitive bi-virus model is monotone \citep{ye2021convergence}. However, it is not known if the coupled bi-virus model is monotone.

Our main contributions in this paper are as follows: 
\begin{enumerate}[label=\roman*)] \itemsep -1ex
    \item 
We derive the coupled bi-virus model 
starting with a $4^n$-state  Markov 
process; see equations~\eqref{x1}-\eqref{z} in Section~\ref{model}.
    \item We provide a sufficient condition which ensures that, irrespective of the initial state of the network (i.e., healthy or sick), both 
    viruses eventually die out; see Theorem~\ref{0global=}.
    \item 
    We provide a sufficient condition for the existence, uniqueness, and asymptotic stability of a single-virus endemic equilibrium; see Theorem~\ref{1epidemic}.
    \item We identify a sufficient condition for local exponential stability of the boundary equilibria (i.e., one virus persists, and the other one dies out); see Theorem~\ref{thm:local:expo:stab:single-virus}.
    \item We show that the coupled bi-virus model is not monotone; see Theorem~\ref{thm:coupled:bivirus:not:monotone}. Consequently, one cannot use the existing tools in the literature on competitive bivirus systems, which are deeply rooted 
in monotone dynamical systems (see \citep{hirsch1988stability,smith1988systems}), to study the limiting behavior of our model.
\end{enumerate}
Additionally, we provide a necessary and sufficient condition for the healthy state to be the unique equilibrium of the system; see Corollary~\ref{cor:healthy:unique}. Assuming both viruses pervade the network, we establish a lower bound on the number of equilibria for the coupled bi-virus system; see Corollary~\ref{prop:multiple:viruses}.
We identify multiple necessary conditions for local exponential convergence to the boundary equilibria; see Proposition~\ref{prop:hatx1:local:expo:stability:nec}. 
Assuming that the healing rates are strictly positive, we show that a point in the $3n$-dimensional state space, where  for each node there is a 
nonzero fraction infected only by virus~1 (resp. virus~2) but no fraction that is infected by both viruses~1 and~2, \emph{cannot} be an equilibrium point; see Proposition~\ref{prop:no:coexistence}. Likewise,  under mild assumptions on the healing rates, a point, where for each node there is a 
nonzero fraction infected by both viruses~1 and~2, but no fraction is infected only by virus~1 (resp. virus~2), \emph{cannot} be an equilibrium; see Proposition~\ref{prop:onlyhatz:notpossible}. We establish a necessary condition for the existence of a coexisting equilibrium wherein the fraction of each node infected by only virus~2  is zero and the rest (i.e., the fraction infected by only virus~1  and the fraction infected by both viruses~1 and~2) are strictly positive; see Proposition~\ref{prop:nece:condn:nox2}. 
Finally, we identify  a 
condition that rules out a \emph{given} point in the state space as a 
coexisting equilibrium where each node has i) a fraction that is infected only by virus~1; ii) a fraction that is infected only by virus~2; and iii) a fraction that is infected by both viruses~1 and~2.; see Proposition~\ref{prop:nece:x1x2z}.




Some of the material in this paper was partially presented earlier in an American Control Conference (ACC) paper \citep{acc_coupled}; the present paper provides a more comprehensive treatment of the work, and considers a more general model. Specifically, the paper provides: 
\begin{enumerate}[label=\roman*)] \itemsep -1ex
\item	an expansion of the model to the case of \break virus-dependent coupling parameters that can be \break greater than 1. However, most of our findings rely on the assumption that $\epsilon^{(m)} \in [0,1]$ for $m=1,2$; 
\item	complete proofs of all the results; 
\item	a derivation of the coupled bi-virus model from a $4^n$-state Markov 
process; see Section~\ref{model}; 
\item	stability results for the boundary equilibria; see Theorem~\ref{thm:local:expo:stab:single-virus}; 
\item	existence results for coexisting equilibria; see Propositions~\ref{prop:no:coexistence},~\ref{prop:onlyhatz:notpossible},~\ref{prop:nece:condn:nox2} and~\ref{prop:nece:x1x2z}. 
\item a result establishing that the coupled bivirus system is not monotone; see Theorem~\ref{thm:coupled:bivirus:not:monotone}; and
\item additional illustrative simulations in Section~\ref{sec:sim}, none of which were included in \citep{acc_coupled}.
\end{enumerate}

 The paper is organized as follows. 
The derivation of the coupled bi-virus model from a $4^n$-state Markov process 
is provided in Section~\ref{model}. The problems of interest and standing assumptions 
are formally stated in Section~\ref{sec:prob:formln}. The main results for the model developed in Section~\ref{model} are split across the next three sections: analysis of the disease-free equilibrium (DFE) is given in Section~\ref{analysis}; persistence of a virus in the population is given in Section~\ref{analysis:persistence}; analysis of various coexisting equilibria are provided in Section~\ref{analysis:endemic-coexistence}; and non-monotoncity of the coupled bivirus model is shown in Section~\ref{sec:no:monotone}. The theoretical findings are illustrated in Section~\ref{sec:sim}. A summary of the results of the paper, and some questions of possible interest to the wider community are given in Section~\ref{con}.
\par We conclude this section by 
 introducing all the notations to be used in the rest of the paper.

{\em Notation:}
For any positive integer $n$, we use $[n]$ to denote the set $\{1,2,\ldots,n\}$.
We use $\0$ and $\1$ to denote the vectors whose entries all equal $0$ and $1$, respectively,
and $I$ to denote the identity matrix,
while the dimensions of the vectors and matrices can be inferred from the context.
For any vector $x\in\R^n$, we use $x^{\top}$ to denote its transpose and ${\rm diag}(x)$ or $X$ to denote the $n\times n$ diagonal matrix
whose $i$th diagonal entry equals $x_i$.
The notation $1_{a=b}$ is used as an indicator function which takes the value one if $a$ equals $b$; and zero otherwise.
For $1_{A=b}$, where $A$ is a matrix, 
the result is a binary matrix of the same dimensions as $A$ with entries $1_{a_{ij}=b}$.
For any two sets $\scr{A}$ and $\scr{B}$,
we use $\scr{A}\setminus \scr{B}$ to denote the set of elements in $\scr{A}$ but not in $\scr{B}$.
For any two real vectors $a,b\in\R^n$, we write $a\geq b$ if
$a_{i}\geq b_{i}$ for all $i\in[n]$,
$a>b$ if $a\geq b$ and $a\neq b$, and $a \gg b$ if $a_{i}> b_{i}$ for all $i\in[n]$. Likewise, for any two real matrices $A, B \in \mathbb{R}^{n_1 \times n_2}$, we write $A \geq B$ if $A_{ij} \geq B_{ij}$ for all $i \in [n_1]$, $j \in [n_2]$, and $A>B$ if $A \geq B$ and $A \neq B$. 
For a real square matrix $M$, we use $s(M)$ to denote the largest real part among the eigenvalues of~$M$, and $\rho(M)$ to denote the spectral radius, i.e., $\rho(M)=\max\{\lvert\lambda\rvert: \lambda \in \sigma(M)\}$,
where $\sigma(M)$ denotes the spectrum of $M$.

A real square matrix $A$ is said to be Metzler if all of its off-diagonal entries are nonnegative. A real square matrix $A$ is said to be a Z-matrix if all of its off-diagonal entries are nonpositive. 
A Z-matrix is an M-matrix if all its eigenvalues have nonnegative real parts. Furthermore, if an M-matrix has an eigenvalue at the origin, then we say that it is singular; if each of its eigenvalues have strictly positive parts, then we say that it is nonsingular.

\section{The Model} \label{model}



Consider two viruses spreading over
a network of $n$ agents.  
Each agent may be infected with either or both 
viruses at the same time.
Specifically, each agent can be infected if one of its neighbors is infected.
The neighbor relationships among the $n$ agents are described by an $n$-vertex directed graph. 
A directed edge from node $j$ to node $i$ means that agent~$i$ can be infected by agent $j$, i.e., agent $j$ is a neighbor of agent~$i$.
We use $\scr N_i$ to denote the set of neighbors of agent~$i$. 
The two viruses may spread through different routes in the network. 
We use $\scr{N}_i^{(m)}$ to denote the set of neighbors of agent~$i$ from which virus $m$ spreads, $m\in\{1,2\}$. Clearly, $\scr N_i^{(1)}\cup \scr N_i^{(2)} = \scr N_i$ for all $i\in[n]$.

\begin{figure}[h!]
\centering
\begin{tikzpicture}[
    knoten/.style={
      circle,
      inner sep=.35cm,
      draw},
    /schriftstueck/.code 2 args={
      \fill[red!50, opacity=#2] #1 rectangle +(.6,.7);
      \foreach \y in {0pt,2pt,4pt,6pt,8pt,10pt,12pt,14pt}
      \draw [yshift=\y, opacity=#2] #1+(0.1,0.1) -- +(0.5,0.1);
      },
    el/.style = {inner sep=4pt},
every label/.append style = {font=\Large}
    ]

  \node at (4,0) (s) [knoten, fill=black, minimum size=0.5cm, draw=black]{\color{white} \Large $S$};
  
  \node at (1,2) (i1) [knoten, fill=red, draw=black] {\Large $I^{(1)}$};




  \node at (4,4) (a) [knoten, fill=green, draw=black] {\Large $I^{(1,2)}$};


  \node at (7,2) (i2) [knoten, fill=blue, draw=black] {\Large $I^{(2)}$};



  \path  [-latex,ultra thick] (i2.north west) edge [bend right] node[el,above] { $\hspace{4ex} \epsilon^{(2)} \sum \beta_{ij}^{(1)}$ } 
  (a.east);
  \path  [-latex,ultra thick] (a.south east) edge [bend right] node[el,above] {$\delta_i^{(1)}$} (i2.west);
  \path  [-latex,ultra thick] (a.south west) edge [bend left] node[el,above] {$\delta_i^{(2)}$} (i1.east);
  \path  [-latex,ultra thick]  (i1.north east) edge [bend left] node[el,above]  {$ \epsilon^{(1)} \sum \beta_{ij}^{(2)} \hspace{6ex}$}
  (a.west);
  \path  [-latex,ultra thick] (i1.east) edge [bend left] node[el,below] {$\delta_i^{(1)}$} (s.north west);
  \path  [-latex,ultra thick] (i2.west) edge [bend right] node[el,below] {$\delta_i^{(2)}$} (s.north east) ;
  \path  [-latex,ultra thick] (s.east) edge [bend right] node[el,below] { $\hspace{4ex} \sum \beta_{ij}^{(2)}$ }
  (i2.south west);
  \path  [-latex,ultra thick] (s.west) edge [bend left] node[el,below] {$\sum \beta_{ij}^{(1)}\hspace{3ex}$}
  (i1.south east);

\end{tikzpicture}
\caption{The possible states and transitions for node $i$. State $S$ stands for being healthy but susceptible. States $I^{(1)}$, $I^{(2)}$, and $I^{(1,2)}$ stand for being infected by virus 1, virus 2, and both virus 1 and virus 2, respectively.
}\label{fig:diag}
\end{figure}
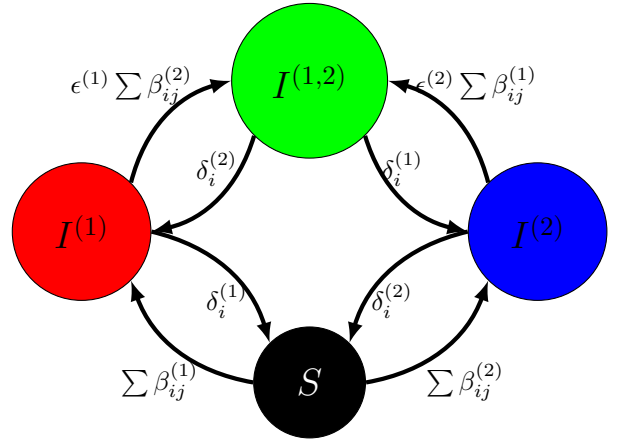

For each virus $m\in\{1,2\}$, each agent $i$ has its curing rate $\delta_i^{(m)}$ and infection rates
$\beta_{ji}^{(m)}$ when $i\in\scr N_j^{(m)}$. 
The former means that if agent $i$ is infected by virus $m$, it is cured with rate $\delta_i^{(m)}$, and the latter means that if agent $i$ is infected by virus $m$ and its neighbor $j$ is not, agent $i$ can infect agent~$j$ at rate $\beta_{ji}^{(m)}$.
The two viruses can simultaneously infect the same node, but not independently. Specifically, they are coupled in the following manner. Let $i$ and $j$ be any pair of integers in $[n]$ such that agent $j$ is a neighbor of agent~$i$.
If agent $j$ has been infected by only one virus, say virus~1, and agent $i$ is infected by the other virus, virus 2, 
then, irrespective of whether (or not) it is infected by virus 1, agent $i$ can infect agent $j$ with virus 2 at a  rate $\epsilon^{(1)} \beta_{ji}^{(2)}$.
See Figure~\ref{fig:diag} for a depiction of the model. 
It is worth noting that there is no transition link from state $I^{(1,2)}$ (infected by both  viruses) to state $S$ (healthy state), as the probability that the two viruses are cured at the same time is zero. 
We call $\epsilon^{(m)}$ the {\em coupling parameter} between the viruses, 
and assume that each $\epsilon^{(m)}$ takes a nonnegative value. 
If $\epsilon^{(m)} \in (0,1)$, that means if a node is infected with virus $m$, it is less susceptible to the other virus. 
If $\epsilon^{(m)}>1$ that means if a node is infected with virus $m$, it is more likely to become infected with the other virus.
If $\epsilon^{(m)}=1$ for all $m$, then the two viruses are independent.  
If $\epsilon^{(m)}=0$ for all $m$, then the two viruses are competitive. 
We will discuss these last two special cases in-depth in Remark~\ref{rem:subsume}. 

Let $B^{(m)}=[\beta_{ij}^{(m)}]$ for $m \in [2]$. The spread of the two viruses across the population can be represented by a two-layer graph, where the vertices of the graph correspond to the population nodes. Each layer contains a set of directed edges, $E^{(m)}$, specific to virus~$k$; there exists a  directed edge from agent $j$ to agent $i$ in $E^{(m)}$ if, assuming agent $j$ is infected with virus~$k$, it can directly infect agent~$i$. Note that there is a one-to-one correspondence between the notations $E^{(m)}$ and $B^{(m)}$ for $m \in [2]$. That is,  $(i,j)\in E^{(m)}$ if, and only if, $[B^{(m)}]_{ji}\neq 0$.
It is worth emphasizing that the two viruses may spread 
along different routes, that is, the 
layers corresponding to 
$B^{(1)}$ and $B^{(2)}$ are not necessarily the same. 
We call the layers corresponding to $B^{(1)}$ and $B^{(2)}$ as the
spreading graphs of viruses 1 and 2, respectively.

For completeness, we provide here a full description of the $4^n$-state  Markov process. 
Each state
, $Y_k(t)$, corresponds to a 
string $s$ of length $n$, where  $s_i=S$, $s_i=I^{(1)}$, $s_i=I^{(2)}$, or $s_i=I^{(1,2)}$ indicate that the $i$th agent is either susceptible, or infected with virus 1, or infected with virus 2, or infected with both viruses 1 and 2, 
 respectively. 
The 
generator
matrix \citep{norris1998markov}, $Q$, is defined by

\scriptsize
\begin{equation}\label{eq:qij}
{q}_{kl}=
\begin{cases}
\delta_i^{(1)}, & \text{ if } s_i = I^{(1)}, k = l + 4^{i-1}\\
\delta_i^{(2)}, & \text{ if } s_i = I^{(2)}, k = l + 2(4^{i-1})\\
\delta_i^{(1)}, & \text{ if } s_i = I^{(1,2)}, k = l + 4^{i-1}\\
\delta_i^{(2)}, & \text{ if } s_i = I^{(1,2)}, k = l + 2(4^{i-1})\\
\displaystyle\sum_{j=1}^n \beta^{(1)}_{ij}  (1_{s_j=I^{(1)}}+1_{s_j=I^{(1,2)}}), &\text{ if } s_i = S, k = l - 4^{i-1} \\[2.5ex]
\displaystyle\sum_{j=1}^n \beta^{(2)}_{ij}  (1_{s_j=I^{(2)}}+1_{s_j=I^{(1,2)}}), &\text{ if } s_i = S, k = l - 2(4^{i-1}) \\[2.5ex]
\epsilon^{(2)} \displaystyle\sum_{j=1}^n \beta^{(1)}_{ij}  (1_{s_j=I^{(1)}}+1_{s_j=I^{(1,2)}}), &\text{ if } s_i = I^{(2)}, k = l - 4^{i-1} \\[2.5ex]
\epsilon^{(1)} \displaystyle\sum_{j=1}^n \beta^{(2)}_{ij}  (1_{s_j=I^{(2)}}+1_{s_j=I^{(1,2)}}), &\text{ if } s_i = I^{(1)}, k = l - 2(4^{i-1}) \\[2.5ex]
-\displaystyle\sum_{j\neq l} {q}_{jl}, & \text{ if } k = l\\
0, & \text{ otherwise,}
\end{cases}
\end{equation}
\normalsize
for $i\in [n]$. Here virus 1 and virus 2 are propagating over a network whose infection rates are given by $\beta^{(1)}_{ij}$ and $\beta^{(2)}_{ij}$, respectively (nonnegative with $\beta^{(1)}_{ii} = \beta^{(2)}_{ii}=0 ,\ \forall j$), 
 $\delta_i^{(1)}$ and 
 $\delta_i^{(2)}$ are the respective healing rates of the $i$th agent, and, again, $s_i=S$, $s_i=I^{(1)}$, $s_i=I^{(2)}$, or $s_i=I^{(1,2)}$ indicate that the $i$th agent is either susceptible, or infected with virus 1, or infected with virus 2, or infected with both viruses 1 and 2, respectively. 
The state vector $y(t)$ is defined as 
\begin{equation}\label{eq:y}
         y_k(t) = Pr[Y_k(t) = k], 
\end{equation}
with $\sum_{k=1}^{4^n} y_k(t) =1$. The Markov 
process evolves as
\begin{equation}\label{eq:3n}
    \frac{dy^{\top}(t)}{dt} = y^{\top}(t){Q}.
\end{equation}


\noindent Let  $v^{(1)}_i(t) = \Pr[X_i(t) = I^{(1)}]$, $v^{(2)}_i(t) = \Pr[X_i(t) = I^{(2)}]$, and $v^{(1,2)}_i(t) = \Pr[X_i(t) = I^{(1,2)}]$,  where $X_i(t)$ is the random variable representing whether the $i$th agent is susceptible or infected with virus 1, or 2, or both. Then, for $i=\{(1),(2),(1,2)\}$
\begin{equation}\label{eq:v}
\begin{split}
    (v^i)^{\top}(t) = y^{\top}(t)M^i,
\end{split}
\end{equation}
where the $i$th columns of $M^{(1)}$, $M^{(2)}$, $M^{(1,2)}$ indicate the states in the Markov 
process where agent $i$ is infected with virus 1, virus 2, and both (all the 
strings where $s_i = I^{(1)}$, 
$s_i = I^{(2)}$, and $s_i = I^{(1,2)}$), respectively, that is,
$M^i = 1_{M=i}$ for $i\in\{(1),(2),(1,2)\}$,
where $M\in \mathbb{R}^{4^n\times n}$ has rows of lexicographically-ordered ternary numbers, bit reversed.\footnote{Matlab code: $M = fliplr(dec2base(0:(4^n)-1,4)-'0')$} 
Therefore, $v^{(1)}_i(t)$, $v^{(2)}_i(t)$, and $v^{(1,2)}_i(t)$ reflect the summation of all probabilities where $s_i = I^{(1)}$, $s_i = I^{(2)}$, and $s_i = I^{(1,2)}$. 
Note that the first state of the process,
which corresponds to $s_i =S$, the healthy state, for $\delta^{(1)}_i,\delta^{(2)}_i>0 \ \forall i$, is the absorbing or sink state of the process. 
That is, once in the healthy state, the Markov 
process will never escape it. Moreover, since the healthy state is the only absorbing state, the system will converge to it with probability one \citep{norris1998markov}. 

We derive the model in \eqref{x1i}-\eqref{zi} using a mean-field type approximation by considering the probability that node $i$ is healthy ($X_i = S$) or infected  with virus 1 ($X_i = I^{(1)}$), or virus~2 
($X_i = I^{(2)}$), or both ($X_i = I^{(1,2)}$) at time $t+\Delta t$.  
From \eqref{eq:qij}, we have
\footnotesize 
\begin{align*}
        &\Pr(X_i(t+\Delta t) = S | X_i(t) = I^{(1)}, X(t)) = \delta^{(1)}\Delta t + o(\Delta t) \\
        &\Pr(X_i(t+\Delta t) = I^{(1)} | X_i(t) = S, X(t)) = o(\Delta t) \\
        & \ \ \ \ \ \ \ \ \ \ \ \ \ \ \ \ \ \ \ \ \ \ \ \ \ \ \  +\textstyle \sum_{j=1}^n \beta^{(1)}_{ij}  (1_{X_j=I^{(1)}}+1_{X_j=I^{(1,2)}}) \Delta t    \\
        &\Pr(X_i(t+\Delta t) = S | X_i(t) = I^{(2)}, X(t)) = \delta^{(2)} \Delta t + o(\Delta t) \\
        &\Pr(X_i(t+\Delta t) = I^{(2)} | X_i(t) = S, X(t)) = o(\Delta t) \\
        & \ \ \ \ \ \ \ \ \ \ \ \ \ \ \ \ \ \ \ \ \ \ \ \ \ \ \  +  \textstyle\sum_{j=1}^n \beta^{(2)}_{ij}  (1_{X_j=I^{(2)}}+1_{X_j=I^{(1,2)}}) \Delta t \\
        &\Pr(X_i(t+\Delta t) = I^{(1)} | X_i(t) = I^{(1,2)}, X(t)) = \delta^{(2)} \Delta t + o(\Delta t) \\
        &\Pr(X_i(t+\Delta t) = I^{(2)} | X_i(t) =  I^{(1,2)}, X(t)) = \delta^{(1)}\Delta t + o(\Delta t) \\
        &\Pr(X_i(t+\Delta t) =  I^{(1,2)} | X_i(t) = I^{(1)}, X(t)) = o(\Delta t) \\
        & \ \ \ \ \ \ \ \ \ \ \ \ \ \ \ \ \ \ \ \ \ \ \ \ \ \ \  +  \epsilon^{(1)} \textstyle\sum_{j=1}^n \beta^{(2)}_{ij}  (1_{X_j=I^{(2)}}+1_{X_j=I^{(1,2)}}) \Delta t  \\
        &\Pr(X_i(t+\Delta t) =  I^{(1,2)} | X_i(t) = I^{(2)}, X(t)) = o(\Delta t) \\
        & \ \ \ \ \ \ \ \ \ \ \ \ \ \ \ \ \ \ \ \ \ \ \ \ \ \ \  +  \epsilon^{(2)} \textstyle\sum_{j=1}^n \beta^{(2)}_{ij}  (1_{X_j=I^{(1)}}+1_{X_j= I^{(1,2)}}) \Delta t   \\
        & \ \ \ \ \ \ \ \ \ \ \ \ \ \ \ \ \ \ \ \ \ \ \ \ \ \ \ \ \ \ \ \ \ \ \ \ \ \ \ \ \ \ \ \ \ \ \vdots 
    \end{align*}
    \normalsize
Letting $\Delta t$ go to zero and taking expectations of $1_{X_{i}(t)=I^{(1)}}$, $1_{X_{i}(t)=I^{(2)}}$, and $1_{X_{i}(t)=I^{(1,2)}}$ gives
\vspace{-0ex} \footnotesize
\begin{align*}
    \frac{d \E(1_{X_{i}(t)=I^{(1)}})}{dt} = - \delta^{(1)}\E(1_{X_{i}(t)=I^{(1)}}) + \delta^{(2)} \E(1_{X_{i}(t)= I^{(1,2)}}) \ \ \ \ \ \ \nonumber \\ 
    + \E \left( 1_{X_{i}(t)=S} \textstyle\sum^{n}_{j=1} \beta^{(1)}_{ij}(1_{X_{j}(t)=I^{(1)}}+1_{X_{j}(t)=I^{(1,2)}})\right) \nonumber \ \ \ \\ 
     - \E \left( 1_{X_{i}(t)=I^{(1)}} \epsilon^{(1)} \textstyle\sum^{n}_{j=1} \beta^{(2)}_{ij}(1_{X_{j}(t)=I^{(2)}}+1_{X_{j}(t)=I^{(1,2)}})\right), \nonumber 
     \\
    \frac{d \E(1_{X_{i}(t)=I^{(2)}})}{dt} = - \delta^{(2)} \E(1_{X_{i}(t)=I^{(2)}}) + \delta^{(1)}E(1_{X_{i}(t)=I^{(1,2)}})  \ \ \ \ \ \ \nonumber \\ 
    + \E \left( 1_{X_{i}(t)=S} \textstyle\sum^{n}_{j=1} \beta^{(2)}_{ij}(1_{X_{j}(t)=I^{(2)}}+1_{X_{j}(t)=I^{(1,2)}})\right) \nonumber \ \ \ \\ 
   - \E \left( 1_{X_{i}(t)=I^{(2)}} \epsilon^{(2)} \textstyle\sum^{n}_{j=1} \beta^{(1)}_{ij}(1_{X_{j}(t)=I^{(1)}}+1_{X_{j}(t)=I^{(1,2)}})\right), \nonumber
   \\
    \frac{d \E(1_{X_{i}(t)=I^{(1,2)}})}{dt} = - (\delta^{(1)}+ \delta^{(2)} ) \E(1_{X_{i}(t)=I^{(1,2)}})  \ \ \ \ \ \ \ \ \ \ \ \ \ \ \ \ \ \ \ \nonumber \\ 
    + \E \left( 1_{X_{i}(t)=I^{(1)}} \epsilon^{(1)} \textstyle\sum^{n}_{j=1} \beta^{(2)}_{ij}(1_{X_{j}(t)=I^{(2)}}+1_{X_{j}(t)=I^{(1,2)}})\right) \nonumber \  \\ 
    + \E \left( 1_{X_{i}(t)=I^{(2)}} \epsilon^{(2)} \textstyle\sum^{n}_{j=1} \beta^{(1)}_{ij}(1_{X_{j}(t)=I^{(1)}}+1_{X_{j}(t)=I^{(1,2)}})\right). \nonumber
\end{align*}
\normalsize
\noindent Using the above equations, $\Pr(z) = E(1_z)$, \break $x^{(1)}_i(t) = \Pr(X_i(t) = I^{(1)})$, $x^{(2)}_i(t) = \Pr(X_i(t) = I^{(2)})$, $z_i(t) = \Pr(X_i(t) = I^{(1,2)})$, $(1 - x^{(1)}_i(t)-x^{(2)}_i(t)-z_i(t)) = \Pr(X_i(t) = S)$, 
 and approximating $\Pr(X_i(t) = I^{(1)},X_j(t) = I^{(1)})\approx x^{(1)}_i(t)x^{(1)}_j(t)$, $\Pr(X_i(t) = I^{(1)},X_j(t) = I^{(2)})\approx x^{(1)}_i(t)x^{(2)}_j(t)$, 
$\Pr(X_i(t) = I^{(1)},X_j(t) = I^{(1,2)})\approx x^{(1)}_i(t)z_j(t)$, 
$\Pr(X_i(t) = I^{(2)},X_j(t) = I^{(2)})\approx x^{(2)}_i(t)x^{(2)}_j(t)$, \break $\Pr(X_i(t) = I^{(2)},X_j(t) = I^{(1,2)})\approx x^{(2)}_i(t)z_j(t)$  \normalsize (which inaccurately assumes independence, as is done in the single-virus \citep{OmicTN09} and bi-virus cases \citep{liu2019analysis})
 gives \\
\footnotesize 
\begin{align}
\dot{x}_i^{(1)}(t) &= -\delta_i^{(1)} x_i^1(t) + \delta_i^{(2)} z_i(t) \label{x1i} \\
& \ \ + (1-x_i^{(1)}(t)-x_i^{(2)}(t)-z_i(t))\textstyle\sum_{j=1}^n \beta_{ij}^{(1)}(x_j^{(1)}(t)+z_j(t)) \nonumber \\
& \ \ -x_i^{(1)}(t)\epsilon^{(1)} \textstyle\sum_{j=1}^n \beta_{ij}^{(2)}(x_j^{(2)}(t)+z_j(t)), \nonumber 
\\
\dot x_i^{(2)}(t) &= -\delta_i^{(2)} x_i^{(2)}(t) + \delta_i^{(1)} z_i(t) \label{x2i}\\
& \ \ + (1-x_i^{(1)}(t)-x_i^{(2)}(t)-z_i(t)) \textstyle\sum_{j=1}^n \beta_{ij}^{(2)}(x_j^{(2)}(t)+z_j(t)) \nonumber \\
& \ \  -x_i^{(2)}(t)\epsilon^{(2)} \textstyle\sum_{j=1}^n \beta_{ij}^{(1)}(x_j^{(1)}(t)+z_j(t)), \nonumber 
\\
\dot z_i(t) &= -(\delta_i^{(1)}+\delta_i^{(2)})z_i(t)   + x_i^{(1)}(t) \epsilon^{(1)} \textstyle \sum_{j=1}^n \beta_{ij}^{(2)}(x_j^{(2)}(t)+z_j(t))  \nonumber \\
& \ \ + x_i^{(2)}(t) \epsilon^{(2)} \textstyle\sum_{j=1}^n \beta_{ij}^{(1)}(x_j^{(1)}(t)+z_j(t)). \label{zi}
\end{align}
\normalsize 
\noindent The above equations can be combined into vector form, as follows:
\footnotesize 
\begin{align}
\dot x^{(1)}(t) &= -D^{(1)}x^{(1)}(t) + D^{(2)}z(t) \label{x1} \\
& \ \ + (I-X^{(1)}(t)-X^{(2)}(t)-Z(t))B^{(1)}(x^{(1)}(t)+z(t)) \nonumber \\
& \ \ - \epsilon^{(1)} X^{(1)}(t)B^{(2)}(x^{(2)}(t)+z(t)), \nonumber\\
\dot x^{(2)}(t) &= -D^{(2)}x^{(2)}(t) + D^{(1)}z(t) \label{x2} \\
& \ \ + (I-X^{(1)}(t)-X^{(2)}(t)-Z(t))B^{(2)}(x^{(2)}(t)+z(t)) \nonumber \\
& \ \ - \epsilon^{(2)} X^{(2)}(t)B^{(2)}(x^{(1)}(t)+z(t)),  \nonumber \\ \nonumber \\
\dot z(t) &= -(D^{(1)}+D^{(2)})z(t) + \epsilon^{(1)} X^{(1)}(t)B^{(2)}(x^{(2)}(t)+z(t))  \label{z} \\
& \ \ + \epsilon^{(2)} X^{(2)}(t)B^{(1)}(x^{(1)}(t)+z(t)),  \nonumber
\end{align}
\normalsize
\noindent where $x^{(1)}(t), \ x^{(2)}(t),\ z(t)$ are the column vectors obtained by stacking $x^{(1)}_i(t), \ x^{(2)}_i(t),$ and  $z_i(t)$, respectively, $B^{(1)}, \ B^{(2)}$ 
are the matrices of $\beta^{(1)}_{ij},\ \beta^{(2)}_{ij}$, 
respectively, $X^{(1)}(t)={\rm diag}(x^{(1)}(t))$, $X^{(2)}(t)={\rm diag}(x^{(2)}t))$, $Z(t)={\rm diag}(z(t))$, $D^{(1)}={\rm diag}(\delta^{(1)})$, and
$D^{(2)}={\rm diag}(\delta^{(2)})$. 
%
%
For completeness, to illustrate the effectiveness of the first-order approximation, we compare \eqref{eq:qij}-\eqref{eq:v} and \eqref{x1}-\eqref{z}  via simulations in Section~\ref{sec:append}. 

\tamer{Note} that an agent $i$ could 
be interpreted as \tamer{either} an individual $i$ or a subpopulation $i$. The two interpretations are equivalent, since the group model that we employ in this paper can also be derived from a group model interpretation of the \seb{original stochastic model}; see \citep{pare2019dtjournal}. \seb{In particular, the former models the probability of each individual being infected over time, while the latter models the fraction of a subpopulation being infected. Thus, there are no abrupt transitions from a susceptible state to one of the infected states for an entire subpopulation. The 
states of each subpopulation are continuously changing variable values between $0$ and $1$. Further, it is assumed that each subpopulation is well mixed/connected, which is the same assumption as in classical single-population epidemic models \citep{kermack1927contribution}.}

\begin{remark}\label{rem:subsume}
We consider two special cases of the model. First, let $\epsilon^{(1)}=\epsilon^{(2)}=0$, and  $\delta_i^{(1)}+\delta_i^{(2)}>0$ for all $i\in[n]$. Then, the system defined by \rep{x1}-\rep{z} simplifies to  \footnotesize
\begin{align}
\dot x^{(1)}(t) &= -D^{(1)}x^{(1)}(t) + D^{(2)}z(t) \label{x10} \\
& \ \ + (I-X^{(1)}(t)-X^{(2)}(t)-Z(t))B^{(1)}(x^{(1)}(t)+z(t)), \nonumber \\
\dot x^{(2)}(t) &= -D^{(2)}x^{(2)}(t) + D^{(1)}z(t) \label{x20} \\
& \ \ + (I-X^{(1)}(t)-X^{(2)}(t)-Z(t))B^{(2)}(x^{(2)}(t)+z(t)),  \nonumber \\
\dot z(t) &= -(D^{(1)}+D^{(2)})z(t) \label{z0}.
\end{align}
\normalsize
In this case, since the matrix $D^{(1)}+D^{(2)}$ is positive definite,
it follows that $z(t)$ converges to $\0$ exponentially fast, and thus the system will eventually become a competitive bi-virus model which has been studied in \citep{prakash2012winner,santos2015bivirus,liu2016onthe,liu2019analysis}.

In the second case, we let $\epsilon^{(1)}=\epsilon^{(2)}=1$. To proceed, we define $y_i^{(1)}(t)=x_i^{(1)}(t)+z_i(t)$ and $y_i^{(2)}(t)=x_i^{(2)}(t)+z_i(t)$ for each $i\in[n]$, 
which represents the total probabilities of agent $i$ being infected by viruses 1 and 2, respectively. 
From \rep{x1i}-\rep{zi}, the dynamics of $y_i^{(1)}$ and $y_i^{(2)}$ are \footnotesize
\begin{align}
\dot{y}_i^{(1)}(t) &= -\delta_i^{(1)} y_i^{(1)}(t) 
+ (1-y_i^{(1)}(t))\textstyle\sum_{j=1}^n \beta_{ij}^{(1)} y_j^{(1)}(t), \nonumber \\
\dot{y}_i^{(2)}(t) &= -\delta_i^{(2)} y_i^{(2)}(t) 
+ (1-y_i^{(2)}(t))\textstyle\sum_{j=1}^n \beta_{ij}^{(2)} y_j^{(2)}(t), \nonumber
\end{align}
\normalsize
which are two independent single SIS dynamics.
Therefore, the system defined by \rep{x1}-\rep{z} subsumes both the single SIS virus (two single, independent viruses) and the competitive 
SIS virus models.
\end{remark}

\section{Problem Formulation}\label{sec:prob:formln}
In this section, we formally state the problems of interest, and the key assumptions needed for ensuring that the model introduced in Section~\ref{model} is well defined. \vspace{-3mm}
\subsection{Problem Statements}\label{sec:prob:stmnts}
With respect to the model in~\eqref{x1}-\eqref{z}, we 
consider the following  questions:
\begin{enumerate}[label=(\roman*)] \itemsep -0.2mm
\item \label{q1} Can we identify 
 a sufficient condition under which, irrespective of the initial state, the dynamics converge asymptotically to the  healthy state?
\item \label{q2} Can we identify  a sufficient condition for virus $m$, such that for any $x^{(m)}(0) \neq \mathbf{0}$
the dynamics asymptotically converge to the single-virus endemic equilibrium of virus $m$, for $m=1,2$?

\item \label{q3} Can we identify  a sufficient condition for local exponential convergence to the boundary equilibrium of virus $m$, for $m=1,2$?

\item \label{q4} Can we provide necessary condition(s) for local exponential convergence to the boundary equilibrium of virus $m$, for $m=1,2$?

\item \label{q5bis} Is it possible for equilibria of the kind (a) $(\hat{x}^{(1)}, \hat{x}^{(2)}, \mathbf{0})$  with $\hat{x}^{(1)}, \hat{x}^{(2)} > \textbf{0}$, and (b) $(\mathbf{0}, \mathbf{0}, \hat{z})$ with   $\hat{z}> \textbf{0}$, to exist?

\item \label{q5}Can we identify  a necessary condition for the existence of the coexisting equilibria (a) $(\hat{x}^{(1)}, \mathbf{0}, \hat{z})$, with $\hat{x}^{(1)},\hat{z}>\mathbf{0}$, 
or (b) $(\mathbf{0}, \hat{x}^{(2)}, \hat{z})$, with  $\hat{x}^{(2)},\hat{z}>\mathbf{0}$? 

\item  \label{q6} Can we identify a condition that rules out an arbitrary point $(\hat{x}^{(1)}, \hat{x}^{(2)}, \hat{z})$  with  $\hat{x}^{(1)}, \hat{x}^{(2)},\hat{z}>\mathbf{0}$ as a coexisting equilibrium?


\item \label{q1bis} Is the system monotone?
\end{enumerate}

\subsection{Key Assumptions and Preliminaries}
We make the following assumptions on the model to ensure that it is well defined. 
\vspace{-3mm}
\begin{assumption}
For all $i\in[n]$, we have $x^{(1)}_i(0),x^{(2)}_i(0),z_i(0)$, $(1-x^{(1)}_i(0)-x^{(2)}_i(0)-z_i(0))\in[0,1]$.
\label{x0}
\end{assumption}
\vspace{-3mm}
\begin{assumption}
For all $i\in[n]$, we have $\delta^{(1)}_i,\delta^{(2)}_i\ge0$. The matrices $B^{(1)}$ and $B^{(2)}$ are nonnegative and irreducible.
\label{para}
\end{assumption}
\vspace{-3ex}
Assumption~\ref{x0} guarantees that the initial infection level with respect to  each virus $m$ ($m \in [2]$) in each node $i \in [n]$ lies in the set $[0,1]$, whereas Assumption~\ref{para} ensures that the healing and infection rates are nonnegative, and that the spreading graphs for virus~1 and~2 are strongly connected.

Define the set \begin{eqnarray}\label{D}
    \scr D: =\{(x^{(1)}, x^{(2)}, z) \; | \; x^{(1)}\ge \0, \; x^{(2)}\ge \0,  \; z\ge \0,\\
      \; x^{(1)}+x^{(2)}+z\le \1\} \nonumber
\end{eqnarray}
The following lemma establishes that the set $\scr D$ is positively invariant with respect to the system 
\rep{x1}-\rep{z}.
\begin{lemma}
Under 
Assumptions  \ref{x0} and \ref{para}, $x^{(1)}_i(t),x^{(2)}_i(t), z(t),$ $x^{(2)}_i(t)+x^{(2)}_i(t)+z(t)\in[0,1]$ for all $i\in[n]$ and $t\ge 0$.
\label{box}
\end{lemma}
\noindent 
{\em Proof:} 
Suppose that at some time $\tau$, $x^{1}_i(\tau),x^{2}_i(\tau), z_i(\tau),$ $x^{1}_i(\tau)+x^{2}_i(\tau)+z_i(\tau)\in[0,1]$ for all $i\in[n]$.
Consider an index $i\in[n]$.
If $x^{1}_i(\tau)=0$, then from \rep{x1i} and Assumption \ref{para}, $\dot x^{1}_i(\tau)\ge 0$.
The same holds for $x^{2}_i(\tau)$, $z_i(\tau)$ and $x^{1}_i(\tau)+x^{2}_i(\tau)+z_i(\tau)$.
If $x^{1}_i(\tau)=1$, then from \rep{x2i} and Assumption \ref{para}, $\dot x^{1}_i(\tau)\le 0$.
The same holds for $x^{2}_i(\tau)$, $z_i(\tau)$, and $x^{1}_i(\tau)+x^{2}_i(\tau)+z_i(\tau)$.
It follows that $x^{1}_i(t),x^{2}_i(t),z_i(t),x^{1}_i(t)+x^{2}_i(t)+z_i(t)$ will be in $[0,1]$ for all times $t\ge \tau$.
Since the above arguments hold for all $i\in[n]$,  $x^{1}_i(t),x^{2}_i(t),z_i(t),x^{1}_i(t)+x^{2}_i(t)+z_i(t)$ will be in $[0,1]$ for
all $i\in[n]$ and $t\ge \tau$.
Since by Assumption \ref{x0}, $x^{1}_i(0),x^{2}_i(0),z_i(0),x^{1}_i(0)+x^{2}_i(0)+z_i(0)\in[0,1]$ for all $i\in[n]$,
it follows that $x^{1}_i(t),x^{2}_i(t),z_i(t),x^{1}_i(t)+x^{2}_i(t)+z_i(t)\in[0,1]$ for all $i\in[n]$ and $t\ge 0$.
\hfill
$\qed$

\noindent
Lemma \ref{box} implies that the set
\begin{eqnarray}\label{D}
    \scr D =\{(x^{(1)}, x^{(2)}, z) \; | \; x^{(1)}\ge \0, \; x^{(2)}\ge \0,  \; z\ge \0,\\
      \; x^{(1)}+x^{(2)}+z\le \1\} \nonumber
\end{eqnarray}
is positively invariant with respect to the system defined by \rep{x1}-\rep{z}. 
Since $x^{(1)}_i$, $x^{(2)}_i$, and $z_i$
denote the probabilities of sickness of agent $i$, or fractions  of group $i$, infected by viruses 1, 2, and both 1 and 2 simultaneously, respectively,
and $1-x^{(1)}_i-x^{(2)}_i-z_i$ denotes the probability of agent $i$, or fraction of group $i$ that is healthy, it is natural to assume that their initial values are in the interval $[0,1]$,
since otherwise the values will 
be devoid of any physical meaning for the spread model considered here.

Let $(\hat x^{(1)},\hat x^{(2)},\hat z)$ be an equilibrium of system \rep{x1}-\rep{z}. 
Then,
the Jacobian matrix of the equilibrium, denoted by $J(\hat x^{(1)},\hat x^{(2)},\hat z)$, with $\hat B^{(i)} = {\rm diag}(B^{(i)}(\hat x^{(i)} + \hat z))$, 
$\hat Z^{(i)}= Z {\rm diag}(B^{(i)}\1)$, $i\in[2]$, and $W = (I - \hat X^{(1)} -\hat X^{(2)}-\hat Z)$, is as given in~\eqref{jacob}, 
\begin{figure*}
\begin{align}\label{jacob}
&J(\hat x^{(1)},\hat x^{(2)},\hat z) = \\
&
\begin{bmatrix}
 J_{1,1}  & -\hat B^{(1)} - \epsilon^{(1)} \hat X^{(1)} B^{(2)}  & W B^{(1)} + D^{(2)} - \hat B^{(1)} -\epsilon^{(1)} \hat X^{(1)} B^{(2)} \\
- \hat B^{(2)} - \epsilon^{(2)}\hat X^{(2)} B^{(1)} & J_{2,2}  & WB^{(2)} + D^{(1)} - \hat B^{(2)} - \epsilon^{(2)} \hat X^{(2)} B^{(1)} \\
\epsilon^{(1)} \hat B^{(2)} + \epsilon^{(2)} \hat X^{(2)} B^{(1)} & \epsilon^{(2)} \hat B^{(1)} + \epsilon^{(1)} \hat X^{(1)} B^{(2)} & J_{3,3}  \end{bmatrix}\normalsize.\nonumber
\end{align}
\end{figure*}
where 
\begin{align}
J_{1,1} &= W B^{(1)} -D^{(1)} - \hat B^{(1)} -\epsilon^{(1)} \hat B^{(2)} \\
J_{2,2} &= W B^{(2)} -D^{(2)} - \hat B^{(2)} -\epsilon^{(2)} \hat B^{(1)} \\
J_{3,3} &= -D^{(1)} - D^{(2)} + \epsilon^{(1)} \hat X^{(1)} B^{(2)} + \epsilon^{(2)} \hat X^{(2)} B^{(1)}.
\end{align}

\section{Analysis of the Disease-Free Equilibrium }\label{analysis}




In this section, we analyze the system defined by \rep{x1}-\rep{z}.
It is easy to see that $(\0,\0,\0)$ is an equilibrium of the system defined by \rep{x1}-\rep{z}. We call it the DFE, or the healthy state. 
We focus on identifying conditions 
under which the healthy state is 
stable. The following proposition provides a necessary and sufficient condition for local exponential convergence to the healthy state.

\begin{proposition}
Consider system~\eqref{x1}-\eqref{z} under Assumptions~\ref{x0} and \ref{para}. The healthy state is locally 
exponentially
stable if, and only if,
$s(-D^{(1)}+B^{(1)})< 0$, $s(-D^{(2)}+B^{(2)})<0$, and $\delta_i^{(1)}+\delta_i^{(2)}>0$ for all $i\in[n]$. If $s(-D^{(1)}+B^{(1)})> 0$ or if $s(-D^{(2)}+B^{(2)})>0$, then the healthy state is unstable.
\label{0global<}
\end{proposition}


\noindent {\em Proof:}
From \eqref{jacob}, we have 
\begin{equation}\label{J0}
J(\0,\0,\0) = \footnotesize\begin{bmatrix}
B^{(1)} -D^{(1)}    & 0 & B^{(1)} + D^{(2)}  \\
0 &  B^{(2)} -D^{(2)} & WB^{(2)} + D^{(1)}  \\
0 & 0 & -D^{(1)} - D^{(2)}  \end{bmatrix}.
\end{equation}
Thus, from \citep[Theorem 4.15 and Corollary~4.3]{khalil2002nonlinear}, the healthy state is locally exponentially stable if, and only if, $s(-D^{(1)}+B^{(1)})< 0$, $s(-D^{(2)}+B^{(2)})<0$, and 
$\delta_i^{(1)}+\delta_i^{(2)}>0$ for all $i\in[n]$. 
Note  that if $s(-D^{(1)}+B^{(1)})> 0$ or if $s(-D^{(2)}+B^{(2)})> 0$, then $s(J(\0,\0,\0))>0$.
The claim on instability then follows from \citep[Theorem~4.7]{khalil2002nonlinear}.~\hfill$\qed$


Note that, on the one hand,  the guarantees provided by Proposition~\ref{0global<} are limited in the sense that they concern trajectories that originate in a small neighborhood of the healthy state. On the other hand, no restrictions, besides nonnegativity, are imposed on $\epsilon^{(m)}$, $m=1,2$. Simulations, as we see in Section~\ref{sec:examples}, 
indicate that the region 
of attraction for the healthy state depends
on the choices of $\epsilon^{(m)}$, $m=1,2$. In particular, if the initial state of system~\eqref{x1}-\eqref{z} is very close to the healthy state, then, even for a larger value of  $\epsilon^{(m)}$, $m=1,2$, the dynamics converge to the healthy state. If   the initial state of system~\eqref{x1}-\eqref{z} is not too close to the healthy state, then for large values of  $\epsilon^{(m)}$, the dynamics do not converge to the healthy state; see Figure~\ref{fig:prop1:divergence} in Section~\ref{sec:examples}.

The following theorem guarantees global  convergence to the healthy state, but with the following caveats: i) the speed of convergence is slower, and ii)  more restrictions on $\epsilon^{(m)}$, $m=1,2$, 
have to be imposed.

\begin{theorem}
Under Assumptions~\ref{x0} and \ref{para},
if $\epsilon^{(1)}, \epsilon^{(2)} \in[0,1]$, $s(B^{(1)}-D^{(1)})\leq 0$ and $s(B^{(2)}-D^{(2)})\leq 0$, then the healthy state is the unique equilibrium of \rep{x1}-\rep{z}, and
the system defined by \rep{x1}-\rep{z} asymptotically converges to the healthy state for any initial state in $\scr D$, as defined in \eqref{D}. 
\label{0global=}\end{theorem}
\noindent 
{\em Proof:} See the Appendix. \hfill \qed\\
Theorem~\ref{0global=}
answers Question~\ref{q1} raised in Section~\ref{sec:prob:stmnts}.
\vspace{-3mm}

\section{Persistence of  Viruses}\label{analysis:persistence}
\vspace{-3mm}

We call an equilibrium $(\hat x^{(1)}, \hat x^{(2)}, \hat z)$ an endemic equilibrium if it is not 
the healthy state, $(\0,\0,\0)$. It turns out that if either (or both) of the spectral abscissa conditions in Theorem~\ref{0global=} are violated, then at least one of the viruses pervades the population. We  detail the same 
in the rest of this section.

\subsection{Existence, Uniqueness and Stability of the Single Virus Endemic Equilibria}
We consider the scenario in which either
$s(B^{(1)}-D^{(1)})$ or $s(B^{(2)} - D^{(2)})$ is greater than zero. Without loss of generality, we assume that $s(B^{(1)}-D^{(1)})>0$, $\epsilon^{(1)} 
\in[0,1]$, and $s(B^{(2)} - D^{(2)})\le 0$. We denote by $\hat{x}^{(1)}$ (resp. $\hat{x}^{(2)}$) the single-virus endemic equilibrium corresponding to virus~$1$ (resp. virus~2). We have the following result.
\begin{theorem}
Under Assumptions~\ref{x0} and \ref{para},
if $s(B^{(1)}-D^{(1)})> 0$, $\epsilon^{(1)} 
\in[0,1]$, and $s(B^{(2)}-D^{(2)})\leq 0$, then system \rep{x1}-\rep{z} has a unique endemic equilibrium $(\hat x^{(1)},\0,\0)$ with $\hat x^{(1)}\gg \0$, and
the system asymptotically converges to the endemic equilibrium for any initial state in $\scr D\setminus \{(\0,x^{(2)}, 
\0) | \0\leq x^{(2)}
\le \1\}$, 
where $\scr D$ is defined in \eqref{D}. 
\label{1epidemic}\end{theorem}

\noindent {\em Proof:} See the Appendix.\hfill$\qed$


\noindent 
Theorem~\ref{1epidemic} establishes the existence, uniqueness, and asymptotic stability of the single-virus endemic equilibrium corresponding to virus~1. An analogous result holds for virus~2; details of which are omitted in the interest of space. Theorem~\ref{1epidemic}
answers Question~\ref{q2} raised in Section~\ref{sec:prob:stmnts}. Note that, for a single virus system, assuming that the existence of the endemic equilibrium is guaranteed, an exact characterization of the same has been provided in \cite[Theorem~4.3, statement~(iii)(b)]{mei2017dynamics}.
\par Combining Theorems~\ref{0global=} and~\ref{1epidemic}, we obtain a necessary and sufficient condition for the healthy state to be the unique equilibrium of the coupled bi-virus system, as stated below. \vspace{-4mm}
\begin{corollary}\label{cor:healthy:unique}
Consider system~\rep{x1}-\rep{z} under \break Assumptions~\ref{x0} and \ref{para}. Suppose further that $\epsilon^{(1)}, \epsilon^{(2)} \in[0,1]$. The healthy state is the unique equilibrium if, and only if, each of the following conditions are satisfied i)
$s(B^{(1)}-D^{(1)})\leq 0$ and ii) $s(B^{(2)}-D^{(2)})\leq 0$.
\end{corollary}
\vspace{-5mm}
\subsection{Both Viruses Pervading the System}
Note that Theorem~\ref{1epidemic} accounts for the case where exactly one of the viruses pervades the system, or, in other words, exactly one of the spectral abscissa conditions in Theorem~\ref{0global=} is violated. 
However, 
what happens when \emph{both} the spectral abscissa conditions in Theorem~\ref{0global=} are violated? The following corollary 
answers 
this question.
\begin{corollary}
\label{prop:multiple:viruses}
Consider system~\eqref{x1}-\eqref{z} under \break Assumptions~\ref{x0} and \ref{para}. Suppose further that $\epsilon^{(1)} =\epsilon^{(2)}=\epsilon \in [0,1]$. If $s(B^{(1)} - D^{(1)}) > 0$ and $s(B^{(2)} - D^{(2)}) > 0$, then system~\eqref{x1}-\eqref{z} has at least three equilibria, namely, the healthy state $(\textbf{0},\textbf{0},\textbf{0})$, which is unstable; the single virus endemic equilibrium corresponding to virus~1 $(\hat x^{(1)},\0,\0)$; and the single virus endemic equilibrium corresponding to virus~2 $(\0, \hat x^{(2)},\0)$. 
\end{corollary}
\textit{Proof:} 
See the Appendix. \hfill \qed

The equilibria  of the kind $(\hat{x}^{(1)}, \mathbf{0}, \mathbf{0})$ and $(\mathbf{0}, \hat{x}^{(2)}, \mathbf{0})$ are hereafter referred to as the \emph{boundary} equilibria. Note that  $\hat{x}^{(1)}$ and $\hat{x}^{(2)}$ are asymptotically stable in the single virus (i.e., one of the two viruses has died out) systems  corresponding to virus~1 and virus~2, respectively; when both 
viruses pervade the network, the stability of $(\hat{x}^{(1)}, \mathbf{0}, \mathbf{0})$ and $(\mathbf{0}, \hat{x}^{(2)}, \mathbf{0})$, in even the local (let alone global) sense, is not guaranteed. As such, in the rest of this section we will focus on identifying a sufficient condition (resp. some necessary conditions) for local exponential stability of the boundary equilibria.
We need the following assumption, which is slightly stronger than Assumption~\ref{para}.
\begin{assumption}
For all $i\in[n]$, we have $\delta^{(1)}_i,\delta^{(2)}_i>0$. The matrices $B^{(1)}$ and $B^{(2)}$ are nonnegative and irreducible.
\label{para-1}
\end{assumption}
\vspace{-3ex}

The following theorem provides a sufficient condition for local exponential stability of the boundary 
equilibrium $(\hat{x}^{(1)}, \mathbf{0}, \mathbf{0})$.
\vspace{-3mm}
\begin{theorem}\label{thm:local:expo:stab:single-virus}
Consider system~\eqref{x1}-\eqref{z} under \break Assumptions~\ref{x0}  and~\ref{para-1}. Suppose that i) $\epsilon^{(1)} = \epsilon^{(2)} = \epsilon \in [0,1]$,  ii) $s(-D^{(1)}+B^{(1)})>0$, and iii) $s(-D^{(2)}+B^{(2)})>0$.  The equilibrium point $(\hat x^{(1)}, \mathbf{0}, \mathbf{0})$ is locally exponentially stable if 
\begin{enumerate}[label=\roman*)]
    \item \footnotesize $s(-D^{(2)} + (I-\hat{X}^{(1)})B^{(2)}) <0$; and 
    \item
        \footnotesize  $s\bigg((-D^{(1)}-D^{(2)}+\epsilon
\hat{X}^{(1)} B^{(2)})-(\epsilon \hat{B}^{(1)}+ \epsilon \hat{X}^{(1)}B^{(2)})  (-D^{(2)} + (I-\hat{X}^{(1)})B^{(2)}-\epsilon \hat{B}^{(1)})^{-1}((I-\hat{X}^{(1)})B^{(2)} + D^{(1)})\bigg) <0$. 
\end{enumerate}

\end{theorem}
\noindent \textit{Proof:} See the Appendix. \hfill \qed\\
Theorem~\ref{thm:local:expo:stab:single-virus}
answers Question~\ref{q3} raised in Section~\ref{sec:prob:stmnts}.

The following proposition provides necessary conditions for local exponential stability of the equilibrium 
$(\hat{x}^{(1)}, \mathbf{0}, \mathbf{0})$.
\begin{proposition}\label{prop:hatx1:local:expo:stability:nec}
Consider system~\eqref{x1}-\eqref{z} under Assumptions~\ref{x0} and~\ref{para-1}. Suppose that $\epsilon^{(1)} = \epsilon^{(2)} = \epsilon \in [0,1]$, and that $s(-D^{(1)}+B^{(1)})>0$. 
The equilibrium point $(\hat x^{(1)}, \mathbf{0}, \mathbf{0})$ is locally exponentially stable   only if each of the following conditions are satisfied
\begin{enumerate}[label=\roman*)]
    \item \footnotesize $s(-D^{(2)} + (I-\hat{X}^{(1)})B^{(2)}-\epsilon\hat{B}^{(1)})<0$; and 
    \item
       \footnotesize   $s\bigg((-D^{(1)}-D^{(2)}+\epsilon
\hat{X}^{(1)} B^{(2)})-(\epsilon \hat{B}^{(1)}+ \epsilon \hat{X}^{(1)}B^{(2)})   (-D^{(2)} + (I-\hat{X}^{(1)})B^{(2)}-\epsilon \hat{B}^{(1)})^{-1}((I-\hat{X}^{(1)})B^{(2)} + D^{(1)})\bigg) <0$.
\end{enumerate}
\end{proposition}

\noindent \textit{Proof:} 
See the Appendix. \hfill \qed \\
Proposition~\ref{prop:hatx1:local:expo:stability:nec}
answers Question~\ref{q4} raised in Section~\ref{sec:prob:stmnts}.

\begin{remark}\label{rem:sicon:result}
Note that, in general, there is a gap between the sufficient condition in Theorem~\ref{thm:local:expo:stab:single-virus} and the necessary conditions in Proposition~\ref{prop:hatx1:local:expo:stability:nec}. However, 
if $\epsilon^{(m)}=0$ for $m=1,2$, the sufficient condition in Theorem~\ref{thm:local:expo:stab:single-virus} and the necessary conditions in Proposition~\ref{prop:hatx1:local:expo:stability:nec} coincide to yield a necessary and sufficient condition for local exponential convergence to $(\hat x^{(1)},\mathbf{0},\mathbf{0})$. Further,  by Assumption~\ref{para-1},  $\delta_i^{(1)} >0$ and $\delta_i^{(2)} >0$ for each $i \in [n]$. Hence, 
if $\epsilon^{(m)}=0$ for $m=1,2$, then
condition ii) in both Theorem~\ref{thm:local:expo:stab:single-virus} 
and 
Proposition~\ref{prop:hatx1:local:expo:stability:nec} is always satisfied. As a consequence, 
condition~i) in Theorem~\ref{thm:local:expo:stab:single-virus} becomes a necessary and sufficient condition, which is consistent with \citep[Theorem~3.10]{ye2021convergence}.
\end{remark}

\begin{remark}
Theorem~\ref{thm:local:expo:stab:single-virus} pertains to local exponential stability of the boundary equilibrium $(\hat x^{(1)},\mathbf{0},\mathbf{0})$. It is of interest to know when 
the boundary equilibrium $(\hat x^{(1)},\mathbf{0},\mathbf{0})$ can be globally stable. A partial answer is as follows: Suppose that the conditions in Theorem~\ref{thm:local:expo:stab:single-virus} are satisfied. Suppose that $\epsilon^{(1)} = \epsilon^{(2)} = 0$. Then if the system~\eqref{x1}-\eqref{z} has no coexistence equilibria (guaranteed by, for instance, $B^{(2)}>B^{(1)}$ \citep[Corollary~2]{axel2020TAC}), then the boundary equilibrium $(\hat x^{(1)},\mathbf{0},\mathbf{0})$ is globally asymptotically stable; see \citep[Corollary~3.16]{ye2021convergence} 
\end{remark}

We next present a result for instability of the boundary equilibrium $(\hat{x}^{(1)}, \mathbf{0}, \mathbf{0})$.\vspace{-3mm}
\begin{corollary}\label{cor:instab:hatx1}
Consider system~\eqref{x1}-\eqref{z} under \break Assumptions~\ref{x0} and~\ref{para-1}. Suppose that $\epsilon^{(1)} = \epsilon^{(2)} = \epsilon \in [0,1]$, and that $s(-D^{(1)}+B^{(1)})>0$.  If $s(-D^{(1)}-D^{(2)}+\epsilon\hat{X}^{(1)}B^{(2)})>0$, then  the equilibrium 
$(\hat x^{(1)}, \mathbf{0}, \mathbf{0})$ is  unstable.
\end{corollary}
\noindent \textit{Proof:} 
See the Appendix. \hfill \qed
\par Note that by suitably changing the notations of 
Theorem~\ref{thm:local:expo:stab:single-virus}, Proposition~\ref{prop:hatx1:local:expo:stability:nec}, and Corollary~\ref{cor:instab:hatx1}, we can obtain a sufficient condition, necessary conditions, and  a condition for instability, respectively, for the boundary equilibrium $(\mathbf{0}, \hat{x}^{(2)}, \mathbf{0})$.
\vspace{-3mm}
\section{Analysis of  Coexisting Equilibria}\label{analysis:endemic-coexistence}

Equilibria  of the kind $(\hat{x}^{(1)}, \hat{x}^{(2)}, \hat{z})$ with at least any two of $\hat{x}^{(1)}, \hat{x}^{(2)}, \hat{z}$ being  non-zero vectors, are referred to as \emph{coexisting equilibria}. In this section,  we show that certain kinds of coexisting equilibria cannot exist, and we focus on identifying necessary conditions 
for the existence of certain other kinds of coexisting equilibria. 

\subsection{Impossibility of Existence of a Certain Kind of Coexisting Equilibria}\label{sec:impossible}
It is well known that for the competitive bivirus case (i.e., $\epsilon^{(m)}=0$ for $m=1,2$), one of the possible equilibria is the so-called coexistence equilibrium, where separate fractions of each node is infected by virus~1 and virus~2 \citep{liu2019analysis,ye2021convergence,axel2020TAC}. More formally, these are equilibria of the kind $(\hat{x}^{(1)}, \hat{x}^{(2)})$ with $\textbf{0} \ll \hat{x}^{(1)} \ll \mathbf{1}$, $\textbf{0} \ll \hat{x}^{(2)} \ll \mathbf{1}$ and $\hat{x}^{(1)}+\hat{x}^{(2)} \ll \mathbf{1}$. It is natural to wonder whether similar equilibria could exist even when $\epsilon^{(m)} >0$ for $m=1,2$. It turns out, however, that such coexistence equilibria do \emph{not} exist for coupled bivirus systems. The following proposition formalizes this. \vspace{-3mm} 
\begin{proposition} \label{prop:no:coexistence}
Consider system~\eqref{x1}-\eqref{z} under Assumptions~\ref{x0} and~\ref{para}. Suppose that $\epsilon^{(m)}>0$ for some $m \in [2]$. There does not exist a coexisting equilibrium of the form $(\hat{x}^{(1)}, \hat{x}^{(2)}, \textbf{0})$ with $\hat{x}^{(1)}, \hat{x}^{(2)} > \textbf{0}$. 
\end{proposition}
\textit{Proof:} 
See the Appendix. \hfill \qed

In a similar vein to Proposition~\ref{prop:no:coexistence}, the following proposition states that the coupled bivirus system cannot have an equilibrium where a fraction of each node is infected by both viruses at the same time, but that no fraction of any node is infected only by one of the viruses.
\begin{proposition}\label{prop:onlyhatz:notpossible}
 Consider system~\eqref{x1}-\eqref{z} under Assumption~\ref{para-1}. There does not exist an equilibrium of the form $(\textbf{0}, \textbf{0}, \hat{z})$, where $\hat{z}>\textbf{0}$.
\end{proposition}
\noindent \textit{Proof:} 
See the Appendix. \hfill \qed

Taken together, Propositions~\ref{prop:no:coexistence} and~\ref{prop:onlyhatz:notpossible} answer, in the negative, Question~\ref{q5bis} raised in Section~\ref{sec:prob:stmnts}. In so doing,  Propositions~\ref{prop:no:coexistence} and~\ref{prop:onlyhatz:notpossible} 
restrict the set of possible endemic equilibria for the coupled bivirus system.
\subsection{Necessary Conditions for Existence of Certain Kinds of Coexisting Equilibria}\label{sec:nec:cond}
While Subsection~\ref{sec:impossible} has dealt with the impossibility of existence of certain kinds of coexisting equilibria, in this subsection we 
are 
interested in identifying some necessary conditions for the existence of certain other kinds of coexisting equilibria.
We begin by presenting a necessary condition for the existence of a coexisting equilibrium where for each node the fraction infected only by virus~1 is non-zero, by both viruses~1 and~2 is non-zero, but only by virus~2 is zero.
\begin{proposition}\label{prop:nece:condn:nox2}
Consider system~\eqref{x1}-\eqref{z} under Assumption~\ref{para-1}. Suppose that i) $\rho((D^{(1)})^{-1}B^{(1)})>1$, and ii) $\epsilon^{(1)} = \epsilon^{(2)} = \epsilon$ with  $\epsilon \in (0,1)$.
Then, there exists an equilibrium 
$(\hat{x}^{(1)}, \mathbf{0}, \hat{z})$ with $\hat{x}^{(1)},\hat{z}>\mathbf{0}$ only if $\rho((D^{(1)})^{-1}(B^{(2)})) \geq 1$.
\end{proposition}
\textit{Proof:} See the Appendix.~$\blacksquare$
By following analogous arguments as in the proof of Proposition~\ref{prop:nece:condn:nox2}, it can be shown that there exists an equilibrium 
$(\mathbf{0},\hat{x}^{(2)},  \hat{z})$ only if $\rho((D^{(2)})^{-1}(B^{(1)})) \geq 1$. Thus, Proposition~\ref{prop:nece:condn:nox2} conclusively answers Question~\ref{q5} raised in Section~\ref{sec:prob:stmnts}.

Next, we present a 
condition that rules out a \emph{given} point in the state space as a 
coexisting equilibrium where each node has i) a fraction that is infected only by virus~1; ii) a fraction that is infected only by virus~2; and iii) a fraction that is infected by both viruses~1 and~2. To this end, we need the following assumption.
\begin{assumption}\label{assum:identical:viruses}
The healing and infection rates are the same for each virus. That is, $\delta_i^{(1)}=\delta_i^{(2)}$ for all $i \in [n]$, and $\beta_{ij}^{(1)}=\beta_{ij}^{(2)}$ for all $i=j\in [n]$ and $(i,j) \in \mathcal E$.
\end{assumption}
In words, Assumption~\ref{assum:identical:viruses} states that two identical heterogeneous viruses spread over the same graph. This implies that $D^{(1)}=D^{(2)}=D$, and $B^{(1)}=B^{(2)}=B$. With Assumption~\ref{assum:identical:viruses} in place, we have the following result.
\begin{proposition}\label{prop:nece:x1x2z}
Consider system~\eqref{x1}-\eqref{z} under Assumptions~\ref{x0},~\ref{para-1} and~\ref{assum:identical:viruses}. Suppose further that $\epsilon^{(1)}=\epsilon^{(2)} = \epsilon \geq 0$. Then, there exists an equilibrium 
$(\hat{x}^{(1)}, \hat{x}^{(2)}, \hat{z})$, where $\hat{x}^{(1)}, \hat{x}^{(2)}, \hat{z} > \textbf{0}$, only if $\rho((I-\hat{X}^{(1)}-\hat{X}^{(2)}-\hat{Z})D^{-1}B)<1$.
\end{proposition}
\textit{Proof:} 
See the Appendix. \hfill \qed \\
Proposition~\ref{prop:nece:x1x2z} answers Question~\ref{q6} raised in Section~\ref{sec:prob:stmnts}.\\
Note that Proposition~\ref{prop:nece:x1x2z} is, in itself, \emph{not} a necessary condition. That is, if a given point $(\hat{x}^{(1)}, \hat{x}^{(2)}, \hat{z})$, where  \break $\hat{x}^{(1)}, \hat{x}^{(2)}, \hat{z} > \textbf{0}$, were to not fulfil the condition in  Proposition~\ref{prop:nece:x1x2z}, it does not mean that there cannot exist another point, say, $(\tilde{x}^{(1)}, \tilde{x}^{(2)}, \tilde{z})$, where $\tilde{x}^{(1)}, \tilde{x}^{(2)}, \tilde{z} > \textbf{0}$, that satisfies the condition in Proposition~\ref{prop:nece:x1x2z}. Of course, if every point in the state space violates the aforementioned condition, then no equilibrium of the form  $(\hat{x}^{(1)}, \hat{x}^{(2)}, \hat{z})$, $\hat{x}^{(1)}, \hat{x}^{(2)}, \hat{z} > \textbf{0}$, can exist.

\section{Monotonicity (or lack thereof) of the coupled bivirus system} \label{sec:no:monotone} \vspace{-3mm}
The discussion 
heretofore has centered around existence, uniqueness and stability of certain specific equilibria. It is natural to seek a more general perspective 
on the coupled bi-virus model, which is the main focus of this section.


With $\epsilon^{(1)}=\epsilon^{(2)}=0$ (i.e., the competitive bivirus case), the system defined by \rep{x1}-\rep{z} is monotone; see \citep[Lemma~3.3]{ye2021convergence}. This section seeks to answer  whether the same holds for the case when $\epsilon^{(1)} >0$ and/or $\epsilon^{(2)}>0$. To this end, first we introduce a graph structure, and then use 
this  graph to provide a conclusive answer.
\vspace{-3mm}
\subsection{Construction of the graph associated with the Jacobian of a non-linear system} \label{sec:graph:jacob}
Consider a system $\dot{x} =f(x)$, and let $J(\cdot)$ denote the Jacobian of 
this system.
It turns out that we can construct a graph associated with $J(\cdot)$; call this graph $\bar{G}$. 
The construction follows the outline provided in \citep{sontag2007monotone}. More specifically, the number of nodes in
the graph $\bar{G}$ equals the number of rows (resp. columns) of $J(\cdot)$, whereas  the edges of  $\bar{G}$ are drawn based on the entries in 
 $J(\cdot)$.
If, independent of the argument of $J(\cdot)$,  $[J(\cdot)]_{ij} \leq 0$ for  $i \neq j$, then we draw an edge from node $j$ to node $i$ labeled with a ``$-$" sign;  if, independent of the argument of $J(\cdot)$,  $[J(\cdot)]_{ij}  \geq 0$ for  $i \neq j$, then we draw an edge labeled with a ``+" sign. If  $[J(\cdot)]_{ij}  \leq 0$ for some argument, and $[J(\cdot)]_{ij}  \geq 0$ for some other argument
then from node $j$ to node $i$ we draw an edge labeled with a ``$-$" sign, and also an edge labeled with a ``+" sign. Furthermore, if from node $j$ to node $i$, there are edges  with a ``$-$" sign and with a ``+" sign, then we introduce a node between $i$ and $j$, say $i^\prime$ such that there is an edge with a ``+" sign from $i$ to $i^\prime$; an edge with ``+" sign from  $i^\prime$  to $j$; the edge with a ``$-$" sign from $j$ to $i$ is retained; see \citep[Figure~3]{sontag2007monotone}. We refer to the graph constructed with the addition of such nodes as $\hat {G}$.  Thus,  $\hat {G}$ is a signed graph. Note that  $\hat {G}$ has no self-loops.

We will also be requiring  the following notion from graph theory. A signed graph is said to be consistent if every undirected cycle in the graph has a net positive sign, i.e., it has an even number of ``$-$" signs \citep{sontag2007monotone}. 

\subsection{The coupled bivirus system is not monotone} \label{sec:couipled:not:monotone}
We now show that the coupled bivirus system is not monotone.
\begin{theorem} \label{thm:coupled:bivirus:not:monotone}
Under Assumptions~~\ref{x0} and \ref{para},
system~\rep{x1}-\rep{z} is not monotone.
\end{theorem}
\textit{Proof:} First, note that the Jacobian $J(x^{(1)}, x^{(2)}, z)$, as in~\eqref{jacob}, has $3n$ rows (resp. $3n$ columns). Hence, the graph $\bar{G}$ constructed with respect to the Jacobian in~\eqref{jacob} has $3n$ nodes.
Next, observe that the 12 and 21 blocks of $J(x^{(1)}, x^{(2)}, z)$ are, due to Assumptions~\ref{x0} and \ref{para}, negative matrices, which implies that for any node $i$, the edge from node $i$ (resp. $i+n$) to node  $i+n$ (resp. $i$) has a  ``$-$" sign. Similarly, the 31 and 32 blocks of $J(x^{(1)}, x^{(2)}, z)$ are, due to Assumptions~\ref{x0} and \ref{para}, positive matrices, which implies that for any node $i$, the edge from node $i$ (resp. $i+n$) to node $i+2n$ has a  ``+" sign.\\
Note that the 13 block 
of $J(x^{(1)}, x^{(2)}, z)$ can change its sign depending on the argument. Hence, 
 it is clear that, for any node $i$, there is an edge from node $i+2n$ to $i$ with  a ``$-$" sign, and an edge with a ``+" sign. Therefore, we introduce $n$ additional nodes, and label these $3n+1$, $3n+2 \hdots 4n$.
Similarly, since the 23 block of $J(x^{(1)}, x^{(2)}, z)$ can change its sign depending on the argument, it follows that for any node $i$, there is an edge from node $i+2n$ to $i+n$ with a ``$-$" sign, and an edge with ``+" sign. Therefore, we introduce further $n$ additional nodes, and label those $4n+1$, $4n+2 \hdots 5n$.
The edges corresponding to the nodes labeled $3n+1$, $3n+2 \hdots 4n$, and $4n+1, 4n+2, \hdots 5n$ are assigned as outlined in Section~\ref{sec:graph:jacob}; thus obtaining the corresponding graph $\hat {G}$.\\
In graph $\hat {G}$, 
the loop starting from node $i$ traversing through node $i+3n$, node $i+2n$, node $i+n$ and back to node $i$ is a 4-length cycle that has an odd number (namely, one) of negative signs. Therefore, from \cite[page 62]{sontag2007monotone}, the signed graph $\hat{G}$ is not consistent. 
Consequently, from \cite[page 63]{sontag2007monotone}, it follows that the system~\rep{x1}-\rep{z} is not monotone.~\hfill \qed

Theorem~\ref{thm:coupled:bivirus:not:monotone}
answers Question~\ref{q1bis} raised in Section~\ref{sec:prob:stmnts}. Furthermore, 
Theorem~\ref{thm:coupled:bivirus:not:monotone} implies that we cannot leverage  the rich literature on monotone dynamical systems \citep{smith1988systems,hirsch1988stability} to study the limiting behavior of system~\rep{x1}-\rep{z}. In general, for non-monotone systems, no dynamical behavior, including chaos, can be definitively ruled out 
\citep{sontag2007monotone}. Therefore, novel tools are needed to study coupled bivirus systems  more in-depth. The development of such tools is beyond the scope of this paper. 
\section{Simulations} \label{sec:sim}

This section presents a comparison of the $4^n$-state  Markov 
process in \eqref{eq:qij}-\eqref{eq:v}  to \eqref{x1}-\eqref{z} via simulation, and also provides a set of 
simulations of \seb{the coupled virus model on a small- and large-scale networks, whose exact parameter values used along with the \MATLAB \ 
code used are available 
via github\footnote{https://github.com/philpare/coupled\_bivirus}.} 


\subsection{Comparison to Full Probabilistic Model}\label{sec:append}


\begin{figure}[h!]
\centering
\includegraphics[width=.8\columnwidth]{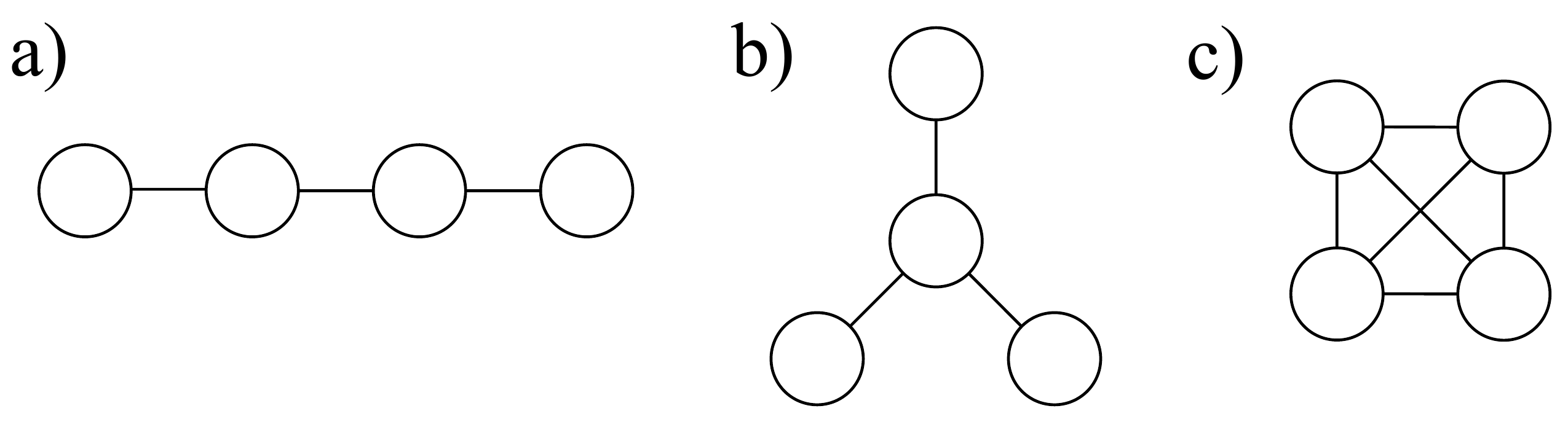}
\caption{Graph structures: a) line, b) star, c) complete.}
\label{fig:graphs}
\end{figure}

\begin{figure}[h!]
    \centering
\begin{overpic}[width=\columnwidth]{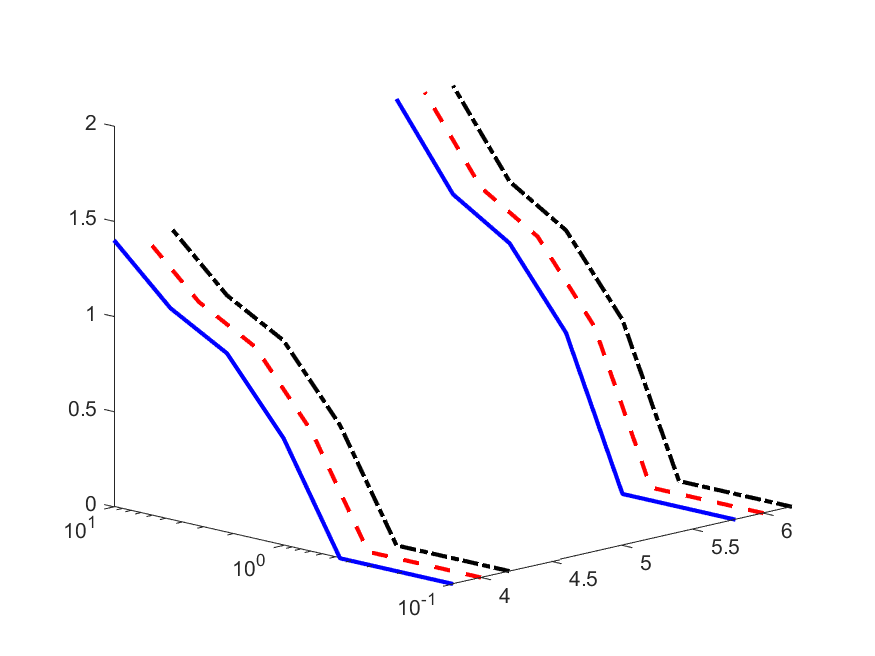}
    \put(76.5,6.5){{\parbox{\linewidth}{%
     $n$}}}
     \put(21,5){{\parbox{\linewidth}{%
     $\displaystyle\frac{\beta}{\delta}$}}}
     \put(4,38.5){{\parbox{\linewidth}{%
     \rotatebox{90}{error}}}}
\end{overpic}
\caption{
A plot of $\|[v^{(1)}(T); v^{(2)}(T); v^{(3)}(T)] -[x^{(1)}(T); x^{(2)}(T); z(T)]\|$ for the line graph, $T=10,000$. Results from using the different initial conditions \eqref{eq:initial1}, \eqref{eq:initial2}, and \eqref{eq:initial3} are depicted by the blue lines, red dashed lines, and black dash-dot lines, respectively. 
}
\label{fig:line}
\end{figure}
We compare the model in \eqref{x1}-\eqref{z} to the full probabilistic $4^n$-state  model in \eqref{eq:qij}-\eqref{eq:v} via simulations to illustrate the effectiveness of the approximation. 
We set $\epsilon^{(1)}=\epsilon^{(2)}=3$, and use line graphs, star (hub--spoke) graphs, and complete graphs. For examples of each type of graph, see Figure \ref{fig:graphs}. 
All adjacency matrices for these graphs are symmetric and binary-valued, and both viruses spread over the same graph. In the star graph, the central node is the first agent. 
Each simulation was run for 10,000 time steps  (final time $T=10,000$), with three initial conditions: 1) the first node is infected by virus 1 and the second node is infected by virus 2, 
\begin{equation}\label{eq:initial1}
    \begin{split}
        x^{(1)}(0) &= [ 1 \  0 \ \cdots\ 0]^{\top}\\
        x^{(2)}(0) &= [ 0 \ 1 \ 0 \ \cdots\ 0]^{\top}\\
        z(0) &= \0 ;
    \end{split}
\end{equation}
\normalsize
2) the first node is infected by virus 1, the second node is infected by virus 2, and the third node is infected by both virus 1 and virus 2, 
\begin{equation}\label{eq:initial2}
    \begin{split}
         x^{(1)}(0) &= [ 1 \  0 \ \cdots\ 0]^{\top}\\
        x^{(2)}(0) &= [ 0 \ 1 \ 0 \ \cdots\ 0]^{\top}\\
        z(0) &= [ 0 \ 0 \ 1 \ 0  \ \cdots\ 0]^{\top} ;
    \end{split}
\end{equation}
\normalsize
and 3) the first node is infected by virus 1, the second node is infected by virus 2, and the remaining  nodes are infected by both virus 1 and virus 2, 
\begin{equation}\label{eq:initial3}
    \begin{split}
         x^{(1)}(0) &= [ 1 \  0 \ \cdots\ 0]^{\top}\\
         x^{(2)}(0) &= [ 0 \ 1 \ 0 \ \cdots\ 0]^{\top}\\
        z(0) &= [ 0 \ 0 \ 1  \ \cdots\ 1]^{\top} .
    \end{split}
\end{equation}
\normalsize
In these tests we  explore identical homogeneous viruses, $(\beta,\delta) = (\beta^{(1)},\delta^{(1)}) = (\beta^{(2)},\delta^{(2)}) = (\beta^{(3)},\delta^{(3)})$. The $(\beta,\delta)$ pairs  are \scriptsize $[ (0.1,1), (0.215,1), (0.464,1),(0.5,0.5), (1,0.464),$ $ (1,0.215), (1, 0.1)]$, \normalsize and the numbers of agents are $n = 4,6$. 
We limited simulations to these $n$ values since mean field approximations are typically worse for small values of $n$ and there is a computational limitation due to the size of the $4^n$-state Markov  model.
\begin{figure}[h!]
    \centering
\begin{overpic}[width=\columnwidth]{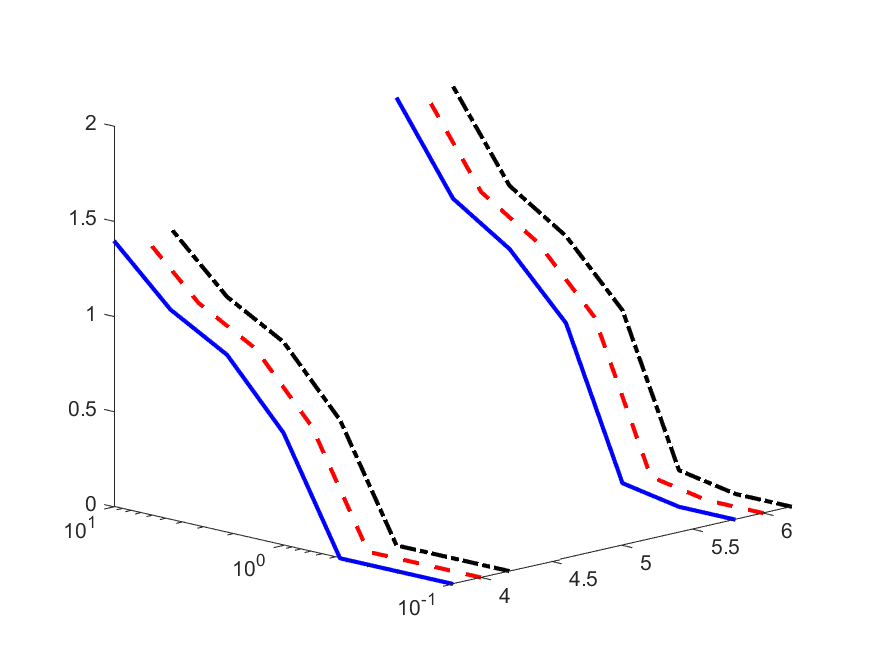}
    \put(76.5,6.5){{\parbox{\linewidth}{%
     $n$}}}
     \put(21,5){{\parbox{\linewidth}{%
     $\displaystyle\frac{\beta}{\delta}$}}}
     \put(4,38.5){{\parbox{\linewidth}{%
     \rotatebox{90}{error}}}}
\end{overpic}
\caption{
A plot of  $\|[v^{(1)}(T); v^{(2)}(T); v^{(3)}(T)] -[x^{(1)}(T); x^{(2)}(T); z(T)]\|$ for the star graph, $T=10,000$. Results from using the different initial conditions \eqref{eq:initial1}, \eqref{eq:initial2}, and \eqref{eq:initial3} are depicted by the blue lines, red dashed lines, and black dash-dot lines, respectively. 
}
\label{fig:star}
\end{figure}

\begin{figure}[h!]
    \centering
    \begin{overpic}[width=\columnwidth]{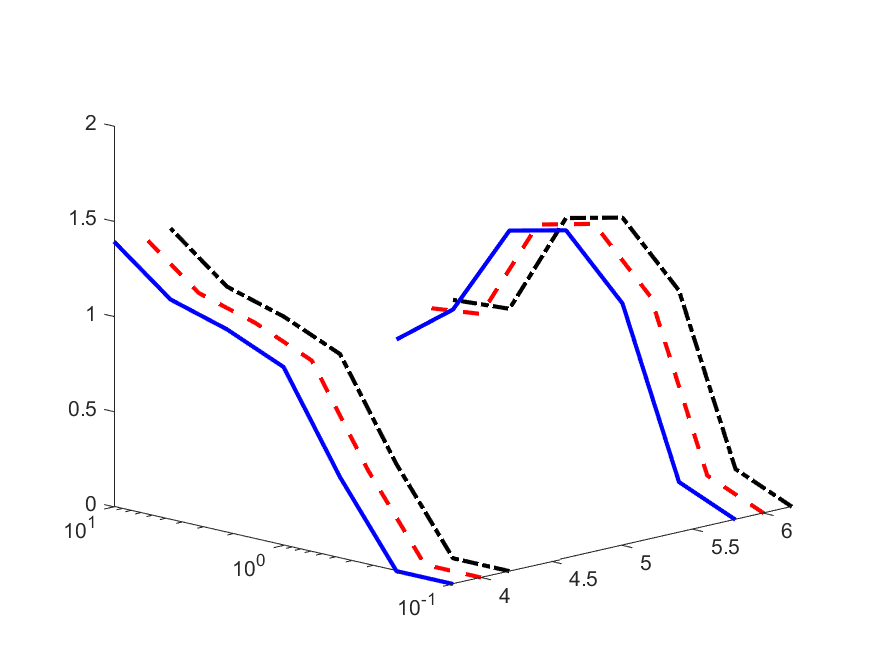}
    \put(76.5,6.5){{\parbox{\linewidth}{%
     $n$}}}
     \put(21,5){{\parbox{\linewidth}{%
     $\displaystyle\frac{\beta}{\delta}$}}}
     \put(4,38.5){{\parbox{\linewidth}{%
     \rotatebox{90}{error}}}}
\end{overpic}
\caption{
A plot of the error $\|[v^1(T); v^2(T); v^3(T)] -[x^{(1)}(T); x^{(2)}(T); z(T)]\|$ for the complete graph, $T=10,000$. Results from using the different initial conditions \eqref{eq:initial1}, \eqref{eq:initial2}, and \eqref{eq:initial3} are depicted by the blue lines, red dashed lines, and black dash-dot lines, respectively. 
}
\label{fig:full}
\end{figure}

The results are given in Figures \ref{fig:line}-\ref{fig:full} in terms of the 2-norm of the difference between the states of \eqref{x1}-\eqref{z} at the final time ($[x^{(1)}(T); x^{(2)}(T); z(T)]$), and the means of the three states in the $4^n$-state  Markov model at the final time ($[v^{(1)}(T); v^{(2)}(T); v^{(3)}(T)]$ as defined by \eqref{eq:v}). 

The accuracy of the approximation appears to be very similar to the single virus case \citep{OmicTN09} and to the two-virus case \citep{liu2019analysis}. 
Since the model in \eqref{x1}-\eqref{z} is an upper bounding approximation, the results show that the two models converge to the healthy state for the smaller values of $\frac{\beta }{\delta}$, resulting in small errors between the two models. For many of the larger values of $\frac{\beta }{\delta}$, the model in \eqref{x1}-\eqref{z} again performs quite well since it is at an epidemic state and the $4^n$-state Markov  model does not appear to reach the healthy state in the finite time considered in the simulations ($T=10,000$). Therefore, for certain values of $\frac{\beta }{\delta}$ and certain time scales, the model in \eqref{x1}-\eqref{z} is a sufficient approximation of the $4^n$-state  Markov model. 
For values of $\frac{\beta }{\delta}$ that are near 1, the models are quite different, similar to the single- and bi-virus cases. The $4^n$-state Markov  model appears, in most cases, to be at or close to the healthy state while the model in \eqref{x1}-\eqref{z} is at an epidemic state, resulting in large errors. 


\subsection{Small-Scale Network}\label{sec:examples}

Virus 1 is depicted by the color red ($r$), virus 2 is depicted by the color blue ($b$), and the state of being infected with both states, $z(t)$, is depicted by the color green ($g$). For all $i\in[n]$, the color at each time $t$ for node $i$ is given~by 
\begin{equation}\label{eq:color} 
\frac{x^{(1)}_i(t)}{s_i(t)}r + \frac{x^{(2)}_i(t)}{s_i(t)}b + \frac{z_i(t)}{s_i(t)}g,
\end{equation}
where $s_i(t) = x^{(1)}_i(t)+x^{(2)}_i(t)+z_i(t)$.
When $x^{(1)}_i(t)+x^{(2)}_i(t)+z_i(t)=0$, the color is black, indicating completely healthy, susceptible.
These color variations are used to facilitate the depiction of the parallel equilibria ($\hat x^{(1)} = \alpha \hat x^{(2)} $), a behavior that is exhibited by the $\epsilon^{(1)}=\epsilon^{(2)} = 0$ case (see \citep{liu2016onthe}) and are illustrated by all the nodes converging to the same color.
For all $i\in[n]$, the diameter of node 
$i$ is given by \vspace{-2mm} 
\begin{equation}\label{eq:diam}
d_0 + s_i(t)r_0,
\end{equation}
\normalsize
with $d_0$ being the default/smallest diameter and $r_0$ being the scaling factor depending on the total sickness of node~$i$. 
Therefore, the color indicates the type of virus(es) each agent has and the diameter indicates the 
degree of sickness for each agent.  
The graph structure is as follows: 
\begin{equation}\label{eq:at}
a_{ij}(t) = \begin{cases}
    e^{-\|y_i(t) - y_j(t)\|^2}, & \text{if } \|z_i(t) - z_j(t)\| < \hat{r}, \\
    0,              & \text{otherwise},
\end{cases}
\end{equation}
where $y_i(t)\in \mathbb{R}^2$ is the position of node $i$, with $\hat{r}=.35$.

\begin{figure}
\centering
\includegraphics[width=.7\columnwidth]{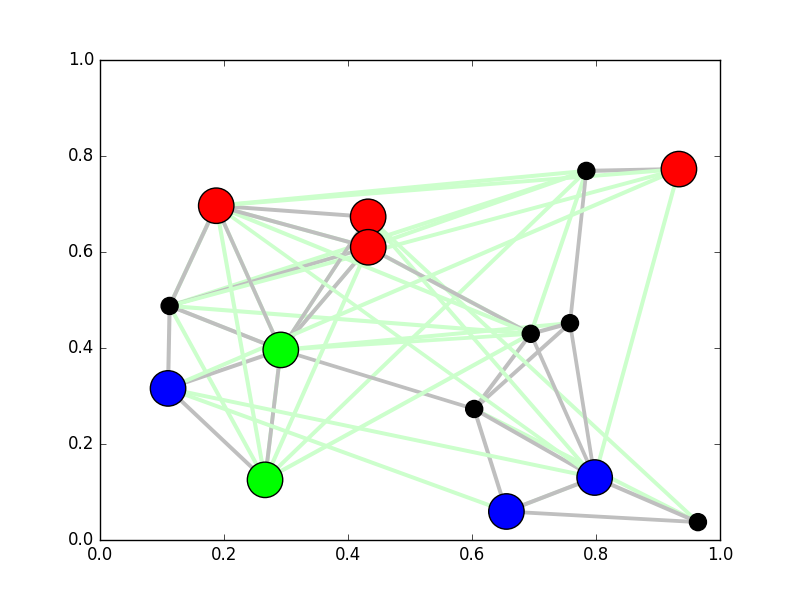}
\caption{Initial condition of the network.
Black: healthy nodes; red: nodes infected only by virus~1; blue: nodes infected only by virus~2; and green: nodes infected by both viruses~1 and~2. Black edges: spreading pattern for virus~1; green edges: spreading pattern for virus~2.}
\label{fig:hetero}
\end{figure}

We consider a network of $n=15$ nodes. The binary matrices $A^{(1)}$ and $A^{(2)}$ are populated in correspondence, respectively, to the black and green edges depicted in Figure~\ref{fig:hetero}. 
The initial condition of the network is 
shown in Figure~\ref{fig:hetero}.
The elements of diagonal matrices $D^{(1)}$ and $D^{(2)}$ are chosen uniformly at random from $[0,1]$.
Let $b^{(1)}$ be a vector whose elements are chosen uniformly at random from $[0,1]$. 
Let $B^{(1)}=\diag(B^{(1)}/6)A^{(1)}$. Vector $b^{(2)}$ 
is
chosen analogous to $b^{(1)}$. 
Then, let $B^{(2)}=\diag(b^{(2)}/11)A^{(2)}$.
Let $\epsilon^{(1)} = \epsilon^{(2)}= \epsilon$, and fix $\epsilon=0.5$. Given that the matrices involved are of dimension $15 \times 15$, in the interest of space, we refrain from providing exact values.
% ...
With these choices of model parameters, it turns out that
$s(-D^{(1)}+B^{(1)})=-0.0037<0$ and $s(-D^{(2)}+B^{(2)})=-0.0013<0$. 
Consistent with Proposition~\ref{0global<} (and since $\epsilon \in (0,1)$, also consistent with Theorem~\ref{0global=}), both the viruses die out; see Figure~\ref{fig:eps0p5}. Next, with the same model parameters and initial condition as before, we set $\epsilon=2$. Again, consistent with Proposition~\ref{0global<}, the dynamics converge to the healthy state, albeit the rate of convergence is slower than that with $\epsilon=0.5$; see Figure~\ref{fig:eps2}. Once again, with the same model parameters and initial condition as before, we set $\epsilon=1000$. Although $s(-D^{(1)}+B^{(1)})=-0.0037<0$ and $s(-D^{(2)}+B^{(2)})=-0.0013<0$, as a consequence of the effect of the large value of $\epsilon$, the dynamics do not die out; see Figure~\ref{fig:eps:opposite}. Next,  using the same set of parameters as for the  simulations in Figure~\ref{fig:eps0p5}, except for $\epsilon^{(1)}=2$ and $\epsilon^{2}=0.5$, we check the conditions in Proposition~1. It seems that, although it needs to be proven rigorously, even with the aforementioned choice of $\epsilon^{(m)}$ for $m=1,2$, the conditions in Proposition~1 guarantee eradication of both the viruses; see Figure~\ref{fig:prop1:divergence}.

For the next simulation, 
$D^{(1)}$, $D^{(2)}$, 
$b^{(1)}$, and $b^{(2)}$ are 
the same as before. 
Let $B^{(1)}=\diag(b^{(1)})A^{(1)}$, and let $B^{(2)}=\diag(b^{(2)}/11)A^{(2)}$.
Choose $\epsilon=0.5$. With these choices of model parameters, it turns out that
$s(-D^{(1)}+B^{(1)}) = 1.4904>0$ and $s(-D^{(2)}+B^{(2)})=-0.0013<0$.
Consistent with Theorem~\ref{1epidemic}, virus~2 dies out,  virus~1 becomes endemic in the population, and no fraction of any node is infected by both viruses~1 and~2 simultaneously; see Figure~\ref{fig:thm2}.

\par Next, 
$D^{(1)}$, $D^{(2)}$, 
$b^{(1)}$, and $b^{(2)}$  take the same
values
as before. Let $B^{(1)}=\diag(b^{(1)}/2)A^{(1)}$, and let $B^{(2)}=\diag(b^{(2)}/10)A^{(2)}$.
We choose $\epsilon=0.5$. With such a choice, we obtain $s(-D^{(1)}+B^{(1)}) = 0.5026>0$, thus ensuring the existence of an endemic equilibrium $\hat{x}^{(1)}$. 
We also have $s(-D^{(2)}+B^{(2)})=0.0063$.
Further, we have that $s(-D^{(2)}+(I-\hat{X}^{(1)})B^{(2)}) = -0.0210<0$, and $s\bigg((-D^{(1)}-D^{(2)}+\epsilon
\hat{X}^{(1)} B^{(2)})-(\epsilon \hat{B}^{(1)}+ \epsilon
\hat{X}^{(1)}B^{(2)})   (-D^{(2)} + (I-\hat{X}^{(1)})B^{(2)}-  \epsilon \hat{B}^{(1)})
^{-1}((I-\hat{X}^{(1)})B^{(2)} + D^{(1)})\bigg) = -0.0259<0$. Hence, in line with the result in Theorem~\ref{thm:local:expo:stab:single-virus}, 
the dynamics converge to $(\hat{x}^{(1)}, \textbf{0}, \textbf{0})$ exponentially fast; see Figure~\ref{fig:thm3}. 

\subsection{Large-Scale Network}\label{sec:examples:2}
\seb{Next, we test the mean-field model on a large-scale network. Specifically, we run  simulations on a graph of adjacent counties in the contiguous United States of America ($n=3109$), using the same adjacency matrix as \citep{pare2019dtjournal} for both infection graphs, $A^{(1)}$ and $A^{(2)}$}. 

\seb{Specifically, 
 the adjacency matrix (dimension $3109 \times 3109$) is calculated using  the adjacency of counties, that is, 
\begin{equation}
    a_{ij} = 
    \begin{cases}
    1, & \text{ if county }i \text{ and county } j \text{ share a border,}\\
    1, &\text{ if } i=j\text{,} \\
    0, & \text{ otherwise,}
    \end{cases}\label{eq:astate}
\end{equation} 
and is depicted in Figure~\ref{fig:hetero_usa}; the colors of the nodes indicate
the initial condition of the network.} 

\begin{figure}
    \centering
    \subfloat[With $\epsilon=0.5$\label{fig:eps0p5}]{
      \includegraphics[width=.48\columnwidth]{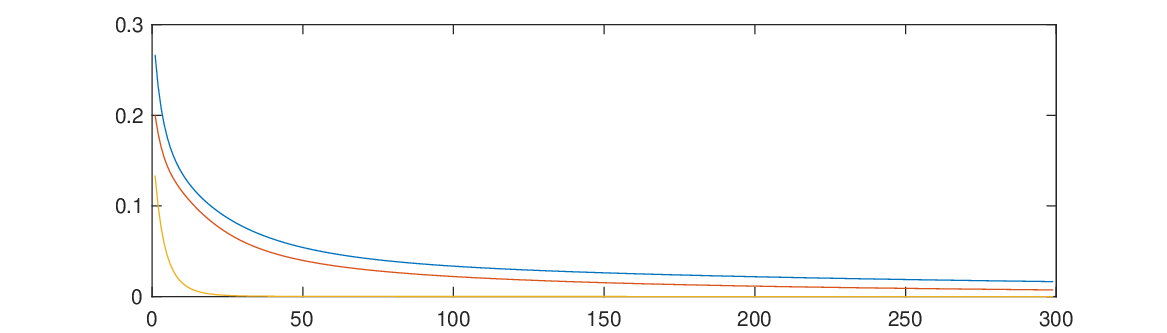}}
    \hfill
    \subfloat[With $\epsilon=2$\label{fig:eps2}]{
      \includegraphics[width=.48\columnwidth]{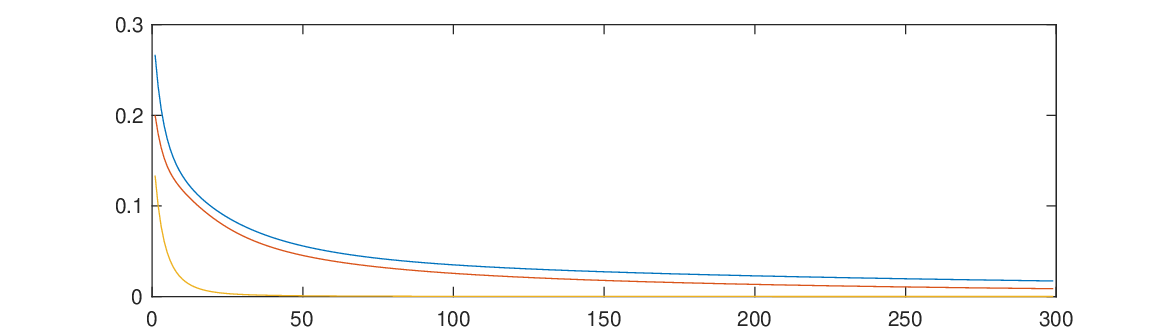}}
    \caption{In both  figures, the red line indicates the average infection level of the population with respect to virus~1; the blue line with respect to virus~2; and the yellow line with respect to both viruses~1 and~2.}
\label{fig:prop1:thm1}
\end{figure}
\begin{figure}[h!]
\centering
\includegraphics[width=\columnwidth]{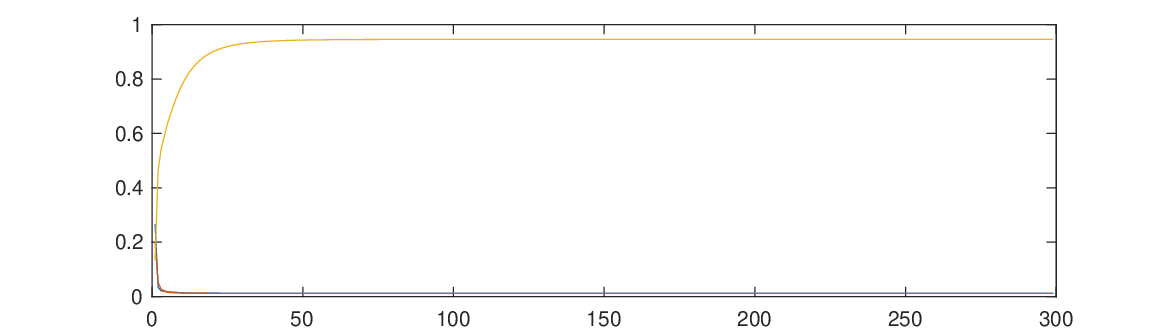}
\caption{The parameters chosen fulfil the conditions in Proposition~\ref{0global<}, but since $\epsilon (=1000)$ is quite large the dynamics do not converge to the healthy state.}
\label{fig:eps:opposite}
\end{figure}
\begin{figure}
\centering
\includegraphics[width=\columnwidth]{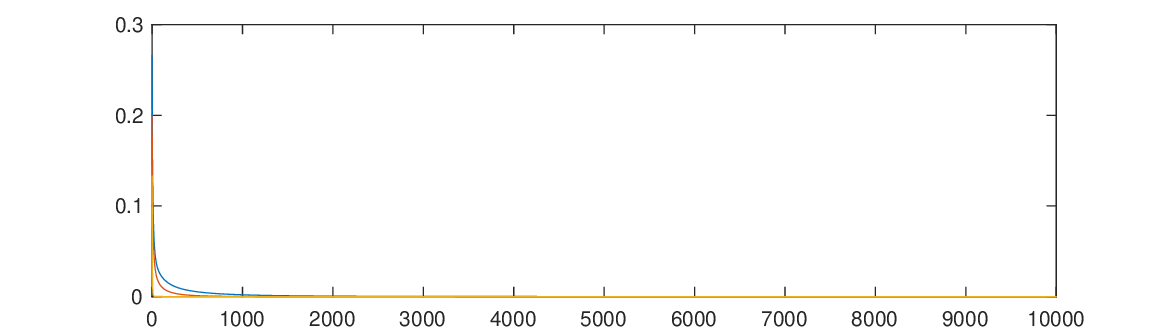}
\caption{The parameters are chosen so as to fulfil the conditions in Proposition~\ref{0global<}, with the exception that $\epsilon^{(1)}=0.5$ and $\epsilon^{(2)}=2$. The dynamics still  converge to the healthy state.}
\label{fig:prop1:divergence}
\end{figure}
\begin{figure}
    \centering
    
      \includegraphics[width=\columnwidth]{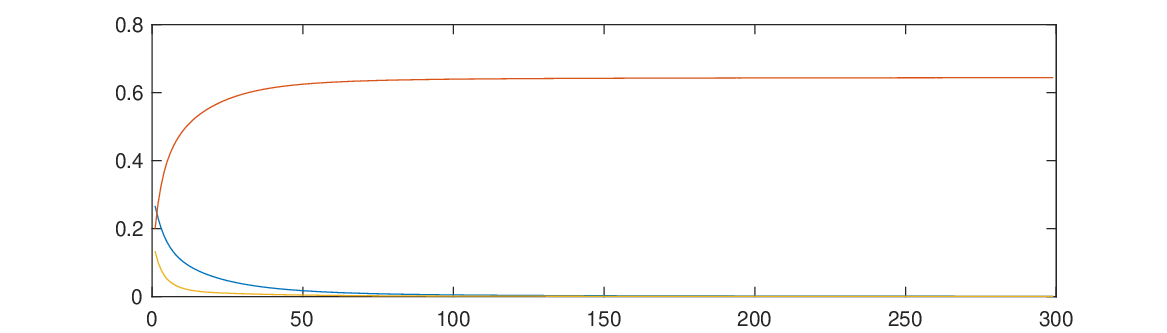}
      
    \caption{Virus~1 becomes endemic; virus~2 has died out completely, and no fraction of any node is infected by both viruses~1 and~2.
}
\label{fig:thm2}
\end{figure}
\begin{figure}
\centering
\includegraphics[width=\columnwidth]{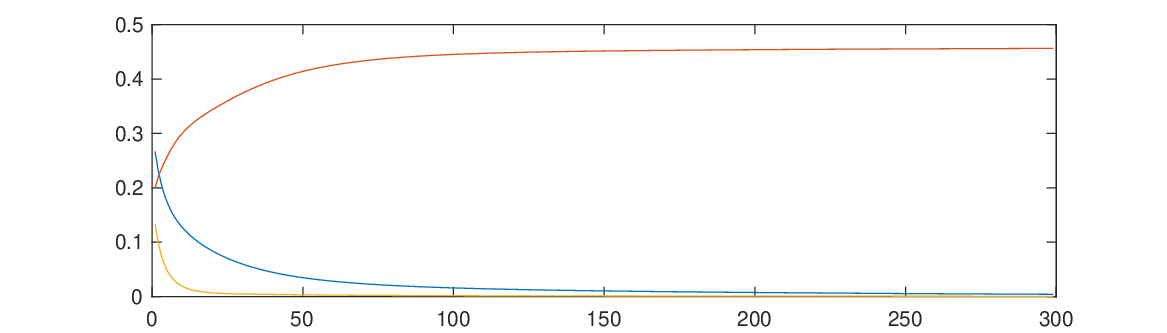}
\caption{Average infection level with respect to virus~1, virus~2, and virus~1 and virus~2  using the initial condition in Figure \ref{fig:hetero}.}
\label{fig:thm3}
\end{figure}
\begin{figure}
 \centering
\includegraphics[width=\columnwidth]{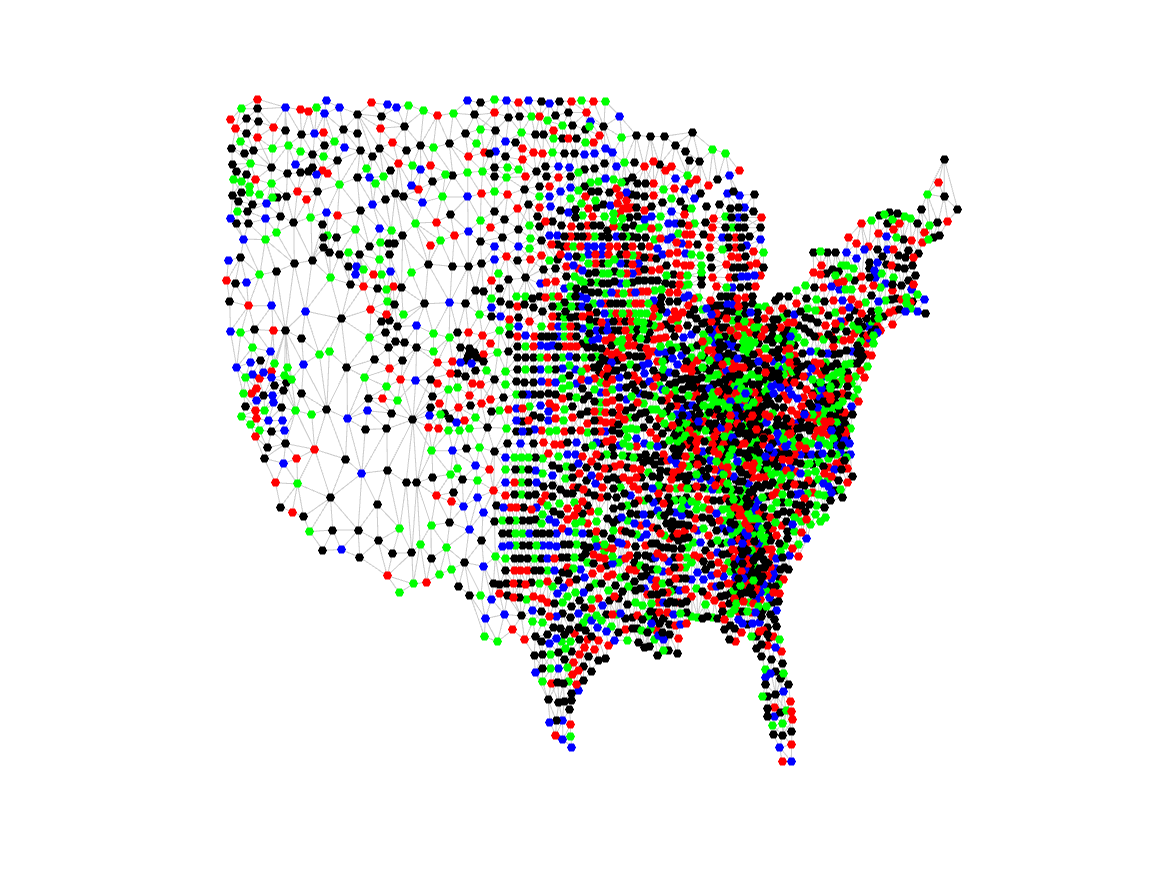}
\caption{ Initial condition of the network. Black: healthy nodes; red: nodes infected only by virus~1; blue: nodes infected only by virus~2; and green: nodes infected by both viruses~1 and~2. Black edges: spreading pattern for virus~1; green edges: spreading pattern for virus~2.}
\label{fig:hetero_usa}
\end{figure}
\begin{figure}
    \centering      \includegraphics[width=\columnwidth]{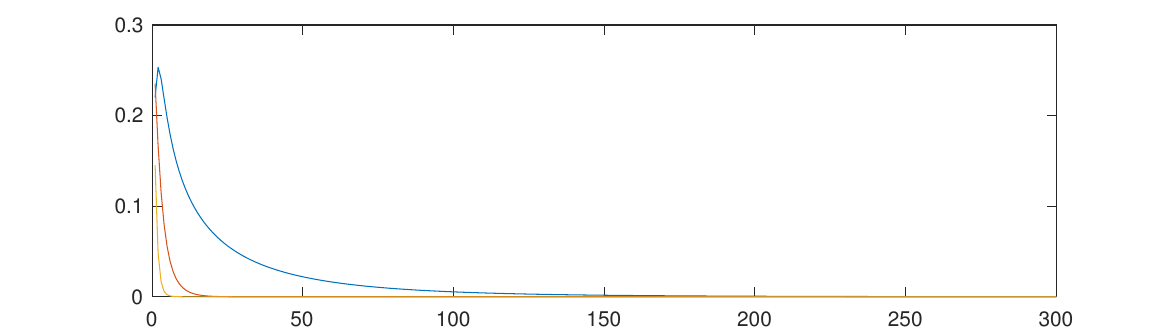}
    \caption{The red line indicates the average infection level of the population with respect to virus~1; the blue line with respect to virus~2; and the yellow line with respect to both viruses~1 and~2. We set with $\epsilon=0.5$\label{fig:eps0p5_usa}}
 \label{fig:prop1:thm1_usa}
\end{figure}

\seb{\seb{The elements of diagonal matrices $D^{(1)}$ and $D^{(2)}$, and that of vector $b^{(1)}$ and $b^{(2)}$ are chosen as in the simulation for Figure~\ref{fig:eps0p5}}.
Let $B^{(1)}=\diag(B^{(1)}/8)A^{(1)}$. 
and, let $B^{(2)}=\diag(b^{(2)}/21)A^{(2)}$.
Let $\epsilon^{(1)} = \epsilon^{(2)}= \epsilon$, and fix $\epsilon=0.5$. 
% ...
With these choices of model parameters, it turns out that
$s(-D^{(1)}+B^{(1)})=-0.0037<0$ and $s(-D^{(2)}+B^{(2)})=-0.0013<0$. 
Consistent with Proposition~\ref{0global<} 
(and since $\epsilon \in (0,1)$, also consistent with Theorem~\ref{0global=}), 
both  
viruses die out; see Figure~\ref{fig:eps0p5_usa}. 
\seb{Next, with the same model parameters and initial condition as used for the simulation in Figure~\ref{fig:eps0p5_usa}},
we set $\epsilon=2$. It turns out that consistent with Proposition~\ref{0global<}, 
the dynamics converge to the healthy state, albeit the rate of convergence is slower than that with $\epsilon=0.5$; see Figure~\ref{fig:eps2_usa}.
\seb{Next,  using the same set of parameters as for the  simulations in Figure~\ref{fig:eps0p5_usa},} 
with the exception that $\epsilon^{(1)}=2$ and $\epsilon^{2}=0.5$, we check the conditions in Proposition~\ref{0global<}. It seems that 
even with the aforementioned choice of $\epsilon^{(m)}$ for $m=1,2$, the conditions in Proposition~\ref{0global<} guarantee eradication of both 
viruses; see Figure~\ref{fig:prop1:divergence_usa}. Once again, with the same model parameters and initial condition used in the simulation for Figure~\ref{fig:eps0p5_usa},
we set $\epsilon=1000$. Even though $s(-D^{(1)}+B^{(1)})=-0.1848<0$ and $s(-D^{(2)}+B^{(2)})=-0.0081<0$, because of the effect of the large value of $\epsilon$, the dynamics do not die out; see Figure~\ref{fig:eps:opposite_usa}.} 

\seb{\seb{For the next simulation, 
$D^{(1)}$, $D^{(2)}$, 
$b^{(1)}$, and $b^{(2)}$ are chosen to be
 the same as in the simulation for Figure~\ref{fig:eps0p5_usa}.}
Let $B^{(1)}=\diag(b^{(1)})A^{(1)}$, and 
$B^{(2)}=\diag(b^{(2)}/21)A^{(2)}$.
Choose $\epsilon=0.5$. With these choices of model parameters, it turns out that
$s(-D^{(1)}+B^{(1)}) = 4.2619>0$ and $s(-D^{(2)}+B^{(2)})=-0.0081<0$.
In line with the result in Theorem~\ref{1epidemic}, 
virus~2 dies out,  virus~1 stays endemic in the population, and no fraction of any node is infected by both viruses~1 and~2. simultaneously; see Figure~\ref{fig:thm2_usa}.}
\par \seb{Next, 
$D^{(1)}$, $D^{(2)}$, 
$b^{(1)}$, and $b^{(2)}$  take the same
values
as before. Let $B^{(1)}=\diag(b^{(1)}/2)A^{(1)}$, and let $B^{(2)}=\diag(b^{(2)}/20)A^{(2)}$.
We choose $\epsilon=0.5$. Such a choice yields $s(-D^{(1)}+B^{(1)}) = 1.7209>0$, thus ensuring the existence of an endemic equilibrium $\hat{x}^{(1)}$. 
We also have $s(-D^{(2)}+B^{(2)})=0.0067$.
Further, we have that $s(-D^{(2)}+(I-\hat{X}^{(1)})B^{(2)}) = -0.2045<0$, and $s\bigg((-D^{(1)}-D^{(2)}+\epsilon
\hat{X}^{(1)} B^{(2)})-(\epsilon \hat{B}^{(1)}+ \epsilon
\hat{X}^{(1)}B^{(2)})   (-D^{(2)} + (I-\hat{X}^{(1)})B^{(2)}-  \epsilon \hat{B}^{(1)})
^{-1}((I-\hat{X}^{(1)})B^{(2)} + D^{(1)})\bigg) = -0.1834<0$. Hence, consistent with Theorem~\ref{thm:local:expo:stab:single-virus}, 
the dynamics of the coupled bivirus system converge to $(\hat{x}^{(1)}, \textbf{0}, \textbf{0})$ exponentially fast; see Figure~\ref{fig:thm3_usa}. }


\begin{figure}
    \centering      \label{fig:eps2_usa}
\includegraphics[width=\columnwidth]{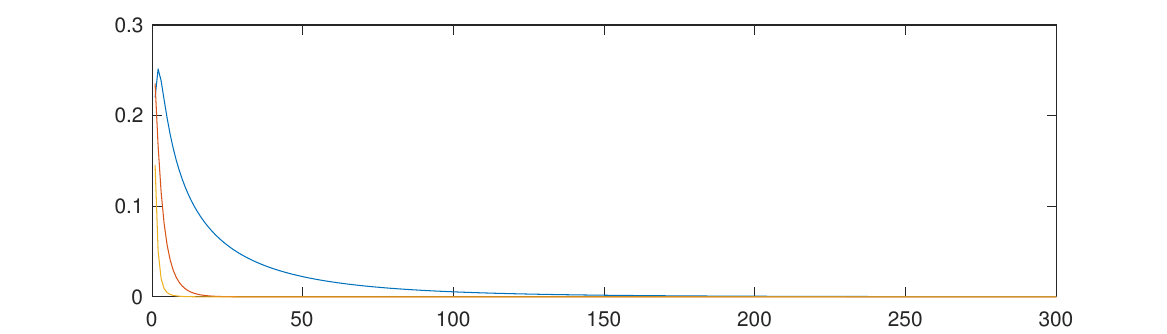}
    \caption{The red line indicates the average infection level of the population with respect to virus~1; the blue line with respect to virus~2; and the yellow line with respect to both viruses~1 and~2. We set with $\epsilon=2$}
\label{fig:eps2_usa}
\end{figure}

\begin{figure}
\centering
\includegraphics[width=\columnwidth]{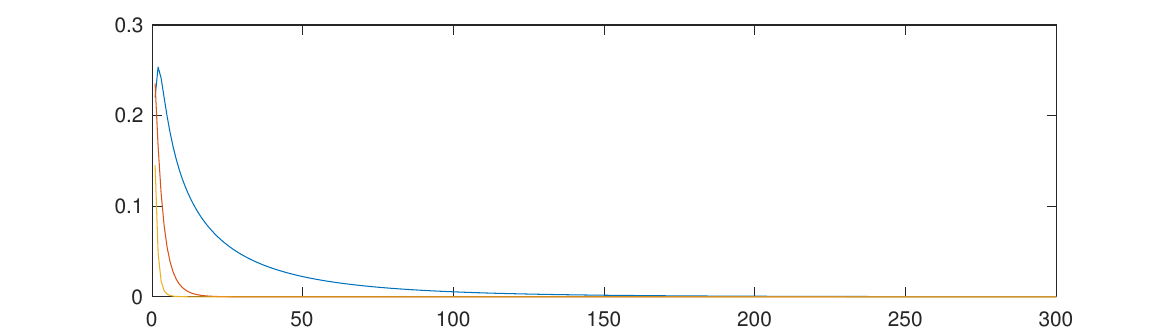}
\caption{The parameters are chosen so as to fulfil the conditions in Proposition~\ref{0global<}, with the exception that $\epsilon^{(1)}=0.5$ and $\epsilon^{(2)}=2$. The dynamics still  converge to the healthy state.}
\label{fig:prop1:divergence_usa}
\end{figure}
 

\begin{figure}
    \centering
    \includegraphics[width=\columnwidth]{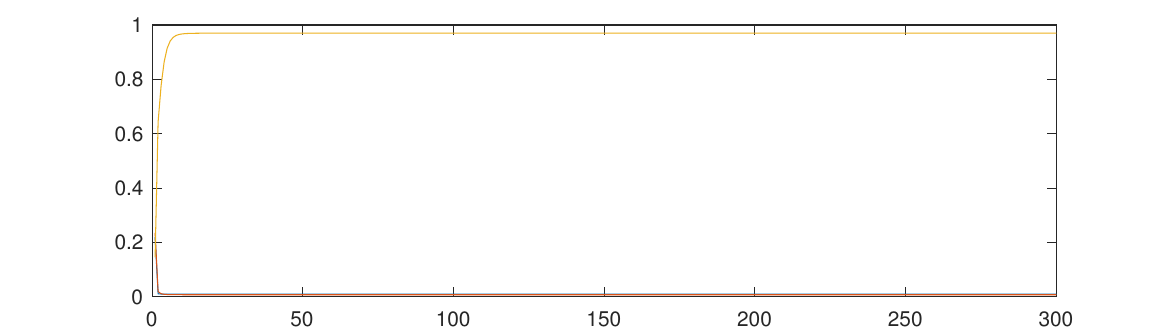}
    \caption{The parameters chosen fulfil the conditions in Proposition~\ref{0global<}, 
 but since $\epsilon (=1000)$ is quite large the dynamics do not converge to the healthy state.}
    \label{fig:eps:opposite_usa}
\end{figure}

\begin{figure}
    \centering
    \includegraphics[width=\columnwidth]{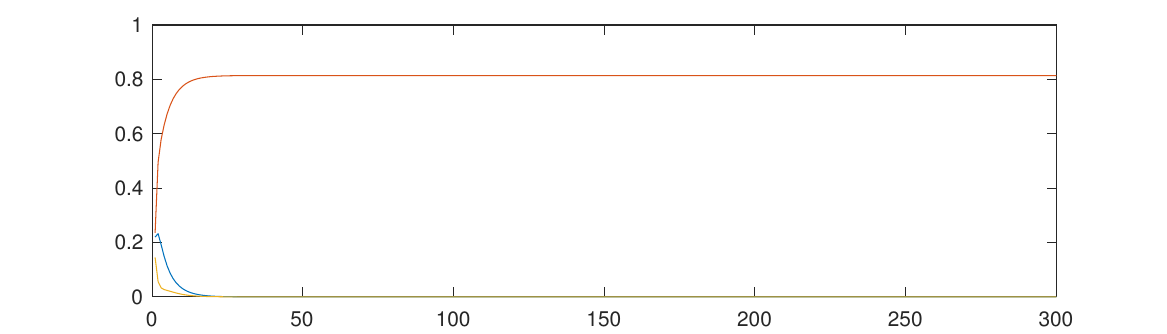}
    \caption{Virus~1 becomes endemic; virus~2 has died out completely, and no fraction of any node is infected by both viruses~1 and~2. }
    \label{fig:thm2_usa}
\end{figure}

\begin{figure}
\centering
\includegraphics[width=\columnwidth]{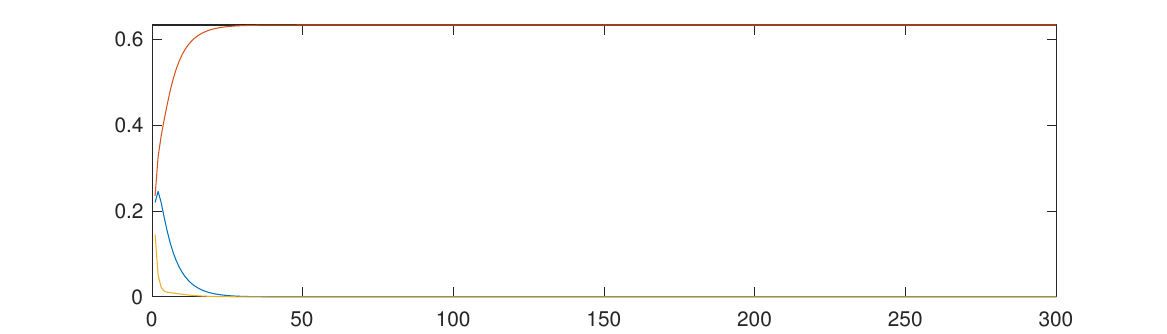}
\caption{Average infection level with respect to virus~1, virus~2, and virus~1 and virus~2  using the initial condition in Figure \ref{fig:hetero_usa}.}
\label{fig:thm3_usa}
\end{figure}
\section{Conclusion}\label{con}
We have addressed the problem of simultaneous infection of an individual (resp. subpopulation) by possibly two viruses. We derived a coupled bi-virus model from a $4^n$-state  Markov process. 
We identified a condition that leads to the extinction of both 
viruses; likewise a condition that causes one of the viruses to become endemic in the population. Subsequently, we provided a sufficient condition and two necessary conditions for 
local exponential convergence of boundary equilibria. With respect to coexistence equilibria, we conclusively ruled out the existence of the following 
types of coexisting equilibria: i) a point in the state space where  for each node there is a non-trivial fraction infected only by virus~1,  a non-trivial fraction infected only by virus~2, but no fraction that is infected by both viruses~1 and~2; and ii) a point in the state space where  for each node there is a fraction that is infected simultaneously by both viruses~1 and 2, but no fraction is infected only by virus~1 (resp. virus~2).  We provided a necessary condition for the existence of certain other kinds of coexisting equilibria. Finally, we showed that the coupled bi-virus model is not monotone.
\par The fact that the coupled bivirus system is not monotone makes its stability analysis harder. However, one could leverage the theory of singular perturbations for monotone systems \citep{wang2006remark} to possibly draw conclusions on the generic convergence of the coupled bivirus system, whereas one could possibly take recourse to the Lyapunov techniques espoused in \citep{shuai2013global} to establish global asymptotic stability of boundary equilibria.
Other problems of further interest include  identifying  condition(s) 
for stability (local or global) 
of  various coexisting equilibria. 
\vspace{-3mm}
\section*{Acknowledgements}
We thank Prof.~Wendy Anne Beauvais (Department of Comparative Pathobiology, College of Veterinary Medicine, Purdue University) for discussions regarding co-infection in animal populations that have contributed to the design of our model. We also thank the anonymous reviewers whose feedback has improved the paper.

\vspace{-3mm}
\bibliographystyle{apalike} 
\bibliography{bib}

\begin{thebibliography}{}

\bibitem[Ahn et~al., 2006]{ahn2006epidemic}
Ahn, Y.-Y., Jeong, H., Masuda, N., and Noh, J.~D. (2006).
\newblock Epidemic dynamics of two species of interacting particles on
  scale-free networks.
\newblock {\em Physical Review E}, 74(6):066113.

\bibitem[Alemu et~al., 2013]{alemu2013effect}
Alemu, A., Shiferaw, Y., Addis, Z., Mathewos, B., and Birhan, W. (2013).
\newblock Effect of malaria on {HIV}/{AIDS} transmission and progression.
\newblock {\em Parasites \& Vectors}, 6(1):1--8.

\bibitem[Arcede et~al., 2020]{arcede2020accounting}
Arcede, J.~P., Caga-Anan, R.~L., Mentuda, C.~Q., and Mammeri, Y. (2020).
\newblock Accounting for symptomatic and asymptomatic in a {SEIR}-type model of
  {COVID-19}.
\newblock {\em Mathematical Modelling of Natural Phenomena}, 15:34.

\bibitem[Arzt et~al., 2021]{arzt2021simultaneous}
Arzt, J., Fish, I.~H., Bertram, M.~R., Smoliga, G.~R., Hartwig, E.~J., Pauszek,
  S.~J., Holinka-Patterson, L., Diaz-San~Segundo, F.~C., Sitt, T., Rieder, E.,
  et~al. (2021).
\newblock Simultaneous and staggered foot-and-mouth disease virus coinfection
  of cattle.
\newblock {\em Journal of Virology}, 95(24):e01650--21.

\bibitem[Bailey et~al., 1975]{bailey1975mathematical}
Bailey, N.~T. et~al. (1975).
\newblock {\em The Mathematical Theory of Infectious Diseases and Its
  Applications}.
\newblock Charles Griffin \& Company Ltd, Bucks, U.K.

\bibitem[Berman and Plemmons, 1994]{berman1994nonnegative}
Berman, A. and Plemmons, R.~J. (1994).
\newblock {\em Nonnegative Matrices in the Mathematical Sciences}.
\newblock SIAM.

\bibitem[Bernoulli, 1760]{bernoulli1760essai}
Bernoulli, D. (1760).
\newblock Essai d'une nouvelle analyse de la mortalit{\'e} caus{\'e}e par la
  petite v{\'e}role et des avantages de l'inoculation pour la pr{\'e}venir.
\newblock {\em Histoire de l'Acad. Roy. Sci.(Paris) avec M{\'e}m. des Math. et
  Phys. and M{\'e}m}, pages 1--45.

\bibitem[Beutel et~al., 2012]{beutel2012interacting}
Beutel, A., Prakash, B.~A., Rosenfeld, R., and Faloutsos, C. (2012).
\newblock Interacting viruses in networks: Can both survive?
\newblock In {\em Proceedings of the 18th ACM SIGKDD International Conference
  on Knowledge Discovery and Data Mining}, pages 426--434. ACM.

\bibitem[Bhat et~al., 2022]{bhat2022coinfection}
Bhat, S., James, J., Sadeyen, J.-R., Mahmood, S., Everest, H.~J., Chang, P.,
  Walsh, S.~K., Byrne, A.~M., Mollett, B., Lean, F., et~al. (2022).
\newblock Coinfection of chickens with {H9N2} and {H7N9} avian influenza
  viruses leads to emergence of reassortant {H9N9} virus with increased fitness
  for poultry and a zoonotic potential.
\newblock {\em Journal of Virology}, 96(5):e01856--21.

\bibitem[Bloom et~al., 2018]{bloom2018epidemics}
Bloom, D.~E., Cadarette, D., and Sevilla, J. (2018).
\newblock Epidemics and {E}conomics.
\newblock {\em Finance \& Development}, 55(002).

\bibitem[Briat, 2017]{briat2017sign}
Briat, C. (2017).
\newblock Sign properties of {M}etzler matrices with applications.
\newblock {\em Linear Algebra and its Applications}, 515:53--86.

\bibitem[Castillo-Chavez et~al., 1989]{castillo1989epidemiological}
Castillo-Chavez, C., Hethcote, H.~W., Andreasen, V., Levin, S.~A., and Liu,
  W.~M. (1989).
\newblock Epidemiological models with age structure, proportionate mixing, and
  cross-immunity.
\newblock {\em Journal of Mathematical Biology}, 27(3):233--258.

\bibitem[Chen et~al., 2013]{chen2013outbreaks}
Chen, L., Ghanbarnejad, F., Cai, W., and Grassberger, P. (2013).
\newblock Outbreaks of coinfections: The critical role of cooperativity.
\newblock {\em EPL (Europhysics Letters)}, 104(5):50001.

\bibitem[Chu and Lee, 2008]{chu2008hepatitis}
Chu, C.-J. and Lee, S.-D. (2008).
\newblock Hepatitis {B} virus/{H}epatitis {C} virus coinfection: Epidemiology,
  clinical features, viral interactions and treatment.
\newblock {\em Journal of Gastroenterology and Hepatology}, 23(4):512--520.

\bibitem[Creighton et~al., 2003]{creighton2003co}
Creighton, S., Tenant-Flowers, M., Taylor, C.~B., Miller, R., and Low, N.
  (2003).
\newblock Co-infection with gonorrhoea and chlamydia: How much is there and
  what does it mean?
\newblock {\em International Journal of STD \& AIDS}, 14(2):109--113.

\bibitem[de~Souza et~al., 2012]{de2012hev}
de~Souza, A. J.~S., Gomes-Gouv{\^e}a, M.~S., Soares, M. d. C.~P., Pinho, J.
  R.~R., Malheiros, A.~P., Carneiro, L.~A., dos Santos, D. R.~L., and Pereira,
  W. L.~A. (2012).
\newblock {HEV} infection in swine from {Eastern Brazilian Amazon}: Evidence of
  co-infection by different subtypes.
\newblock {\em Comparative Immunology, Microbiology and Infectious Diseases},
  35(5):477--485.

\bibitem[Dupont-Rouzeyrol et~al., 2015]{dupont2015co}
Dupont-Rouzeyrol, M., O’Connor, O., Calvez, E., Daures, M., John, M.,
  Grangeon, J.-P., and Gourinat, A.-C. (2015).
\newblock Co-infection with {Z}ika and {D}engue viruses in 2 patients, {New}
  {Caledonia}, 2014.
\newblock {\em Emerging Infectious Diseases}, 21(2):381.

\bibitem[Fall et~al., 2007]{FallMMNP07}
Fall, A., Iggidr, A., Sallet, G., and Tewa, J.~J. (2007).
\newblock Epidemiological models and {L}yapunov functions.
\newblock {\em Mathematical Modelling of Natural Phenomena}, 2(1):55--73.

\bibitem[Hethcote, 2000]{hethcote2000mathematics}
Hethcote, H.~W. (2000).
\newblock The mathematics of infectious diseases.
\newblock {\em SIAM Review}, 42(4):599--653.

\bibitem[Hirsch, 1988]{hirsch1988stability}
Hirsch, M.~W. (1988).
\newblock Stability and convergence in strongly monotone dynamical systems.
\newblock {\em Journal fur die reine und angewandte Mathmatik}.

\bibitem[Janson et~al., 2020]{axel2020TAC}
Janson, A., Gracy, S., Par{\'e}, P.~E., Sandberg, H., and Johansson, K.~H.
  (2020).
\newblock Networked multi-virus spread with a shared resource: Analysis and
  mitigation strategies.
\newblock {\em arXiv preprint arXiv:2011.07569}.

\bibitem[Kermack and McKendrick, 1927]{kermack1927contribution}
Kermack, W.~O. and McKendrick, A.~G. (1927).
\newblock A contribution to the mathematical theory of epidemics.
\newblock {\em Proceedings of the royal society of london. Series A, Containing
  papers of a mathematical and physical character}, 115(772):700--721.

\bibitem[Khalil, 2002]{khalil2002nonlinear}
Khalil, H. (2002).
\newblock {\em Nonlinear Systems}.
\newblock Prentice Hall.

\bibitem[Khanafer et~al., 2016]{khanafer2016stability}
Khanafer, A., Ba{\c{s}}ar, T., and Gharesifard, B. (2016).
\newblock Stability of epidemic models over directed graphs: A positive systems
  approach.
\newblock {\em Automatica}, 74:126--134.

\bibitem[Krauland et~al., 2022]{krauland2022impact}
Krauland, M.~G., Galloway, D.~D., Raviotta, J.~M., Zimmerman, R.~K., and
  Roberts, M.~S. (2022).
\newblock Impact of low rates of influenza on next-season influenza infections.
\newblock {\em American Journal of Preventive Medicine}, 62(4):503--510.

\bibitem[Liu et~al., 2016]{liu2016onthe}
Liu, J., Par\'{e}, P.~E., Nedi\'{c}, A., Tang, C.~Y., Beck, C.~L., and
  Ba\c{s}ar, T. (2016).
\newblock On the analysis of a continuous-time bi-virus model.
\newblock In {\em Proceedings of the 55th IEEE Conference on Decision and
  Control}, pages 290--295.

\bibitem[Liu et~al., 2019]{liu2019analysis}
Liu, J., Par{\'e}, P.~E., Nedi{\'c}, A., Tang, C.~Y., Beck, C.~L., and
  Ba{\c{s}}ar, T. (2019).
\newblock Analysis and control of a continuous-time bi-virus model.
\newblock {\em IEEE Transactions on Automatic Control}, 64(12):4891--4906.

\bibitem[Matouk, 2020]{matouk2020complex}
Matouk, A. (2020).
\newblock Complex dynamics in susceptible-infected models for {COVID-19} with
  multi-drug resistance.
\newblock {\em Chaos, Solitons \& Fractals}, 140:110257.

\bibitem[Mei et~al., 2017]{mei2017dynamics}
Mei, W., Mohagheghi, S., Zampieri, S., and Bullo, F. (2017).
\newblock On the dynamics of deterministic epidemic propagation over networks.
\newblock {\em Annual Reviews in Control}, 44:116--128.

\bibitem[Meyer, 2000]{meyer2000matrix}
Meyer, C. (2000).
\newblock {\em Matrix Analysis and Applied Linear Algebra}.
\newblock Other Titles in Applied Mathematics. SIAM.

\bibitem[Mieghem et~al., 2009]{OmicTN09}
Mieghem, P.~V., Omic, J., and Kooij, R. (2009).
\newblock Virus spread in networks.
\newblock {\em IEEE/ACM Transactions on Networking}, 17(1):1--14.

\bibitem[Munster and Fouchier, 2009]{munster2009avian}
Munster, V. and Fouchier, R. (2009).
\newblock Avian influenza virus: Of virus and bird ecology.
\newblock {\em Vaccine}, 27(45):6340--6344.

\bibitem[Newman and Ferrario, 2013]{newman2013interacting}
Newman, M.~E. and Ferrario, C.~R. (2013).
\newblock Interacting epidemics and coinfection on contact networks.
\newblock {\em PloS one}, 8(8):e71321.

\bibitem[Newman et~al., 2002]{newman}
Newman, M. E.~J., Forrest, S., and Balthrop, J. (2002).
\newblock Email networks and the spread of computer viruses.
\newblock {\em Physical Review E}, 66(3):035101.

\bibitem[Norris, 1998]{norris1998markov}
Norris, J.~R. (1998).
\newblock {\em Markov Chains}.
\newblock Cambridge University Press.

\bibitem[Par{\'e} et~al., 2020a]{pare2020modeling}
Par{\'e}, P.~E., Beck, C.~L., and Ba{\c{s}}ar, T. (2020a).
\newblock Modeling, estimation, and analysis of epidemics over networks: An
  overview.
\newblock {\em Annual Reviews in Control}, 50:345--360.

\bibitem[Par\'{e} et~al., 2018]{acc_coupled}
Par\'{e}, P.~E., Liu, J., Beck, C.~L., and Ba\c{s}ar, T. (2018).
\newblock A coupled bi-virus spread model in networked systems.
\newblock In {\em Proceedings of the American Control Conference}, pages
  4414--4419.

\bibitem[Par{\'e} et~al., 2020b]{pare2019dtjournal}
Par{\'e}, P.~E., Liu, J., Beck, C.~L., Kirwan, B.~E., and Ba{\c{s}}ar, T.
  (2020b).
\newblock Analysis, estimation, and validation of discrete-time epidemic
  processes.
\newblock {\em IEEE Transactions on Control Systems Technology}, 28(1):79--93.

\bibitem[Pawlowski et~al., 2012]{pawlowski2012tuberculosis}
Pawlowski, A., Jansson, M., Sk{\"o}ld, M., Rottenberg, M.~E., and
  K{\"a}llenius, G. (2012).
\newblock Tuberculosis and {HIV} co-infection.
\newblock {\em PLoS Pathogens}, 8(2):e1002464.

\bibitem[Plemmons, 1977]{plemmons1977m}
Plemmons, R.~J. (1977).
\newblock M-matrix characterizations. {I}--nonsingular {M}-matrices.
\newblock {\em Linear Algebra and its Applications}, 18(2):175--188.

\bibitem[Prakash et~al., 2012]{prakash2012winner}
Prakash, B.~A., Beutel, A., Rosenfeld, R., and Faloutsos, C. (2012).
\newblock Winner takes all: Competing viruses or ideas on fair-play networks.
\newblock In {\em Proceedings of the 21st {I}nternational {C}onference on World
  Wide Web}, pages 1037--1046.

\bibitem[Qu, 2009]{qu2009cooperative}
Qu, Z. (2009).
\newblock {\em Cooperative control of dynamical systems: Applications to
  autonomous vehicles}.
\newblock Springer Science \& Business Media.

\bibitem[Rantzer, 2011]{rantzer}
Rantzer, A. (2011).
\newblock Distributed control of positive systems.
\newblock In {\em Proceedings of the 50th IEEE Conference on Decision and
  Control}, pages 6608--6611.

\bibitem[Roberts, 2021]{bbccovid}
Roberts, M. (2021).
\newblock Covid: Woman aged 90 died with double variant infection.
\newblock {\em BBC}.

\bibitem[Rothe et~al., 2020]{rothe2020transmission}
Rothe, C., Schunk, M., Sothmann, P., Bretzel, G., Froeschl, G., Wallrauch, C.,
  Zimmer, T., Thiel, V., Janke, C., Guggemos, W., et~al. (2020).
\newblock Transmission of 2019-{nCoV} infection from an asymptomatic contact in
  {Germany}.
\newblock {\em New England Journal of Medicine}, 382(10):970--971.

\bibitem[Sahneh and Scoglio, 2014]{sahneh2014competitive}
Sahneh, F.~D. and Scoglio, C. (2014).
\newblock Competitive epidemic spreading over arbitrary multilayer networks.
\newblock {\em Physical Review E}, 89(6):062817.

\bibitem[Santos et~al., 2015]{santos2015bivirus}
Santos, A., Moura, J. M.~F., and Xavier, J. M.~F. (2015).
\newblock Bi-virus {SIS} epidemics over networks: Qualitative analysis.
\newblock {\em IEEE Transactions on Network Science and Engineering},
  2(1):17--29.

\bibitem[Shuai and van~den Driessche, 2013]{shuai2013global}
Shuai, Z. and van~den Driessche, P. (2013).
\newblock Global stability of infectious disease models using {Lyapunov}
  functions.
\newblock {\em SIAM Journal on Applied Mathematics}, 73(4):1513--1532.

\bibitem[Smith, 1988]{smith1988systems}
Smith, H.~L. (1988).
\newblock Systems of ordinary differential equations which generate an order
  preserving flow. a survey of results.
\newblock {\em SIAM review}, 30(1):87--113.

\bibitem[Sontag, 2007]{sontag2007monotone}
Sontag, E.~D. (2007).
\newblock Monotone and near-monotone biochemical networks.
\newblock {\em Systems and Synthetic Biology}, 1(2):59--87.

\bibitem[Souza et~al., 2017]{souza2017note}
Souza, M., Wirth, F.~R., and Shorten, R.~N. (2017).
\newblock A note on recursive {S}chur complements, block {H}urwitz stability of
  {M}etzler matrices, and related results.
\newblock {\em IEEE Transactions on Automatic Control}, 62(8):4167--4172.

\bibitem[Van~Mieghem et~al., 2009]{van2009virus}
Van~Mieghem, P., Omic, J., and Kooij, R. (2009).
\newblock Virus spread in networks.
\newblock {\em IEEE/ACM Transactions on Networking}, 17(1):1--14.

\bibitem[Varga, 1999]{varga1999matrix}
Varga, R. (1999).
\newblock {\em Matrix Iterative Analysis}.
\newblock Springer Series in Computational Mathematics. Springer Berlin
  Heidelberg.

\bibitem[Varga, 2009]{varga2009matrix}
Varga, R.~S. (2009).
\newblock Matrix properties and concepts.
\newblock In {\em Matrix Iterative Analysis}, pages 1--30. Springer.

\bibitem[Wang and Sontag, 2006]{wang2006remark}
Wang, L. and Sontag, E.~D. (2006).
\newblock A remark on singular perturbations of strongly monotone systems.
\newblock In {\em Proceedings of the 45th IEEE Conference on Decision and
  Control}, pages 989--994. IEEE.

\bibitem[Wang et~al., 2020]{wang2020clinical}
Wang, M., Wu, Q., Xu, W., Qiao, B., Wang, J., Zheng, H., Jiang, S., Mei, J.,
  Wu, Z., Deng, Y., et~al. (2020).
\newblock Clinical diagnosis of 8274 samples with 2019-novel coronavirus in
  {Wuhan}.
\newblock {\em MedRxiv}.

\bibitem[Wang et~al., 2003]{wang2003epidemic}
Wang, Y., Chakrabarti, D., Wang, C., and Faloutsos, C. (2003).
\newblock Epidemic spreading in real networks: An eigenvalue viewpoint.
\newblock In {\em Proceedings of the 22nd International Symposium on Reliable
  Distributed Systems}, pages 25--34.

\bibitem[Xu et~al., 2012]{xu2012multi}
Xu, S., Lu, W., and Zhan, Z. (2012).
\newblock A stochastic model of multivirus dynamics.
\newblock {\em IEEE Transactions on Dependable and Secure Computing},
  9(1):30--45.

\bibitem[Ye et~al., 2022]{ye2021convergence}
Ye, M., Anderson, B.~D., and Liu, J. (2022).
\newblock Convergence and equilibria analysis of a networked bivirus epidemic
  model.
\newblock {\em SIAM Journal on Control and Optimization}, 60(2):S323--S346.

\bibitem[Zhang et~al., 2022]{zhang2022networked}
Zhang, C., Gracy, S., Ba{\c{s}}ar, T., and Par{\'e}, P.~E. (2022).
\newblock A networked competitive multi-virus {SIR} model: Analysis and
  observability.
\newblock {\em IFAC-PapersOnLine}, 55(13):13--18.

\bibitem[Zhao et~al., 2020]{zhao2020dynamics}
Zhao, L., Wang, Z.-C., and Ruan, S. (2020).
\newblock Dynamics of a time-periodic two-strain {SIS} epidemic model with
  diffusion and latent period.
\newblock {\em Nonlinear Analysis: Real World Applications}, 51:102966.

\end{thebibliography}
\vspace{-3mm}
\section*{Appendix}
\vspace{-2mm}
\noindent
\textbf{Proof of Theorem~\ref{0global=}:}
Let $y_i^{(1)}(t)=x_i^{(1)}(t)+z_i(t)$ and $y_i^{(2)}(t)=x_i^{(2)}(t)+z_i(t)$ for each $i\in[n]$.
From \rep{x1i}-\rep{zi}, the dynamics of $y_i^{(1)}(t)$ and $y_i^{(2)}(t)$ are \vspace{-2mm}
\footnotesize
\begin{align}
\dot{y}_i^{(1)}(t) &= -\delta_i^{(1)} y_i^{(1)}(t) 
+ (1-y_i^{(1)}(t)-(1-\epsilon^{(2)})x_i^{(2)}(t))\textstyle\sum_{j=1}^n \beta_{ij}^{(1)} y_j^1(t), \nonumber \\
\dot{y}_i^{(2)}(t) &= -\delta_i^{(2)} y_i^{(2)}(t) 
+ (1-y_i^{(2)}(t)-(1-\epsilon^{(1)})x_i^{(1)}(t))\textstyle\sum_{j=1}^n \beta_{ij}^{(2)} y_j^{(2)}(t). \label{mmm1}
\end{align}
\normalsize
Since $\epsilon^{(1)}, \epsilon^{(2)} \in[0,1]$, it follows that  \vspace{-3mm}
\begin{align}
\footnotesize
\dot{y}_i^{(1)}(t) &\le -\delta_i^{(1)} y_i^{(1)}(t) 
+ (1-y_i^{(1)}(t))\textstyle\sum_{j=1}^n \beta_{ij}^{(1)} y_j^{(1)}(t), \nonumber \\
\dot{y}_i^{(2)}(t) &\le -\delta_i^{(2)} y_i^{(2)}(t) 
+ (1-y_i^{(2)}(t))\textstyle\sum_{j=1}^n \beta_{ij}^{(2)} y_j^{(2)}(t), \nonumber
\end{align} 
which implies that the trajectories of $y_i^{(1)}(t)$ and $y_i^{(2)}(t)$ in \rep{mmm1} are both bounded above by the trajectory of 
a single-virus SIS model. From 
\citep[Proposition~3]{liu2016onthe}, in the case when $s(B^{(1)}-D^{(1)})\leq 0$ and $s(B^{(2)}-D^{(2)})\leq 0$, then 
all $y_i^{(1)}(t)$ and $y_i^{(2)}(t)$ asymptotically converge to $0$ for any initial condition, which implies that system \rep{x1}-\rep{z} asymptotically converges to 
the healthy state for any initial state in $\scr D$. Therefore, the healthy state is the unique equilibrium of the system.\\

\noindent
\textbf{Proof of Theorem~\ref{1epidemic}:}
Let $y_i^{(2)}(t)=x_i^{(2)}(t)+z_i(t)$ for each $i\in[n]$. From the proof of Theorem~\ref{0global=}, since $s(B^{(2)}-D^{(2)})\leq 0$ and $\epsilon^{(1)} 
\in[0,1]$, all $y_i^{(2)}(t)$ converge to zero, which implies all $x_i^{(2)}(t)$ and $z_i(t)$ converge to zero as well. 
From \rep{x1i}, the dynamics of all $x_i^{(1)}(t)$, $i\in[n]$ can be viewed as a cascade system with $x_i^{(2)}(t)$, $z_i(t)$, $i\in[n]$ as inputs. Using similar arguments to those in the proof of 
\citep[Theorem~4]{khanafer2016stability}, this cascade system is input-to-state stable. Note that with $x_i^{(2)}(t)=z_i(t)=0$ for all $i\in[n]$, the dynamics of all $x_i^{(1)}(t)$ in \rep{x1i} simplify to the single-virus networked SIS model. The theorem is then a direct consequence of \citep[Lemma 4.7]{khalil2002nonlinear} and \citep[Proposition~3]{liu2019analysis}.\\

\noindent
\textbf{Proof of Corollary~\ref{prop:multiple:viruses}:} Consider system~\eqref{x1}-\eqref{z}. 
Note that, by the definition of equilibrium, the healthy state $(\0,\0,\0)$ is an equilibrium point, regardless of whether (or not), for $m \in [2]$, $s(B^{(m)}-D^{(m)})\leq 0$. Since, by assumption, for $m \in [2]$, $s(B^{(m)}-D^{(m)})> 0$, instability of $(\0,\0,\0)$ follows from the proof of Proposition~\ref{0global<}. The existence and uniqueness of the equilibrium points $(\hat{x}^{(1)},\0,\0)$ and $(\0, \hat{x}^{(2)}, \0)$ follow from the proof of Theorem~\ref{1epidemic}.\\

We need
the following lemmas 
to prove
the remaining main results in this paper.
\vspace{-3mm}

\begin{lemma} \label{lem:eigspec}
\citep[Proposition~1]{liu2019analysis} Suppose that $\Lambda$ is a negative diagonal matrix and $N$ is an irreducible nonnegative matrix. 
Let $M$ be the irreducible Metzler matrix $M = \Lambda+N$. 
Then, $s(M) < 0$ if and only if \break $\rho(-\Lambda^{-1} N) < 1, s(M)=0$ if and only if $\rho(-\Lambda^{-1} N) = 1$, and $s(M)>0$ if and only if, $\rho(-\Lambda^{-1} N) > 1$.
\end{lemma}

\vspace{-5mm}
\begin{lemma} \label{lem:perron_frob}
\citep[Chapter 8.3]{meyer2000matrix} \citep[Theorem~2.7]{varga1999matrix} 
Suppose that $N$ is an irreducible nonnegative matrix. Then, 
\vspace{-2mm}
\begin{enumerate}[label=(\roman*)]\itemsep-0.4mm
    \item $r = \rho(N)$ is a simple eigenvalue of $N$. \label{item:perfrob_simpleeig}
    \item There is an eigenvector $\zeta \gg \textbf{0}$ corresponding to the eigenvalue $r$. \label{item:perfrob_pos_exists}
    \item $x > \textbf{0}$ is an eigenvector only if $Nx = rx$ and $x \gg \textbf{0}$. 
    \label{item:perfrob_pos_necess}
    \vspace{-2.5ex}
    \item If $A$ is a nonnegative matrix such that $A < N$, then $\rho(A) < \rho(N)$. \label{item:perfrob_matrix_ineq}
\end{enumerate}

\end{lemma}

\begin{lemma} \label{lem:perron_frob_metz}
\citep[Lemma~2.3]{varga1999matrix} Suppose that $M$ is an irreducible Metzler matrix. Then $r = s(M)$ is a simple eigenvalue of $M$, and if $M\zeta = r \zeta$, then $\zeta ~\gg~\textbf{0}$.
\end{lemma}
\vspace{-4mm}

\begin{lemma}\label{lem:phil}
Suppose that Assumptions~\ref{x0} and \ref{para} hold and that $\delta_i^{(1)},\delta_i^{(2)}>0$ for all $i\in[n]$. 
If $(\hat x^{(1)},\hat x^{(2)},\hat z)$ is an equilibrium of system \rep{x1}-\rep{z} such that $\hat z_i>0$ for at least one $i\in[n]$, then $\hat x^{(1)},\hat x^{(2)},\hat z \gg \0$. 
\end{lemma}
\textit{Proof:} 
Suppose that, to the contrary, there exists an $i\in[n]$ such that $\hat x_i^1=0$.
From~\rep{x1i}, it holds that $\hat z_i=0$. It follows from~\rep{zi} that 
$\hat x_i^2 \epsilon^{(2)} \textstyle\sum_{j=1}^n\beta_{ij}^1(\hat x_j^1+\hat z_j) = 0,$
which implies that $\hat x_i^2= 0$ or $ \epsilon^{(2)} \sum_{j=1}^n\beta_{ij}^1(\hat x_j^1+\hat z_j) = 0$. 
If the latter is true, since $\beta_{ij}^1>0$ whenever $j\in\scr N_i^1$, it holds that $\hat x_j^1=0$ and $\hat z_j=0$ for all $j\in\scr N_i^1$. If $\hat x_i^2= 0$, it follows from~\rep{x1i} that $\sum_{j=1}^n \beta_{ij}^1(\hat x_j^1+\hat z_j) = 0$, which also implies that  $\hat x_j^1=0$ and $\hat z_j=0$ for all $j\in\scr N_i^1$.
By repeating this argument, since the graph of $B^{(1)}$ is strongly connected, $\hat z_i=0$ for all $i\in[n]$, which is a contradiction. Therefore, $\hat x_i^1>0$ for all $i\in[n]$.
Similarly, it holds that $\hat x_i^2>0$ for all $i\in[n]$.

Now suppose that, to the contrary, there exists one $i\in[n]$ for which $\hat z_i=0$. From~\rep{zi}, it holds that \break 
$\hat x_i^1\epsilon^{(1)} \sum_{j=1}^n \beta_{ij}^2(\hat x_j^2+\hat z_j) =0$ and  
$\hat x_i^2\epsilon^{(2)} \sum_{j=1}^n \beta_{ij}^1(\hat x_j^1+\hat z_j)=0$,
which respectively imply that $\hat x_i^1=0$ or $\hat x_j^2=0$ for all $j\in\scr N_i$, and $\hat x_i^2=0$ or $\hat x_j^1=0$ for all $j\in\scr N_i$.
From the preceding discussion, each of these four cases implies that $\hat z_i=0$ for all $i\in[n]$, which is impossible. Therefore, $\hat z_i>0$ for all $i\in[n]$.
This completes the proof.
\vspace{-3mm}
\begin{lemma}\label{lem:endemic:equilibria:characteristics}
Consider system~\eqref{x1}-\eqref{z} under Assumptions~\ref{x0} and~\ref{para-1}. If $(\hat{x}^{(1)}, \hat{x}^{(2)}, \hat{z}) \in \mathcal D$ such that $\hat{x}^{(1)}, \hat{x}^{(2)}, \hat{z} > \textbf{0}$, 
then
$\textbf{0} \ll \hat{x}^{(1)} \ll \textbf{1}$ or $\textbf{0} \ll \hat{x}^{(2)} \ll \textbf{1}$ or $\textbf{0} \ll \hat z \ll \textbf{1}$. Furthermore, $\hat{x}^{(1)}+ \hat{x}^{(2)} + \hat{z}\ll \textbf{1}$.
\end{lemma}
\textit{Proof:} Suppose that $(\hat{x}^{(1)}, \hat{x}^{(2)}, \hat{z}) \in \mathcal D$ is an endemic equilibrium of~\eqref{x1}-\eqref{z} such that $\hat{x}^{(1)}>\textbf{0}$, $\hat{x}^{(2)}>\textbf{0}$, and $\hat{z}>\textbf{0}$. Since $(\hat{x}^{(1)}, \hat{x}^{(2)}, \hat{z})$ is an equilibrium point, from \eqref{x1i}-\eqref{zi}, we have the following: \vspace{-3mm}
\begin{align}\label{eq:sum:less:than:1}
    &(1-\hat{x}_i^{(1)}-\hat{x}_i^{(2)}-\hat{z}_i)  \nonumber \\ &\times\big(\textstyle \sum_{j=1}^{n}\beta^{(1)}_{ij}(\hat{x}_j^{(1)}+\hat{z}_j) + \textstyle  \sum_{j=1}^{n}\beta^{(2)}_{ij}(\hat{x}_j^{(2)}+\hat{z}_j) \big) \nonumber \\=& \delta_i^{(1)} \hat{x}_i^{(1)} + \delta_i^{(2)} \hat{x}_i^{(2)}.
\end{align}
Suppose, by way of contradiction, that, for some $i \in [n]$, $\hat{x}^{(1)}_i+ \hat{x}^{(2)}_i + \hat{z}_i\geq 1$. Hence, $1-(\hat{x}^{(1)}_i+ \hat{x}^{(2)}_i + \hat{z}_i) \leq 0$. Since, by Assumption~\ref{para-1}, the matrices $B^{(1)}$ and $B^{(2)}$ are nonnegative irreducible, and  from Lemma~\ref{box} we know that $x_i^{(1)}(t), x_i^{(2)}(t), z_i(t) \in [0,1]$ for all $i \in [n]$ and for all $t \in \mathbb{R}_{\geq 0}$, it follows that \vspace{-3mm}\scriptsize
\begin{align} \label{eq:lessthanorequaltozero}
     &(1-\hat{x}_i^{(1)}-\hat{x}_i^{(2)}-\hat{z}_i)  \nonumber \\
     & \times \big(\textstyle \sum_{j=1}^{n}\beta^{(1)}_{ij}(\hat{x}_j^{(1)}+\hat{z}_j) + \textstyle  \sum_{j=1}^{n}\beta^{(2)}_{ij}(\hat{x}_j^{(2)}+\hat{z}_j) \big)  \leq 0.
 \end{align}
\normalsize
 Hence from~\eqref{eq:sum:less:than:1} and~\eqref{eq:lessthanorequaltozero}, it follows that 
 $\delta_i^{(1)} \hat{x}_i^{(1)} + \delta_i^{(2)} \hat{x}_i^{(2)} \leq 0$, which, because i) by Assumption~\ref{para-1}, $\delta_i^{(1)}, \delta_i^{(2)}>0$,
and 
ii) $\hat{x}^{(1)}, \hat{x}^{(2)} \in \mathcal D$, further implies that $\hat{x}_i^{(1)}=0$ 
and  $\hat{x}_i^{(2)}=0$. Since by assumption, $\hat{x}^{(1)}_i+ \hat{x}^{(2)}_i + \hat{z}_i\geq 1$, 
we have
 that $\hat{z}_i\geq 1$. 
Given that, by assumption,
$\hat{z}_i \in \mathcal D$, it must follow that  $\hat{z}_i = 1$. 
Using $\hat{x}_i^1=0$,  $\hat{x}_i^2=0$, and $\hat{z}_i = 1$ in~\eqref{zi} yields \vspace{-3mm}
 \begin{align} \label{eq:contra:sum:greaterthan:1}
     0&=-(\delta_i^{(1)} + \delta_i^{(2)}).
 \end{align}
By Assumption~\ref{para-1}, $\delta_i^{(1)}>0$ and $\delta_i^{(2)}>0$, it must be that $-(\delta_i^{(1)} + \delta_i^{(2)})<0$, which contradicts~\eqref{eq:contra:sum:greaterthan:1}. Hence, for each $i \in [n]$, $\hat{x}^{(1)}_i+ \hat{x}^{(2)}_i + \hat{z}_i< 1$, which further implies that $\hat{x}^{(1)}+ \hat{x}^{(2)} + \hat{z}\ll \textbf{1}$. Consequently, since $(\hat{x}^{(1)}, \hat{x}^{(2)}, \hat{z}) \in \mathcal D$, it follows that $\hat{x}^{(1)} \ll \textbf{1}$, $\hat{x}^{(2)} \ll \textbf{1}$, and $\hat{z} \ll \textbf{1}$.
 
It remains to be shown that if $(\hat{x}^{(1)}, \hat{x}^{(2)}, \hat{z}) \in \mathcal D$ is an endemic equilibrium of~\eqref{x1}-\eqref{z}, then $\textbf{0} \ll \hat{x}^{(1)} $, $\textbf{0} \ll \hat{x}^{(2)} $, and $\textbf{0} \ll \hat{z}$.
 Suppose, by way of contradiction, that $(\hat{x}^{(1)}, \hat{x}^{(2)}, \hat{z}) \in \mathcal D$ is an endemic equilibrium, but yet there exist $i,k, \ell \in [n]$ such that $\hat{x}_i^{(1)}=0$, $\hat{x}_k^{(2)}=0$, and $\hat{z}_\ell=0$. Since $\hat{x}_i^{(1)}=0$, then from~\eqref{x1i} we have
 $(1-\hat{x}_i^{(2)}-\hat{z}_i)
     \textstyle \sum_{j=1}^{n}\beta_{ij}^{(1)}(\hat{x}_j^{(1)}+\hat{z}_j)=-\delta_i^{(2)}\hat{z}_i  \leq 0$,
where the inequality is obtained by noting that $\delta_i^{(2)}>0$ and $(\hat{x}^{(1)}, \hat{x}^{(2)}, \hat{z}) \in \mathcal D$. 
 Consequently, since $\hat{x}^{(1)}+ \hat{x}^{(2)} + \hat{z}\ll \textbf{1}$, it follows that $  \textstyle\sum_{j=1}^{n}\beta_{ij}^{(1)}(\hat{x}_j^{(1)}+\hat{z}_j)\leq 0$. 
 Given that,
 by Assumption~\ref{para-1}, $B^{(1)}$ is nonnegative irreducible, and 
 because $(\hat{x}^{(1)}, \hat{x}^{(2)}, \hat{z}) \in \mathcal D$, it follows that $\hat{x}_j^{(1)}=0$ and $\hat{z}_j=0$. By relabeling $j$ as $i$, and by repeating the same arguments, it can be shown iteratively that $\hat{x}_i^{(1)}=0$ for all $i \in [n]$. This implies that $\hat{x}^{(1)}=\textbf{0}$, which 
leads to a contradiction 
to the assumption that $\hat{x}^1 > \textbf{0}$. Hence, $\textbf{0}\ll \hat{x}^{(1)}$. Analogously, it can be shown that $\textbf{0}\ll \hat{x}^{(2)}$, and $\textbf{0}\ll \hat{z}$.\\

\noindent
\textbf{Proof of Theorem~\ref{thm:local:expo:stab:single-virus}:}
In order to prove the claim in Theorem~\ref{thm:local:expo:stab:single-virus}, we need the following lemma.\vspace{-3mm}
\begin{lemma}\citep[Corollary~1]{souza2017note}\label{lem:souza}
Let $A \in \mathbb{R}^{n \times n}$ be a Metzler matrix partitioned in blocks 
as
\begin{align}
    A = \scriptsize \begin{bmatrix}
     A_{11} && A_{12}\\
      A_{21} && A_{22}
    \end{bmatrix} \nonumber
\end{align}
in which $A_{11}$ and $A_{22}$ are square matrices. Define $A/A_{11}: =A_{22}-A_{21}A_{11}^{-1}A_{12}$. 
Then, $A$ is Hurwitz if, and only if, 
$A_{11}$ and $A/A_{11}$ are Hurwitz Metzler matrices.
\end{lemma}
{\bf Proof of Theorem~\ref{thm:local:expo:stab:single-virus}:}
Consider the equilibrium point \break $(\hat{x}^{(1)}, \mathbf{0}, \mathbf{0})$. The Jacobian matrix of 
this equilibrium point is as given in~\eqref{Jacobian:hatx1}.
\begin{figure*}[h]
	{\noindent}
\begin{equation} \label{Jacobian:hatx1}
J(\hat x^{(1)},\mathbf{0},\mathbf{0})  = 
 \scriptsize
\left[
 \begin{matrix} 
  -D^{(1)} +(I-\hat{X}^{(1)})B^{(1)} - \hat{B}^{(1)}
  && -\hat{B}^{(1)}-\epsilon \hat{X}^{(1)}B^{(2)} && D^{(2)}-\hat{B}^{(1)}+ (I-\hat{X}^{(1)})B^{(1)}-\epsilon\hat{X}^{(1)}B^{(2)} \\
  \mathbf{0} && -D^{(2)}+(I-\hat{X}^{(1)})B^{(2)}-\epsilon\hat{B}^{(1)} && (I-\hat{X}^{(1)})B^{(2)}+D^{(1)} \\
    \mathbf{0} && \epsilon\hat{B}^{(1)} + \epsilon \hat{X}^{(1)}B^{(2)} && -D^{(1)}-D^{(2)}+\epsilon \hat{X}^{(1)}B^{(2)}
 \end{matrix}
 \right]
\end{equation}
\vspace{-3ex}
\end{figure*}
Hence, we can rewrite $J(\hat x^{(1)},\mathbf{0},\mathbf{0})$ as \scriptsize
\begin{align}\label{J:forhatx1:partitioned}
    J(\hat x^{(1)},\mathbf{0},\mathbf{0})
    =
    & 
    \begin{bmatrix} 
      -D^{(1)} +(I-\hat{X}^{(1)})B^{(1)} - \hat{B}^{(1)}
      && \hat{J} \\
      \mathbf{0} && \tilde{J}
    \end{bmatrix}. 
\end{align}
\normalsize
where $\hat{J} = [\begin{smallmatrix}-\hat{B}^{(1)}
-\epsilon \hat{X}^{(1)}B^{(2)} && D^{(2)}-\hat{B}^{(1)}
+ (I-\hat{X}^{(1)})B^{(1)}-\epsilon\hat{X}^{(1)}B^{(2)} \end{smallmatrix}]$,
while \scriptsize 
\begin{align}\label{jacob:hatx1:22submatrix} 
&\tilde{J} = 
&
\begin{bmatrix} 
-D^{(2)}+(I-\hat{X}^{(1)})B^{(2)}-\epsilon\hat{B}^{(1)} && (I-\hat{X}^{(1)})B^{(2)}+D^{(1)} \\
\epsilon\hat{B}^{(1)}
+ \epsilon \hat{X}^{(1)}B^{(2)} && -D^{(1)}-D^{(2)}+\epsilon \hat{X}^{(1)}B^{(2)} \end{bmatrix}. \end{align}
\normalsize From~\eqref{J:forhatx1:partitioned} it is clear that the matrix $J(\hat x^{(1)},\mathbf{0},\mathbf{0})$ is block upper triangular. Therefore, $s(J(\hat x^{(1)},\mathbf{0},\mathbf{0}))<0$ if, and only if, the following conditions are satisfied: i) $s(-D^{(1)} +(I-\hat{X}^{(1)})B^{(1)} - \hat{B}^{(1)})<0$, and ii) $s(\tilde{J})<0$. 
Since $(\hat{x}^{(1)}, \mathbf{0}, \mathbf{0})$ is an equilibrium point, from~\eqref{x1} we obtain: \vspace{-3mm} \scriptsize
\begin{align}\label{Q:defn}
    (-D^{(1)} + (I-\hat{X}^{(1)})B^{(1)})\hat{x}^{(1)} = \mathbf{0}.
\end{align}
\normalsize
Define $Q: = D^{(1)} -(I-\hat{X}^{(1)})B^{(1)}$. 
Since $-Q$ is an irreducible Metzler matrix, and since $\hat{x}^{(1)} \gg \mathbf{0}$, 
applying Lemma~\ref{lem:perron_frob_metz} to~\eqref{Q:defn} 
yields $s(-Q) =0$. Note that $-Q$ being Meztler implies that $Q$ is an M-matrix, and since $s(-Q) =0$ it follows that $Q$ is an irreducible singular M-matrix. Since $B^{(1)}$ is nonnegative irreducible, and $\hat{x}^{(1)} \gg \mathbf{0}$, it follows that the matrix 
$\hat{B}^{(1)}$ has at least one diagonal element that is strictly positive. Therefore, due to \citep[Corollary~4.33]{qu2009cooperative}, the matrix 
$Q+\hat{B}^{(1)}$ is a non-singular M-matrix, which further implies that $s(-D^{(1)} +(I-\hat{X}^{(1)})B^{(1)} - \hat{B}^{(1)}) <0$.\\ 
From~\eqref{jacob:hatx1:22submatrix}, it is immediate that $\tilde{J}$ is Metzler.
We now prove, by using Lemma~\ref{lem:souza},  that the matrix $\tilde{J}$ is also Hurwitz. Observe that $\tilde{J}_{11} = -D^{(2)}+(I-\hat{X}^{(1)})B^{(2)}-\epsilon
\hat{B}^{(1)}$. Since by Lemma~\ref{lem:endemic:equilibria:characteristics}, $\hat{x}^{(1)} \ll \mathbf{1}$, it follows that $(I-\hat{X}^{(1)})$ is positive diagonal. Since by Assumption~\ref{para}, $B^{(2)}$ is nonnegative irreducible, 
it is clear that $(I-\hat{X}^{(1)})B^{(2)}$ is nonnegative irreducible, and hence
$\tilde{J}_{11}$ is Metzler. By assumption, $s(-D^{(2)}+(I-\hat{X}^{(1)})B^{(2)}) <0$. Hence, by using similar arguments involved in showing that $s(-D^{(1)} +(I-\hat{X}^{(1)})B^{(1)} -\hat{B}^{(1)}) 
< 0$, we can also show that $s(\tilde{J}_{11})<0$. Therefore, $\tilde{J}_{11}$ is a Hurwitz 
Metzler matrix. 

Note that \footnotesize \begin{align}
    \tilde{J}/\tilde{J}_{11} = &(-D^{(1)}-D^{(2)}+\epsilon
\hat{X}^{(1)} B^{(2)}) \nonumber\\&-(\epsilon \hat{B}^{(1)}+ \epsilon \hat{X}^{(1)}B^{(2)})   (-D^{(2)} + (I-\hat{X}^{(1)})B^{(2)} \nonumber\\&-\epsilon \hat{B}^{(1)})^{-1}((I-\hat{X}^{(1)})B^{(2)} + D^{(1)}). \nonumber
\end{align}

\normalsize
\noindent Note that the existence of $(-D^{(2)} + (I-\hat{X}^{(1)})B^{(2)}-\epsilon \hat{B}^{(1)})^{-1}$ 
is a consequence of
Hurwitzness of 
$(-D^{(2)} + (I-\hat{X}^{(1)})B^{(2)}-\epsilon \hat{B}^{(1)})$ \citep{briat2017sign,berman1994nonnegative}.  Furthermore, observe that since 
$(-D^{(2)} + (I-\hat{X}^{(1)})B^{(2)}-\epsilon \hat{B}^{(1)})^{-1}$ is Metzler, it follows that 
$-(-D^{(2)} + (I-\hat{X}^{(1)})B^{(2)}-\epsilon \hat{B}^{(1)})^{-1}$ is an M-matrix. Since $-(-D^{(2)} + (I-\hat{X}^{(1)})B^{(2)}-\epsilon \hat{B}^{(1)})^{-1}$ exists, it follows that 
$-(-D^{(2)} + (I-\hat{X}^{(1)})B^{(2)}-\epsilon \hat{B}^{(1)})^{-1}$ is a nonsingular M-matrix. Hence, $-(-D^{(2)} + (I-\hat{X}^{(1)})B^{(2)}-\epsilon \hat{B}^{(1)})^{-1}$ is a nonnegative matrix \citep[F.15]{plemmons1977m}. 
Define
\scriptsize
\begin{align}
 Q_1:=   &-(\epsilon \hat{B}^{(1)}+ \epsilon \hat{X}^{(1)}B^{(2)}) \times\nonumber \\&  (-D^{(2)} + (I-\hat{X}^{(1)})B^{(2)}-\epsilon \hat{B}^{(1)})^{-1}((I-\hat{X}^{(1)})B^{(2)} + D^{(1)}). \nonumber
\end{align}
\normalsize 
Since
$\epsilon \geq 0$, from Assumption~\ref{para-1} it is clear that $\hat{B}^{(1)}$ is  nonnegative diagonal matrix, $B^{(2)}$ is nonnegative, and  $D^{(1)}$ and $D^{(2)}$ are positive diagonal matrices. Hence, since from Lemma~\ref{lem:endemic:equilibria:characteristics} the matrix $(I-\hat{X}^{(1)})$ is also positive,
it is immediate  that  $Q_1$
 is nonnegative. Therefore, since $(-D^{(1)}-D^{(2)}+\epsilon
\hat{X}^{(1)} B^{(2)})$ is Metzler, it follows that the matrix 
$(-D^{(1)}-D^{(2)}+\epsilon
\hat{X}^{(1)} B^{(2)})+Q_1$ is Metzler. Consequently,  $\tilde{J}/\tilde{J}_{11}$ is Metzler. Since, by assumption, $s\Big((-D^{(1)}-D^{(2)}+\epsilon
\hat{X}^{(1)} B^{(2)})-(\epsilon \hat{B}^{(1)}+ \epsilon \hat{X}^{(1)}B^{(2)})   (-D^{(2)} + (I-\hat{X}^{(1)})B^{(2)}-\epsilon \hat{B}^{(1)})^{-1}((I-\hat{X}^{(1)})B^{(2)} + D^{(1)})\Big) <0$, it means that $s((-D^{(1)}-D^{(2)}+\epsilon
\hat{X}^{(1)} B^{(2)})+Q_1)<0$,
thereby implying that $s(\tilde{J}/\tilde{J}_{11}) <0$.  Thus, $\tilde{J}/\tilde{J}_{11}$ is Hurwitz 
Metzler. Therefore, from Lemma~\ref{lem:souza}, it follows that the matrix $\tilde{J}$ is Hurwitz.
Hence, we can conclude that $s(J(\hat{x}^{(1)}, \mathbf{0}, \mathbf{0})) <0$. Local exponential  stability of $(\hat{x}^{(1)}, \mathbf{0}, \mathbf{0})$ then follows  from \citep[Theorem 4.15 and Corollary~4.3]{khalil2002nonlinear}, thus concluding the proof.~$\blacksquare$\\

\noindent
\textbf{Proof of Proposition~\ref{prop:hatx1:local:expo:stability:nec}:} Consider the matrix $\tilde J$ given in~\eqref{jacob:hatx1:22submatrix}. Suppose that $s(-D^{(2)} + (I-\hat{X}^{(1)})B^{(2)}-\epsilon\hat{B}^{(1)}) \geq 0$.  
This implies that $[\tilde J]_{11}$ (i.e., the $11$-block of $\tilde J$) is not Hurwitz. 
Hence, from Lemma~\ref{lem:souza}, it follows that $\tilde J$ is not Hurwitz. 
That is, $s(\tilde J) \geq 0$. Therefore, from~\eqref{Jacobian:hatx1} it follows that
$J(\hat x^{(1)},\mathbf{0},\mathbf{0})$ is not Hurwitz, and, consequently, the equilibrium point $(\hat x^{(1)},\mathbf{0},\mathbf{0})$ is not locally exponentially stable. \\
Note that if $s\Big((-D^{(1)}-D^{(2)}+\epsilon
\hat{X}^{(1)} B^{(2)})-(\epsilon \hat{B}^{(1)}+ \epsilon \hat{X}^{(1)}B^{(2)})   (-D^{(2)} + (I-\hat{X}^{(1)})B^{(2)}-\epsilon \hat{B}^{(1)})^{-1}((I-\hat{X}^{(1)})B^{(2)} + D^{(1)})\Big)  \geq 0$, then the matrix $\tilde{J}/\tilde{J}_{11}$ is not Hurwitz.
The rest of the proof is analogous to that of necessity of item i).~$\blacksquare$ \\

\noindent
\textbf{Proof of Corollary~\ref{cor:instab:hatx1}:} Assume that $s(-D^1-D^2+\epsilon\hat{x}^{(1)}B^{(2)})>0$. Note that, from the proof of Theorem~\ref{thm:local:expo:stab:single-virus}, 
$(-D^1-D^2+\epsilon\hat{x}^{(1)}B^{(2)}) 
      <  \tilde{J}/\tilde{J}_{11}$,
where $\tilde{J}/\tilde{J}_{11}$ is as given in the proof of Theorem~\ref{thm:local:expo:stab:single-virus}. Since both $ (-D^1-D^2+\epsilon\hat{x}^{(1)}B^{(2)})$ and $\tilde{J}/\tilde{J}_{11}$ are Metzler matrices, due to the assumption that $s(-D^1-D^2+\epsilon\hat{x}^{(1)}B^{(2)})>0$, it follows from \citep[Theorem~2.1]{varga2009matrix} that $s(\tilde{J}/\tilde{J}_{11}) 
      > 0$.
Consequently, it follows from Lemma~\ref{lem:souza} that the matrix $\tilde{J}$ 
in~\eqref{jacob:hatx1:22submatrix} is not Hurwitz. Hence,  $s(J(\hat{x}^{(1)}, \mathbf{0}, \mathbf{0})) >0$, which further implies from \citep[Theorem~4.7]{khalil2002nonlinear} that the boundary equilibrium $(\hat{x}^{(1)}, \mathbf{0}, \mathbf{0})$ is unstable.~$\blacksquare$\\

\noindent
\textbf{Proof of Proposition~\ref{prop:no:coexistence}:} Suppose, to the contrary, that there exists an equilibrium of the form $(\hat{x}^{(1)}, \hat{x}^{(2)}, \textbf{0})$ with $\hat{x}^{(1)}, \hat{x}^{(2)} > \textbf{0}$ for system~\eqref{x1}-\eqref{z}. By assumption $(\hat{x}^{(1)}, \hat{x}^{(2)}, \textbf{0})$ is a non-zero equilibrium point. Hence, from Lemma~\ref{lem:endemic:equilibria:characteristics}, it follows that $\textbf{0} \ll \hat{x}^{(1)}$ and $\textbf{0} \ll \hat{x}^{(2)}$.
Since  $(\hat{x}^{(1)}, \hat{x}^{(2)}, \textbf{0})$ is an equilibrium point, the equilibrium version of equation~\eqref{z} reads as follows:
\begin{equation}\label{eq:nocoexistence2}
    \mathbf{0} = \epsilon^{(1)} \hat{x}^{(1)}B^{(2)}\hat{x}^{(2)} + \epsilon^{(2)} \hat{x}^{(2)}B^{(1)}\hat{x}^{(1)}.
\end{equation}
Note that, by assumption, $\epsilon^{(m)}>0$ for some $m \in [2]$. By Assumption~\ref{para}, the matrices $B^{(1)}$ and $B^{(2)}$ are nonnegtaive irreducible, thus, since $\hat{x}^{(1)}$ and $\hat{x}^{(2)}$ are strictly positive vectors, implying that either  $\epsilon^{(1)} \hat{x}^{(1)}B^{(2)}\hat{x}^{(2)}> \textbf{0}$ 
or $\epsilon^{(2)} \hat{x}^{(2)}B^{(1)}\hat{x}^{(1)}> \textbf{0}$. As a consequence, $\epsilon^{(1)} \hat{x}^{(1)}B^{(2)}\hat{x}^{(2)} + \epsilon^{(2)} \hat{x}^{(2)}B^{(1)}\hat{x}^{(1)}>0$, which contradicts~\eqref{eq:nocoexistence2}. Therefore, there does not exist a coexisting equilibrium of the form $(\hat{x}^{(1)}, \hat{x}^{(2)}, \textbf{0})$ with $\hat{x}^{(1)}, \hat{x}^{(2)} > \textbf{0}$.\\

\noindent
\textbf{Proof of Proposition~\ref{prop:onlyhatz:notpossible}} Suppose that, by way of contradiction, system~\eqref{x1}-\eqref{z} has an equilibrium of the form $(\textbf{0}, \textbf{0}, \hat{z})$, where $\hat{z}>\textbf{0}$. 
 Then, from \eqref{x1}-\eqref{z}, and,
from the definition of equilibrium, we have the following: 
\begin{align}
  \mathbf{0} &= -(D^1+D^2)\hat{z}. \label{onlyzhat:3} 
\end{align}
Note that by Assumption~\ref{para-1}, $D^1$ and $D^2$ are positive diagonal matrices, thus implying that all eigenvalues of $-(D^1+D^2)$ are strictly negative, and therefore non-zero. Hence, the matrix $-(D^1+D^2)$ has full rank, which further implies that~\eqref{onlyzhat:3} is 
satisfied only if $\hat{z} =\textbf{0}$. By assumption, $\hat{z} >\textbf{0}$, and hence we have a contradiction. Therefore, there does not exist an equilibrium of the form $(\textbf{0}, \textbf{0}, \hat{z})$, where $\hat{z}>\textbf{0}$.\\

\noindent
\textbf{Proof of Proposition~\ref{prop:nece:condn:nox2}:} Suppose, by way of contradiction, that  $\rho((D^{(1)})^{-1}(B^{(2)}))~<~1$ and yet there exists an equilibrium point $(\hat{x}^{(1)}, \mathbf{0}, \hat{z})$.

Note that $B^{(2)}$ is nonnegative. Therefore, by noting that $\epsilon \in (0,1)$, and $\mathbf{0} \ll \hat{x}^{(1)} \ll \mathbf{1}$, it is clear that $B^{(2)} > \epsilon \hat{X}^{(1)}B^{(2)}$. Since $B^{(2)} > \epsilon \hat{X}^{(1)}B^{(2)}$, and since $(D^{(1)}+D^{(2)})^{-1}$ is a positive diagonal matrix, it follows that \vspace{-2mm }
\begin{align}\label{ineq:prop3:clearer:expln:1} 
    (D^{(1)}+D^{(2)})^{-1}B^{(2)} > (D^{(1)}+D^{(2)})^{-1}\epsilon \hat{X}^{(1)}B^{(2)}.
\end{align}
\normalsize  
By Assumption~\ref{para-1}, 
$D^{(1)}$ and $D^{(2)}$ are positive diagonal, thus implying that $(D^{(1)})^{-1}$ and $(D^{(1)}+D^{(2)})^{-1}$ exist. Note further that Assumption~\ref{para-1} also implies that $(D^{(1)})^{-1}> (D^{(1)}+D^{(2)})^{-1}$. Hence, we obtain \vspace{-2mm} 
\begin{align}\label{ineq:prop3:clearer:expln:2} 
    (D^{(1)})^{-1}B^{(2)} >(D^{(1)}+D^{(2)})^{-1}B^{(2)}.
\end{align}
\normalsize
From inequalities~\eqref{ineq:prop3:clearer:expln:1} and~\eqref{ineq:prop3:clearer:expln:2}, it is immediate that \vspace{-2mm} 
\begin{align}\label{ineq:prop3:clearer:expln:3} 
    (D^{(1)})^{-1}B^{(2)} >(D^{(1)}+D^{(2)})^{-1}\epsilon \hat{X}^{(1)}B^{(2)}.
\end{align}
\normalsize
Since $(D^{(1)})^{-1}B^{(2)}$ and $(D^{(1)}+D^{(2)})^{-1}\epsilon \hat{X}^{(1)}B^{(2)}$ are irreducible nonnegative matrices, it follows, from Lemma~\ref{lem:perron_frob}, that \vspace{-2mm} 
\begin{align} \label{ineq:nox2} 
    \rho((D^{(1)})^{-1}B^{(2)}) > \rho((D^{(1)}+D^{(2)})^{-1}\epsilon \hat{X}^{(1)}B^{(2)}).
\end{align}
\normalsize
The assumption $\rho((D^{(1)})^{-1}(B^{(2)})) < 1$ and \eqref{ineq:nox2} imply that \vspace{-2mm} 
\begin{align}\label{ineq:nohatx2:1}
   \rho((D^{(1)}+D^{(2)})^{-1}\epsilon \hat{X}^{(1)}B^{(2)}) <1. 
\end{align}
\normalsize
Since $(\hat{x}^{(1)}, \mathbf{0}, \hat{z})$ is an equilibrium of system~\eqref{x1}-\eqref{z}, and since $\epsilon^{(1)} = \epsilon$ by assumption, from~\eqref{z}, and from the definition of an equilibrium, we have the following:  \vspace{-2mm} 
\begin{align}
 \mathbf{0} &= \big(-(D^{(1)}+D^{(2)}) + \epsilon \hat{X}^{(1)}B^{(2)}\big)\hat{z}. \label{x2zero:1} 
\end{align}
\normalsize
Since $\epsilon \in (0,1)$, and $\mathbf{0} \ll \hat{x}^{(1)} \ll \mathbf{1}$, and $B^{(2)}$ is nonnegative irreducible (by Assumption~\ref{para}), it follows that the matrix $\epsilon \hat{X}^{(1)}B^{(2)}$ is nonnegative irreducible. Since $-(D^{(1)}+D^{(2)})$ is a negative diagonal matrix, it follows that $-(D^{(1)}+D^{(2)}) + \epsilon \hat{X}^{(1)}B^{(2)}$ is an irreducible Metzler matrix. Consider~\eqref{x2zero:1}, and note that $\hat{z} \gg \mathbf{0}$. Hence, from Lemma~\ref{lem:perron_frob_metz}, it follows that $s(-(D^{(1)}+D^{(2)}) + \epsilon \hat{X}^{(1)}B^{(2)})=0$, and, consequently, from 
Lemma~\ref{lem:eigspec}, 
it is clear that \vspace{-2mm} 
\begin{align} \label{x2zero:key:equality}
\rho((D^{(1)}+D^{(2)})^{-1}\epsilon \hat{X}^{(1)}B^{(2)})=1. 
\end{align}
\normalsize
Note that~\eqref{x2zero:key:equality} contradicts~\eqref{ineq:nohatx2:1}, meaning that if we assume that $\rho((D^{(1)})^{-1}(B^{(2)}))<1$, then there cannot exist a coexisting equilibrium $(\hat{x}^{(1)}, \mathbf{0}, \hat{z})$,  which concludes the proof.~$\blacksquare$\\

\noindent
\textbf{Proof of Proposition~\ref{prop:nece:x1x2z}:}
In order to prove Proposition~\ref{prop:nece:x1x2z}, we need the following lemma.

\begin{lemma}\citep[Proposition~2]{rantzer}\label{lem:ratzner}
Let $A \in \mathbb{R}^{n\times n}$ be Metzler. Then, $A$ is Hurwitz if, and only if, there exists an $x \in \mathbb{R}^n$ such that $ x \gg \textbf{0}$ and $Ax \ll 0$.
\end{lemma}

\noindent
\textbf{Proof of Proposition~\ref{prop:nece:x1x2z}:} Suppose that by way of contradiction, 
there exists an equilibrium point $(\hat{x}^{(1)}, \hat{x}^{(2)}, \hat{z})$. yet $\rho((I-\hat{x}^{(1)}-\hat{x}^{(2)}-\hat{Z})D^{-1}B) \geq1$. 
Therefore, since by assumption $\epsilon^{(1)}=\epsilon^{(2)} = \epsilon$, the equilibrium version of~\eqref{x1}-\eqref{z} can be written as follows:
\begin{align}
 \mathbf{0} &= -D^1\hat{x}^{(1)} + D^2\hat{z}+ (I-\hat{x}^{(1)}-\hat{x}^{(2)}-\hat{Z})B^{(1)}(\hat{x}^{(1)}+\hat{z}) \nonumber \\&~~~~~~~~~~~~~~~~~~~~~~~~~~~~~~~~~ - \epsilon \hat{x}^{(1)}B^{(2)}(\hat{x}^{(2)}+\hat{z}) \label{x1x2zhat:1} \\
  \mathbf{0} &= -D^2\hat{x}^{(2)} + D^1\hat{z}+ (I-\hat{x}^{(1)}-\hat{x}^{(2)}-\hat{Z})B^{(2)}(\hat{x}^{(2)}+\hat{z})  \nonumber \\&~~~~~~~~~~~~~~~~~~~~~~~~~~~~~~~~- \epsilon \hat{x}^{(2)}B^{(1)}(\hat{x}^{(1)}+\hat{z}) \label{x1x2zhat:2} \\
  \mathbf{0} &=-(D^1+D^2)\hat{z}+ \epsilon \hat{x}^{(1)}B^{(2)}(\hat{x}^{(2)}+\hat{z}) + \epsilon \hat{x}^{(2)}B^{(1)}(\hat{x}^{(1)}+\hat{z}). \label{x1x2zhat:3} 
\end{align}
By Assumption~\ref{assum:identical:viruses}, $B^{(1)}=B^{(2)}$ and $D^{(1)}=D^{(2)}$. Hence, by summing equations~\eqref{x1x2zhat:1}-\eqref{x1x2zhat:3}, we obtain
\begin{align}\label{eq:sumofviruses}
    -D(\hat{x}^{(1)}&+\big(\hat{x}^{(2)})+(I-\hat{x}^{(1)}-\hat{x}^{(2)}-\hat{Z})B \nonumber\\ &~~(\hat{x}^{(1)}+\hat{x}^{(2)}+2\hat{z})\big) = \textbf{0}.
    \end{align}
   Note that, by Assumption~\ref{para-1}, $D$ is a positive diagonal matrix. Furthermore, since we know by Lemma~\ref{lem:endemic:equilibria:characteristics} that $\hat{z} \gg \textbf{0}$, it follows that $2D\hat{z} \gg \textbf{0}$. 
    Hence, we have the following:
    \begin{equation} \label{ineq:x1x2z}
        -D(\hat{x}^{(1)}+\hat{x}^{(2)}) \gg -D(\hat{x}^{(1)}+\hat{x}^{(2)}+2\hat{z}).
    \end{equation}
    Therefore, adding the term $(I-\hat{x}^{(1)}-\hat{x}^{(2)}-\hat{Z})B(\hat{x}^{(1)}+\hat{x}^{(2)}+2\hat{z})$ to both sides of~\eqref{ineq:x1x2z}, and recalling~\eqref{eq:sumofviruses}, yields:
    \begin{align}\label{ineq:sumofviruses:new}
    (-D+(I-\hat{x}^{(1)}-\hat{x}^{(2)}-\hat{Z})B)(\hat{x}^{(1)}+\hat{x}^{(2)}+2\hat{z}) 
    \ll
    \textbf{0}.
    \end{align}
We now show that $(-D+(I-\hat{x}^{(1)}-\hat{x}^{(2)}-\hat{Z})B)$ is irreducible and Metzler. To this end, observe that
since, by assumption,  $(\hat{x}^{(1)}, \hat{x}^{(2)}, \hat{z})$ is a nonzero equilibrium point, it follows from Lemma~\ref{lem:endemic:equilibria:characteristics}
that $\mathbf{0} \ll 
\hat{x}^{(1)}, \hat{x}^{(2)}, \hat{z}, \hat{x}^{(1)}+ \hat{x}^{(2)} + \hat{z}
\ll \mathbf{1}$. 
Hence, $(I-\hat{x}^{(1)}-\hat{x}^{(2)}-\hat{Z})$ is a positive diagonal matrix with every nonzero element in it being strictly less than one.   Therefore, since $B$ is nonnegative irreducible and $D$ is positive diagonal, it is clear that $(-D+(I-\hat{x}^{(1)}-\hat{x}^{(2)}-\hat{Z})B)$ is an irreducible Metzler matrix. Note that, by Lemma~\ref{lem:endemic:equilibria:characteristics}, 
$\hat{x}^{(1)}+\hat{x}^{(2)}+2\hat{z} \gg \textbf{0}.$
Hence, applying Lemma~\ref{lem:ratzner} to vector inequality~\eqref{ineq:sumofviruses:new} yields
    $s(-D+(I-\hat{x}^{(1)}-\hat{x}^{(2)}-\hat{Z})B)<0$, which from Lemma~\ref{lem:eigspec} further implies that $\rho((I-\hat{x}^{(1)}-\hat{x}^{(2)}-\hat{Z})D^{-1}B)<1$.
 However, notice that by assumption, $\rho((I-\hat{x}^{(1)}-\hat{x}^{(2)}-\hat{Z})D^{-1}B) \geq 1$, which leads to a contradiction. Hence, if $(\hat{x}^{(1)}, \hat{x}^{(2)}, \hat{z})$ is an equilibrium point then $\rho((I-\hat{x}^{(1)}-\hat{x}^{(2)}-\hat{Z})D^{-1}B)<1$, thus completing the proof. \hfill \qed


\end{document}